\documentclass[10pt,letterpaper,oneside]{book}

\usepackage{amsmath}      
\usepackage{amssymb}      
\usepackage{amsfonts}     
\usepackage{amscd}        
\usepackage{graphicx}     
\usepackage{subfigure}    
\usepackage{longtable}    
\usepackage{url}          
\usepackage{bm}           
\usepackage{setspace}     

\newcommand{\be}{\begin{eqnarray}}
\newcommand{\ee}{\end{eqnarray}}
\newcommand*{\half}{\frac{1}{2}}
\newcommand*{\bra}[1]{\langle #1 |}
\newcommand*{\ket}[1]{| #1 \rangle}
\newcommand*{\braket}[2]{\langle #1 | #2 \rangle}
\newcommand*{\ketbra}[2]{| #1 \rangle\langle #2 |}

\DeclareMathOperator{\sech}{sech}

\begin{document}

\title{ {\normalsize COHERENT MANIPULATION OF MULTILEVEL ATOMS\\
                     FOR QUANTUM INFORMATION PROCESSING}}
\author{}
\date{December 2005}
\maketitle

\frontmatter
\begin{center}
\thispagestyle{empty}
\vspace*{2cm}

COHERENT MANIPULATION OF MULTILEVEL ATOMS\\
FOR QUANTUM INFORMATION PROCESSING
\vspace*{3cm}

A dissertation submitted in partial fulfillment\\
of the requirements for the degree of\\
Doctor of Philosophy\\
\vspace*{3cm}

By
\vspace*{3cm}

JUAN DAVID SERNA, B.S., M.S.\\
Universidad de Antioquia, 1997\\
University of Arkansas, 2004
\vspace*{3cm}

December 2005\\
University of Arkansas
\end{center}

\newpage
\doublespacing
\thispagestyle{empty}

In quantum information processing, quantum cavities play an important role by
providing the mechanisms to transfer information between atom qubits and photon
qubits, or to couple single atoms with the optical modes of the cavity field. We
explore numerically the population transfer in an atom + cavity system by using
the $\pi$-pulse and adiabatic passage methods. While the first method is very
efficient transferring the atomic population for no radiative decay of the
intermediate level, the second method shows very interesting nonadiabatic,
resonance-like properties that can be used to achieve very large transfer
efficiencies without needing very large Rabi frequencies or very long
interaction times. We introduce a simple analytical model to explore the origin
of these properties and describe ``qualitatively'' the power-law dependence of
the failure probability on the product of the pulse amplitude and the
interaction time. We also examine numerically the transfer of interatomic
coherence in a two-atom + cavity system by using adiabatic methods. For some
specific symmetry conditions, we show that the dynamics of the original system
can be studied as the individual evolution of a symmetric and an antisymmetric
system, interacting separately with the classical field and the cavity mode, but
mutually exchanging the atomic coherence.

\newpage
\singlespace
\thispagestyle{empty}

\vspace*{0.5in}
\noindent This dissertation is approved for\\
Recommendation to the\\
Graduate Council\\
\vspace{1.0cm}

\noindent
\begin{tabular}{ @{}l p{3cm}}
\noindent Dissertation Director: & \\[1.5cm]

\cline{1-2}
Dr. Julio Gea-Banacloche  & \\[2cm]

Dissertation Committee:   & \\[1.5cm]

\cline{1-2}
Dr. Luca Capogna    & \\[1.5cm]
\cline{1-2}
Dr. Michael Lieber  & \\[1.5cm]
\cline{1-2}
Dr. Surendra Singh  & \\[1.5cm]
\cline{1-2}
Dr. Min Xiao        &
\vspace{1cm}
\end{tabular}

\doublespacing
\tableofcontents

\mainmatter
\chapter{\label{Ch:01}Fundamental Concepts}

\section{Introduction}

Population transfer in atoms and molecules is one of the most intriguing
problems of quantum optics. To carry out successful multilevel excitations is an
important task for theoretical and practical purposes. Thus, figuring out
different ways of improving the efficiency of those methods used to transfer
population between multilevel systems has become an important subject of
research during the past years.

In this work, we examine numerically an atom + cavity system with two of the
most important methods for transferring population between atomic levels: the
$\pi$-pulses method and the adiabatic passage scheme. In particular, we explore
the possibility of using the nonadiabatic characteristics of the second method
to achieve very high transfer efficiencies without using large values of the
pulse amplitudes or interaction times.

The $\pi$-pulse method takes advantage of the Rabi population oscillations that
characterize coherent excitation. By adjusting the laser intensity and the pulse
duration so that the time integral of the Rabi frequency (the pulse area) has
the value $\pi$, it is possible to produce complete population transfer between
two states~\cite{Shore:PRA45}. If the system undergoes no spontaneous emission,
transfer efficiencies of $100$\% are possible to obtain with this method.
However, a very accurate control of the pulse area is required, which is a
really disadvantage.

The adiabatic passage scheme is an alternative for transferring population
between selected quantum states. The population can be transferred between two
states by driving the system sufficiently slowly with the appropriate external
fields, with the population remaining in an approximate energy eigenstate. This
method is quite insensitive to changes in parameters like the laser pulse
shapes, intensity, and frequency modulation, as long as certain easily
controllable experimental conditions are satisfied. It is important to note that
the adiabatic passage method tends to minimize the population of the
intermediate level 2 in a three level system. Normally this level undergoes
radiative decay. Therefore, the effects of spontaneous emission are largely
reduced. For the adiabatic following solution to be valid, the detuning from the
two photon resonance should be small compared to the Rabi frequency $\Omega_0$.

The dissertation opens with a mathematical and physical introduction of two- and
three-level systems. Concepts like Bloch equations, rotating-wave approximation,
the $\pi$-pulse method, and the adiabatic following are introduced here. This
chapter is based mainly in the Allen and Eberly~\cite{Allen:Book}, and Scully
and Zubairy~\cite{Scully:book} books.

In the second chapter, we explore numerically an atom + cavity system with four
atomic levels and a quantized coupling field. Adiabatic passage and $\pi$-pulse
methods were used to drive the system classically. We find that for a very
particular set of parameters, like Rabi frequency, pulse width, and time delay,
we get very large transfer probabilities.

In chapter three, we introduce a simple analytical model based in nonlinear
differential equations, that can help to understand why for some nonadiabatic
processes we still achieve very high transfer efficiencies. We found that the
nonlinear system can be converted into a system of equations relating the
angular coordinates of the state vector in the Hilbert space. In this way, we
can explore how the state vector follows the adiabatic states of the Hamiltonian
governing the evolution of the system.

In the last chapter we consider the problem of coherence transfer of ground
state levels, between two atoms inside a quantum microcavity. We proved that
adiabatic methods can be used for transferring such coherence. In addition, we
examined the system by using an alternative model based on symmetric and
antisymmetric eigenstates, finding a connection between the original system and
electromagnetic induced transparency and $2\pi$-processes.

\section{\label{2levelAtom}The two-level atom}

Consider the solution to the Schr\"{o}dinger equation for a two-level atom
interacting with a classical coherent driving field. The state of the system is
described in terms of the vector $\ket{\Psi(t)}$, which obeys the time-dependent
Schr\"{o}dinger equation
\begin{equation}
    i \hbar \frac{d}{dt} \ket{\Psi(t)} = H \ket{\Psi(t)},
    \label{Eq:Schrodinger01}
\end{equation}
with the Hamiltonian operator $H$ given by
\begin{equation}
    H = H_0 - \boldsymbol{\mu} \cdot \mathbf{E}(\mathbf{r}_0,t).
    \label{Eq:Hamiltonian1}
\end{equation}
Here $H_0$ is the unperturbed Hamiltonian, $\boldsymbol{\mu}$ is the atom's
dipole moment operator, and $\mathbf{E}(\mathbf{r}_0,t)$ is the electric field
operator evaluated at the position of the dipole.

\begin{figure}[h]
    \centering
    \includegraphics[scale=0.7]{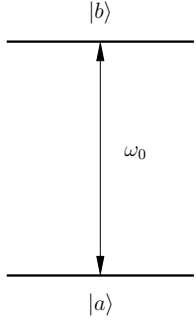}
    \caption{The two-level atom coupled by a near-resonant transition.}
    \label{Fig:2levelAtom}
\end{figure}

We assume that the applied electric field is quasi-monochromatic with a
frequency nearly coincident with the transition frequency connecting the atomic
ground state $\ket{a}$ and some other level $\ket{b}$, as shown in
Fig.~\ref{Fig:2levelAtom}.

\noindent Because of the effect of the interaction is to mix states $\ket{a}$
and $\ket{b}$, we are only concerned with electric dipole transitions between
these two levels. Thus the state vector of the system in the presence of the
applied field can be written as
\begin{equation}
    \ket{\Psi(t)} = C_a(t) e^{-i\omega_a t}\ket{a} +
                    C_b(t) e^{-i\omega_b t}\ket{b},
    \label{Eq:psi2(t)}
\end{equation}
where $C_a(t)$ and $C_b(t)$ represent the probability amplitudes that at time
$t$ the atom is in state $\ket{a}$ or state $\ket{b}$, respectively. Assuming
that these states are eigenvectors of the Hamiltonian $H_0$ with eigenvalues
$\hbar\omega_a$ and $\hbar\omega_b$, the matrix elements of the atomic
operators can be written as
\begin{alignat}{2}
    \bra{a} H_0 \ket{a} & = \hbar\omega_a
        & \qquad \bra{a} H_0 \ket{b} & = 0 \notag \\
    \bra{b} H_0 \ket{a} & = 0
        & \qquad \bra{b} H_0 \ket{b} & = \hbar\omega_b,\\
\intertext{and}
    \bra{a} \boldsymbol{\mu} \ket{a} & = 0 & \qquad
        \bra{a} \boldsymbol{\mu} \ket{b} & = \boldsymbol{\mu}_{ab} \notag \\
    \bra{b} \boldsymbol{\mu} \ket{a} & = \boldsymbol{\mu}_{ab}^* & \qquad
    \bra{b} \boldsymbol{\mu} \ket{b} & = 0.
\end{alignat}
We note that there are no off-diagonal elements of $H_0$, because $\ket{a}$ and
$\ket{b}$ are considered to be \textit{orthonormal} eigenvectors of $H_0$; and
there are no diagonal elements of $\boldsymbol{\mu}$, because it is a
\textit{vector operator} and then has odd parity (we assume that $\ket{a}$ and
$\ket{b}$ have definite parity).

In general, the dipole matrix elements are complex vectors that can be expressed
as
\begin{equation}
    \boldsymbol{\mu}_{ab}   = \boldsymbol{\mu}_r + i \boldsymbol{\mu}_i, \qquad
    \boldsymbol{\mu}_{ab}^* = \boldsymbol{\mu}_r - i \boldsymbol{\mu}_i;
    \label{Eq:complex_dipole1}
\end{equation}
where $\boldsymbol{\mu}_r$ and $\boldsymbol{\mu}_i$ are real vectors. Therefore,
we can represent the \textit{Hermitian} operator $\boldsymbol{\mu}$ by the
two-dimensional off-diagonal matrix:
\begin{equation}
    \boldsymbol{\mu} = \begin{bmatrix}
                           0 & \boldsymbol{\mu}_r + i \boldsymbol{\mu}_i \\
                           \boldsymbol{\mu}_r - i \boldsymbol{\mu}_i & 0 \\
                       \end{bmatrix}.
    \label{Eq:dipoleMatrix}
\end{equation}
By introducing the two-dimensional Pauli matrix operators
\begin{equation}
    \sigma_1 =
    \begin{pmatrix}
        0 & 1 \\ 1 & 0
    \end{pmatrix}, \qquad
    \sigma_2 =
    \begin{pmatrix}
        0 & -i \\ i & 0
    \end{pmatrix}, \qquad
    \sigma_3 =
    \begin{pmatrix}
        1 & 0 \\ 0 & -1
    \end{pmatrix},
    \label{Eq:PauliMatrices}
\end{equation}
the unperturbed Hamiltonian and the atom's dipole moment operators can be
written as
\begin{align}
    H_0 & = \half (\omega_b + \omega_a) I +
        \half(\omega_b-\omega_a) \sigma_3,
    \label{Eq:H0_operator1} \\
    \boldsymbol{\mu} & = \boldsymbol{\mu}_r \sigma_1
        - \boldsymbol{\mu}_i \sigma_2.
    \label{Eq:dipoleOperator}
\end{align}
In this way, the Hamiltonian of the system takes the form
\begin{equation}
    H = \half (\omega_b+\omega_a)I + \half (\omega_b-\omega_a)\sigma_3
        - (\boldsymbol{\mu}_r\cdot\mathbf{E})\sigma_1
        + (\boldsymbol{\mu}_i\cdot\mathbf{E})\sigma_2.
    \label{Eq:H_operator}
\end{equation}
Here $I$ represents the $2 \times 2$ identity operator.

In the Heisenberg picture, the equation of motion for an operator that does not
depend explicitly on time is given by
\begin{equation}
    i \hbar \frac{d}{dt} A = [A,H].
    \label{Eq:Heisenberg}
\end{equation}
Then, by plugging the Pauli matrix operators and Eq.~\eqref{Eq:H_operator} into
the Heisenberg equation~\eqref{Eq:Heisenberg}, we may obtain
\begin{subequations}
\label{Eq:PauliMatrices_dots}
\begin{align}
    \dot{\sigma_1}(t) & = -\omega_0 \sigma_2(t) + \frac{2}{\hbar}
        \bigl[\boldsymbol{\mu}_i\cdot\mathbf{E}(t)\bigr] \sigma_3(t),
    \label{Eq:sigma1_dot} \\
    \dot{\sigma_2}(t) & =  \omega_0 \sigma_1(t) + \frac{2}{\hbar}
        \bigl[\boldsymbol{\mu}_r\cdot\mathbf{E}(t)\bigr] \sigma_3(t),
    \label{Eq:sigma2_dot} \\
    \dot{\sigma_3}(t) & = -\frac{2}{\hbar} \bigl[
        \boldsymbol{\mu}_r\cdot\mathbf{E}(t) \bigr] \sigma_2(t)
                          -\frac{2}{\hbar} \bigl[
        \boldsymbol{\mu}_i\cdot\mathbf{E}(t) \bigr] \sigma_1(t),
    \label{Eq:sigma3_dot}
\end{align}
\end{subequations}
where $\omega_0 = (\omega_b - \omega_a)/\hbar$ represents the atomic transition
frequency, and $\mathbf{E}(t)$ has been taken in the Heisenberg picture.

As we see from Eqs.~\eqref{Eq:PauliMatrices_dots}, the operator nature of the
atom and field variables makes the system very difficult to solve and no general
solutions are known. Moreover, if the operator Maxwell equations that govern the
electric field are included, the system becomes even more complicated. One way
to overcome this problem is by considering an alternative system of equations
for which the quantum correlations between field and atom can safely be
ignored~\cite{Allen:Book}. As a result, the expectation value of any product of
two operators of the form $\mathbf{E}(t)\sigma_i(t)$ can be expressed as the
product of the individual expectation values:
\begin{equation}
    \langle \mathbf{E}(t) \sigma_i(t) \rangle = \langle \mathbf{E}(t) \rangle
        \langle \sigma_i(t) \rangle.
\end{equation}
In this way, we can reformulate the semi-classical radiation theory of two-level
atoms by applying this factorization systematically to
Eqs.~\eqref{Eq:PauliMatrices_dots}. Now, according to the following notation
\begin{align}
    s_i(t) & \equiv \langle \sigma_i(t) \rangle, \qquad i=1,2,3
    \label{Eq:si_exp}\\
    \overline{\mathbf{E}}(t) & \equiv \langle \mathbf{E}(t) \rangle,
    \label{Eq:E_exp}
\end{align}
the set of equations~\eqref{Eq:PauliMatrices_dots}, that represents the general
interaction of a two-level atom with an electric field in the semiclassical
theory, takes the form
\begin{subequations}
\label{Eq:PauliMatrices_exp}
\begin{align}
    \dot{s_1}(t) & = -\omega_0 s_2(t) + \frac{2}{\hbar}
        \bigl[\boldsymbol{\mu}_i\cdot\overline{\mathbf{E}}(t)\bigr] s_3(t),
    \label{Eq:sigma1_exp} \\
    \dot{s_2}(t) & =  \omega_0 s_1(t) + \frac{2}{\hbar}
        \bigl[\boldsymbol{\mu}_r\cdot\overline{\mathbf{E}}(t)\bigr] s_3(t),
    \label{Eq:sigma2_exp} \\
    \dot{s_3}(t) & = -\frac{2}{\hbar} \bigl[
        \boldsymbol{\mu}_r\cdot\overline{\mathbf{E}}(t) \bigr] s_2(t)
                          -\frac{2}{\hbar} \bigl[
        \boldsymbol{\mu}_i\cdot\overline{\mathbf{E}}(t) \bigr] s_1(t).
    \label{Eq:sigma3_exp}
\end{align}
\end{subequations}

It is well known that the dynamical evolution of an $N$-level quantum system can
be described by the rotations of a coherent
vector~\cite{Rabi:RMP26,Feynman:JAP57} which is constrained by the existence of
high-order constants of motion~\cite{Elgin:PL80A,Hioe:PRL81,Hioe:PRA82}. For the
two-dimensional Hilbert space, some of these conservation laws come directly
from the properties of the Pauli matrix operators, reflecting the intrinsic
symmetry of the vector space. For example $\sigma_1^2 = \sigma_2^2 = \sigma_3^2
= I$. Another important constant of motion is derived from the fact that
\begin{equation}
    \biggl(\frac{d}{dt}\biggr) \sigma_1(t)^2 = 0, \qquad
        \text{so that} \quad \sigma_1^2(t) = \sigma_1^2(0) = I.
\end{equation}
Extending all these ideas to the system of
equations~\eqref{Eq:PauliMatrices_exp}, we may obtain their associated
conservation law, which is written as
\begin{equation}
    s_1^2(t) + s_2^2(t) + s_3^2(t) = 1.
\end{equation}
This expression means that the probability of the system is conserved over the
time, or equivalently, the state of the atom remains normalized in time.

For electric dipole $\Delta m = 0$ transitions, we can adjust the arbitrary
phases of the connected levels so that $\boldsymbol{\mu}_i$ vanishes. By
denoting
\begin{equation}
    \frac{2}{\hbar} \,\boldsymbol{\mu}_r \cdot \overline{\mathbf{E}}
        = \frac{2\mu E}{\hbar}
        = \Omega,
\end{equation}
we can write the semiclassical atomic equations~\eqref{Eq:PauliMatrices_exp} in
the simpler form:
\begin{subequations}
\label{Eq:PauliMatrices_reduced}
\begin{align}
    \dot{s}_1(t) & = - \omega_0 s_2(t), \\
    \dot{s}_2(t) & =   \omega_0 s_1(t) + \Omega(t) s_3(t), \\
    \dot{s}_3(t) & = - \Omega(t) s_2(t).
\end{align}
\end{subequations}
Because these equations are the electric-dipole analogues of equations of a
spin-1/2 magnetic dipole undergoing precession in a magnetic
field~\cite{Bloch:PR70}, the vector $\mathbf{s}(t)$ is called the
electric-dipole ``pseudospin.''

The physical meaning of the expectation values $s_1(t)$, $s_2(t)$, and $s_3(t)$
can be interpreted as follows. From Eqs.~\eqref{Eq:si_exp}
and~\eqref{Eq:H0_operator1}, it is clear that $s_3(t)$ represents the internal
energy of the atom in $\half\hbar\omega_0$ units; and from
Eqs.~\eqref{Eq:si_exp} and~\eqref{Eq:dipoleOperator}, we see that $s_1(t)$ and
$s_2(t)$ are both manifestations of the atom's dipole moment
operator~\cite{Feynman:JAP57,Allen:Book}.

The pseudospin Eqs.~\eqref{Eq:PauliMatrices_reduced} can be rewritten as if they
were the equations for the precession of a solid body upon which a known torque
$\mathbf{N}^F$ is acting. The superscript $F$ stands for the coordinate system
of fixed unit vectors $\ket{1}$, $\ket{2}$, and $\ket{3}$. Thus the set of three
equations~\eqref{Eq:PauliMatrices_reduced} can be expressed as the single
equation:
\begin{equation}
    \dot{\mathbf{s}}(t) = \mathbf{N}^F(t) \times \mathbf{s}(t),
    \label{Eq:Pseudospin}
\end{equation}
where the vector $\mathbf{s}$ has components $s_1$, $s_2$, $s_3$, and the torque
vector $\mathbf{N}^F$ has components
\begin{subequations}
\label{Eq:torqueComponents}
\begin{align}
    \mathbf{N}_1^F(t) & = -\Omega(t), \\
    \mathbf{N}_2^F(t) & = 0, \\
    \mathbf{N}_3^F(t) & = \omega_0.
\end{align}
\end{subequations}
We note that the pseudospin precession is originated by the first and third
components of the torque vector.

To simplify the mathematics, we define a coordinate reference frame which
rotates at the same frequency $\omega$ of the field. In this way we reduce the
number of rapidly oscillating variables of the system and consider only those
which change slowly with time. The torque vector is then rewritten as the sum of
three torques, one $\mathbf{N}^0$ along the $\ket{3}$ direction, and two much
smaller torques that lie completely in the $\ket{1}$--$\ket{2}$ plane:
\begin{equation}
    \mathbf{N}^F = \mathbf{N}^+(t) + \mathbf{N}^-(t) + \mathbf{N}^0,
    \label{Eq:torqueExpansion}
\end{equation}
where
\begin{subequations}
\label{Eq:torqueVectors}
\begin{align}
    \mathbf{N}^0 &= \bigl( 0, \,0, \,\omega_0 \bigr), \\
    \mathbf{N}^- &= \bigl( -\Omega\,\cos\omega t,
                        \,-\Omega\,\sin\omega t,
                        \,0 \bigr), \\
    \mathbf{N}^+ &= \bigl( -\Omega\,\cos\omega t,
                        \,+\Omega\,\sin\omega t,
                        \,0 \bigr),
\end{align}
\end{subequations}
and
\begin{equation}
    E(t) = \mathcal{E}(t) \bigl[ e^{i \omega t} + \text{c.c.} \bigr].
    \label{Eq:Efield}
\end{equation}
As we see, $\mathbf{N}^+$ rotates counterclockwise as $t$ increases, while
$\mathbf{N}^-$ rotates clockwise. In a coordinate system following $\mathbf{s}$
and moving to the right at angular velocity $\omega$, the vector $\mathbf{N}^+$
remains constant, and $\mathbf{N}^-$ is counter-rotating at angular velocity
$2\omega$. In such a coordinate frame the effect of the torque $\mathbf{N}^+$ on
a spin is steady and cumulative over long times. On the other hand, the effect
of the torque $\mathbf{N}^-$ reverses itself $10^{15}$--$10^{16}$ times/s, and
is almost completely ineffective~\cite{Shirley:JAP34}.

The \textit{rotating-wave approximation} (RWA) consists of ignoring
$\mathbf{N}^-$ for this reason, and writing the pseudospin equations using
$\mathbf{N}^+$ and $\mathbf{N}^-$ in place of
$\mathbf{N}^F$~\cite{Bloch:PR57,Einwohner:PRA14}. It then follows that
\begin{subequations}
\label{Eq:PseudospinRWA}
\begin{align}
    \dot{s}_1 & = -\omega_0 s_2 - \Omega s_3\sin\omega t, \\
    \dot{s}_2 & =  \omega_0 s_1 + \Omega s_3\cos\omega t, \\
    \dot{s}_3 & = -\Omega
        \bigl[ s_2\cos\omega t - s_1\sin\omega t \bigr].
\end{align}
\end{subequations}

By introducing an appropriate rotation matrix for the vector $\mathbf{s}$ and
defining a nearly stationary vector $\boldsymbol{\rho}$ in the rotating frame
with components $u$, $v$, and $w$, it is possible to determine what the observer
sees in the rotating frame. Thus
\begin{equation}
    \begin{bmatrix}
        u \\ v \\ w
    \end{bmatrix}
    =
    \begin{bmatrix}
         \cos\omega t & \sin\omega t & 0 \\
        -\sin\omega t & \cos\omega t & 0 \\
              0       &      0       & 1
    \end{bmatrix}
\end{equation}

The equations of motion obeyed by the components of the pseudospin
$\boldsymbol{\rho}$ in the rotating frame are
\begin{subequations}
\label{Eq:uvwSystem}
\begin{align}
    \dot{u} & = -(\omega_0 - \omega) v, \\
    \dot{v} & = +(\omega_0 - \omega) u + \Omega w, \\
    \dot{w} & = -\Omega v,
\end{align}
\end{subequations}
which are the same as the single vector equation
\begin{equation}
    \frac{d}{dt} \,\boldsymbol{\rho} = \mathbf{N} \times \boldsymbol{\rho},
    \label{Eq:rho1}
\end{equation}
if the rotating frame torque vector $\mathbf{N}$ has the components
\begin{equation}
    \mathbf{N} \equiv (-\Omega, \,0, \,\omega_0-\omega).
\end{equation}
%

\section{\label{pi-pulses} $\pi$-pulses}

By defining a ``dimensionless'' quantity $\theta(t)$ as
\begin{equation}
    \theta(t) = \int_{-\infty}^t \Omega(t') \,dt',
    \label{Eq:theta(t)}
\end{equation}
equations~\eqref{Eq:uvwSystem} can be integrated to give
\begin{subequations}
\label{Eq:uvwSolutions}
\begin{align}
    u(t;0) & = u_0,                                    \label{Eq:u(0)} \\
    v(t;0) & = w_0 \sin\theta(t) + v_0 \cos\theta(t),  \label{Eq:v(0)} \\
    w(t;0) & = -v_0 \sin\theta(t) + w_0 \cos\theta(t), \label{Eq:w(0)}
\end{align}
\end{subequations}
where $u_0 = u(0;0)$, and so on. The second zero in the labels $v(0;0)$ and
$w(0;0)$ makes reference to the detuning frequency $\Delta = \omega_0 - \omega$.

In the special case when the applied field envelope has a steady state value
between $t_1$ and $t_2$, Eq.~\eqref{Eq:theta(t)} can be integrated to give
\begin{equation}
    \theta = \Omega_0 (t_2 - t_1),
    \label{Eq:thetaConstant}
\end{equation}
where $\Omega_0$ is called the Rabi frequency on resonance.

The Rabi frequency gives the rate at which transitions are coherently induced
between the two atomic levels. If the atom is initially in the ground state
($w_0=-1$, $v_0=0$) then after a time $\delta t$ such that $\Omega_0\,\delta t =
\pi$, Eq.~\eqref{Eq:w(0)} shows that $w=+1$, and the atom is in its upper state.
In other words, a ``$\pi$ pulse'' of electromagnetic radiation inverts the atom
population from the ground state to the excited state. Now, in the spins
terminology, the $\pi$ pulse turns a spin from alignment to anti-alignment with
a static magnetic field.

The quantity $\Omega_0 (t_2 - t_1)$ is exactly the area under the curve pulse
amplitude-time, and represents the well-known ``area theorem'' written as
\begin{equation}
    A(t) = \int_{-\infty}^t \Omega(t') \,dt' = \theta(t).
    \label{Eq:AreaTheorem}
\end{equation}
Resonant pulses with areas $\pi$, $2\pi$, $3\pi$, and so on, invert the atomic
population one, two, three, and so on, times.

We can also note that solutions~\eqref{Eq:uvwSolutions} are the result of a
rotation. If the rotating frame Eqs.~\eqref{Eq:uvwSystem} are written also as a
single vector precession equation
\begin{equation}
    \frac{d}{dt} \boldsymbol{\rho} = \mathbf{N} \times
        \boldsymbol{\rho}
    \label{Eq:rho2}
\end{equation}
where
\begin{equation}
    \boldsymbol{\rho} = (u,v,w),
\end{equation}
and
\begin{equation}
    \mathbf{N} = (-\Omega,0,\Delta),
\end{equation}
then Eq.~\eqref{Eq:rho2} can be represented as
\begin{equation}
    \frac{d}{dt}
    \begin{bmatrix}
        u \\ v \\ w
    \end{bmatrix}
    =
    \begin{bmatrix}
           0   &      -\Delta       &    0   \\
        \Delta &          0         & \Omega \\
           0   &      -\Omega       &    0
    \end{bmatrix}
    \begin{bmatrix}
        u \\ v \\ w
    \end{bmatrix}.
\end{equation}
%

\section{\label{Sec:3levelAtom} The three-level atom}

\begin{figure}[h]
    \centering
    \includegraphics[scale=0.8]{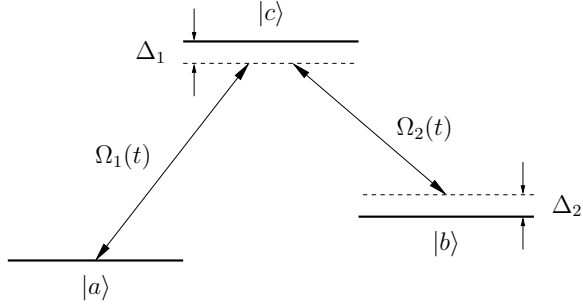}
    \caption{The three-level atom in a $\Lambda$-configuration.}
    \label{Fig:3levelAtom}
\end{figure}

Consider the interaction of a two-mode radiation field with the three-level atom
shown schematically in Fig.~\ref{Fig:3levelAtom}. Let $\ket{a}$, $\ket{c}$, and
$\ket{b}$ represent the initial, excited, and final states of the atom in a
$\Lambda$ configuration. They are eigenstates of the unperturbed part of the
Hamiltonian $H_0$ with the eigenvalues $\hbar\omega_a$, $\hbar\omega_c$, and
$\hbar\omega_b$, respectively ($\omega_a<\omega_b<\omega_c$).

\noindent We are only interested in electric dipole transitions between these
three levels, so we can concern ourselves exclusively with the three-dimensional
Hilbert space spanned by these eigenstates. Then, the state vector of the
system (in the Sch\"{o}dinger picture) can be written as
\begin{align}
    \ket{\psi(t)} = C_a(t) e^{-i\omega_a t}\ket{a} +
                    C_b(t) e^{-i\omega_b t}\ket{b} +
                    C_c(t) e^{-i\omega_c t}\ket{c},
    \label{Eq:psi3(t)}
\end{align}
where $C_a$, $C_b$, and $C_c$ are the \textit{slowly} varying amplitudes of
finding the atom in states $\ket{a}$, $\ket{b}$, and $\ket{c}$, respectively.
The corresponding time-dependent Schr\"{o}dinger equation is
\begin{align}
    i \hbar \,\frac{d}{dt} \ket{\psi(t)} & = H(t) \ket{\psi(t)},
    \label{Eq:Schrodinger2(t)}
\end{align}
with
\begin{align}
    H(t) & = H_0 + H_1(t),
    \label{Eq:Hamiltonian2}
\end{align}
where $H_0$ and $H_1$ represent the unperturbed and interaction parts of the
Hamiltonian, respectively. By using the completeness relation
\begin{equation}
    \ketbra{a}{a} + \ketbra{b}{b} + \ketbra{c}{c} = I,
    \label{Eq:completeness}
\end{equation}
we can write $H_0$ as follows
\begin{align}
    H_0 & = I H_0 I \notag \\
        & = \hbar\omega_a\ketbra{a}{a} +
            \hbar\omega_b\ketbra{b}{b} +
            \hbar\omega_c\ketbra{c}{c},
    \label{Eq:H0_operator2}
\end{align}
where we have used $H_0\ket{j}=\hbar\omega_j\ket{j}$, with $j=a,b,c$. Similarly,
the part of the Hamiltonian $H_1$ that represents the interaction of the atom
with the radiation field is described by
\begin{equation}
    H_1(t) = -\boldsymbol{\mu} \cdot \mathbf{E}(\mathbf{r_0},t),
    \label{Eq:H1}
\end{equation}
where $\boldsymbol{\mu}$ is the atom's dipole moment operator, and
$\mathbf{E}(\mathbf{r_0},t)$ is the two-mode electric field operator. In the
dipole approximation, the electric field is evaluated at the dipole position
$\mathbf{r_0}=\boldsymbol{0}$, and the operator can be written in the form
\begin{equation}
    \mathbf{E}(\mathbf{0},t) = \sum_{k=1}^2 \mathrm{Re} \big[
        \hat{\boldsymbol{\epsilon}}_{k}\mathcal{E}_k(t)
        e^{i\omega_k t} \big],
    \label{Eq:E_operator}
\end{equation}
where $\hat{\boldsymbol{\epsilon}}_k$ and $\mathcal{E}_k(t)$ represent the
normalized polarization vector and the amplitude of the electric field,
respectively. We can also write $H_1$ in Dirac notation by using
Eq.~\eqref{Eq:completeness}
\begin{align}
    H_1(t) & = -I \bigl[\boldsymbol{\mu} \cdot \mathbf{E}(\mathbf{0},t)
                  \bigr] I \notag \\
           & = -\sum_{i,j}\sum_{k=1}^2 \ketbra{i}{i} \boldsymbol{\mu}
                \cdot \hat{\boldsymbol{\epsilon}}_k     \biggl[
                \frac{\mathcal{E}_k(t)   e^{i\omega_k t} +
                      \mathcal{E}_k^*(t) e^{-i\omega_k t}}{2}
                                                        \biggr] \ketbra{j}{j}.
    \label{Eq:H1_operator1}
\end{align}
Assuming that the electric field is linearly polarized along the dipole moment
direction, we have
\begin{equation}
    \bra{i} \boldsymbol{\mu} \cdot \hat{\boldsymbol{\epsilon}}_k \ket{j}
        = \bra{i}\mu\ket{j}
        = \mu_{ij},
    \label{Eq:mu.E}
\end{equation}
and Eq.~\eqref{Eq:H1_operator1} becomes
\begin{equation}
    H_1(t) = -\frac{1}{2} \sum_{i,j}\sum_{k=1}^2 \mu_{ij} \bigl[
        \mathcal{E}_k(t) e^{i\omega_k t} +
        \mathcal{E}_k^*(t) e^{-i\omega_k t}               \bigr]
        \ketbra{i}{j}, \quad j=a,b,c.
    \label{Eq:H1_operator2}
\end{equation}
In the general case the dipole matrix elements are complex numbers that might be
written simply as (see \ref{Eq:complex_dipole1} for an alternative notation)
\begin{equation}
    \mu_{ij} = \big| \mu_{ij} \big| e^{i\alpha}.
    \label{Eq:complex_dipole2}
\end{equation}

In this problem we consider only the relevant dipole transitions $\ket{a}
\leftrightarrow \ket{c}$ and $\ket{b} \leftrightarrow \ket{c}$, coupled by
the electric fields $\mathbf{E}_1$ and $\mathbf{E}_2$, respectively. The
transition $\ket{a} \leftrightarrow \ket{b}$ is forbidden by the dipole
selection rules. Thus, the only matrix elements of the electric dipole moment
that survive are: $\mu_{ac}$, $\mu_{bc}$, with their respective complex
conjugates. Plugging Eq.~\eqref{Eq:complex_dipole2} into
Eq.~\eqref{Eq:H1_operator2}, and considering only the allowed dipole
transitions, we get
\begin{multline}
    H_1(t) = -\frac{1}{2} \bigl[
        |\mu_{ac}| \mathcal{E}_1
        e^{i(\omega_1 t + \alpha)} \ketbra{a}{c} +
        |\mu_{ca}| \mathcal{E}_1
        e^{i(\omega_1 t - \alpha)} \ketbra{c}{a}  \\
      + |\mu_{bc}| \mathcal{E}_2
        e^{i(\omega_2 t + \beta)}  \ketbra{b}{c} +
        |\mu_{cb}| \mathcal{E}_2
        e^{i(\omega_2 t - \beta)} \ketbra{c}{b} + \text{H.c.}  \bigr]
\end{multline}

We next derive the equations of motion for the probability amplitudes $C_a$,
$C_b$, and $C_c$. By introducing Eq.~\eqref{Eq:psi3(t)} into
Eq.~\eqref{Eq:Schrodinger2(t)} and multiplying the resulting equation from the
left by $\bra{a}$, we find that
\begin{equation}
    \dot{C}_a = \frac{i}{2} \,\Omega_1(t) e^{i\alpha}
        e^{i(\omega_1-\omega_{ca})t} C_c,
    \label{Eq:Ca_dot1}
\end{equation}
where the Rabi frequency $\Omega_1(t)$ is defined as
\begin{equation}
    \Omega_1(t) = \frac{|\mu_{ac}| \mathcal{E}_1(t)}{\hbar},
    \label{Eq:Rabi1}
\end{equation}
and $\omega_{ca} = \omega_c - \omega_a$ is the atomic transition frequency. In
deriving Eq.~\eqref{Eq:Ca_dot1}, we have ignored counter-rotating terms
proportional to $\text{exp}[\pm i(\omega_1+\omega_{ca})]$ on the right-hand side
in the \textit{rotating-wave approximation} (RWA). Similarly, by multiplying
instead by $\bra{b}$ and $\bra{c}$, we find
\begin{align}
    \dot{C}_b & = \frac{i}{2} \,\Omega_2(t) e^{i\beta}
        e^{i(\omega_2-\omega_{cb})t} C_c,                 \label{Eq:Cb_dot1} \\
    \dot{C}_c & = \frac{i}{2} \,\Omega_1^*(t) e^{-i\alpha}
        e^{-i(\omega_1-\omega_{ca})t} C_a
                + \frac{i}{2} \, \Omega_2^*(t) e^{-i\beta}
        e^{-i(\omega_2 - \omega_{cb})t} C_b.              \label{Eq:Cc_dot1}
\end{align}
with the Rabi frequency
\begin{equation}
    \Omega_2(t) = \frac{|\mu_{bc}| \mathcal{E}_2(t)}{\hbar},
    \label{Eq:Rabi2}
\end{equation}
and the atomic transition frequency $\omega_{cb} = \omega_c - \omega_b$.
Introducing the \textit{detuning} factors
\begin{subequations}
\label{Eq:Deltas}
\begin{align}
    \Delta_1 & = \omega_1 - \omega_{ca},            \label{Eq:Delta1} \\
    \Delta_2 & = \omega_2 - \omega_{cb} - \Delta_1, \label{Eq:Delta2}
\end{align}
\end{subequations}
the coupled Eqs.~\eqref{Eq:Ca_dot1}, \eqref{Eq:Cc_dot1}, and~\eqref{Eq:Cb_dot1}
then reduce to the set
\begin{subequations}
\label{Eq:C_dots2}
\begin{align}
    \dot{C}_a & = \frac{i}{2} \,\Omega_1(t)
        e^{i\alpha}e^{i\Delta_1 t} C_c,                   \label{Eq:Ca_dot2} \\
    \dot{C}_b & = \frac{i}{2} \,\Omega_2(t)
        e^{i\beta} e^{i(\Delta_1 + \Delta_2)t} C_c,       \label{Eq:Cb_dot2} \\
    \dot{C}_c & = \frac{i}{2} \,\Omega_1^*(t)
        e^{-i\alpha} e^{-i\Delta_1 t} C_a + \frac{i}{2} \,\Omega_2^*(t)
        e^{-i\beta} e^{-i(\Delta_1 + \Delta_2)t} C_b.     \label{Eq:Cc_dot2}
\end{align}
\end{subequations}
By making a change of variables and choosing values for the phases (see
Appendix~\ref{App:changeVariables}), it is possible to eliminate the
exponentials from Eqs.~\eqref{Eq:C_dots2}. It then follows that
\begin{subequations}
\begin{align}
    i\dot{C}_a(t) & = \Delta_1 C_a(t) +
                      \frac{\Omega_1(t)}{2} \,C_c(t),   \\
    i\dot{C}_b(t) & = (\Delta_1 + \Delta_2) C_b(t) +
                      \frac{\Omega_2(t)}{2} \,C_c(t),   \\
    i\dot{C}_c(t) & = \frac{\Omega_1^*(t)}{2} \,C_a(t) +
                      \frac{\Omega_2^*(t)}{2} \,C_b(t).
\end{align}
\end{subequations}
These are the equations of motion for the probability amplitudes of the
three-level $\Lambda$-system shown in Fig.~\ref{Fig:3levelAtom}. The
Schr\"{o}dinger equation for these amplitudes in the rotating-wave approximation
reads:
\begin{equation}
    i\hbar \frac{d}{dt} \mathbf{C}(t) = H(t) \mathbf{C}(t),
\end{equation}
where
\begin{equation}
    H(t) = \frac{\hbar}{2} \,
        \begin{bmatrix}
              2\Delta_1   &            0           & \Omega_1(t) \\
                  0       & 2(\Delta_1 + \Delta_2) & \Omega_2(t) \\
            \Omega_1^*(t) &    \Omega_2^*(t)       &      0
        \end{bmatrix},
    \label{Eq:Hmatrix01}
\end{equation}
and $\mathbf{C}(t) = \bigl[ C_a(t), C_b(t), C_c(t) \bigr]^T$.

For the case of two-photon resonance ($\Delta_1 = \Delta_2 = 0$) and real matrix
elements of the dielectric dipole moment, Eq.~\eqref{Eq:Hmatrix01} takes the
simple form
\begin{equation}
    H(t) = \frac{\hbar}{2} \,
        \begin{bmatrix}
                 0      &      0      & \Omega_1(t) \\
                 0      &      0      & \Omega_2(t) \\
            \Omega_1(t) & \Omega_2(t) &       0
        \end{bmatrix}.
    \label{Eq:Hmatrix02}
\end{equation}
It is easy to verify (see Appendix~\ref{App:eigensystem}) that the following
linear combination of bare states $\ket{a}$, $\ket{c}$, and $\ket{b}$ are
eigenstates of the instantaneous RWA Hamiltonian
\begin{subequations}
\label{Eq:eigenstates_trig}
\begin{align}
    \ket{W^+} & = \frac{1}{\sqrt{2}} \bigl[ \sin\Phi(t) \ket{a}
        + \cos\Phi(t) \ket{b} + \ket{c} \bigr],
    \label{Eq:W+trig} \\
    \ket{W^0} & = \cos\Phi(t) \ket{a} - \sin\Phi(t) \ket{b},
    \label{Eq:W0trig}\\
    \ket{W^-} & = \frac{1}{\sqrt{2}} \bigl[ \sin\Phi(t) \ket{a}
        + \cos\Phi(t) \ket{b} - \ket{c} \bigr].
    \label{Eq:W-trig}
\end{align}
\end{subequations}
where the (time-varying) \textit{mixing angle} $\Phi$ is defined by the
relationship
\begin{equation}
    \tan \Phi(t) = \frac{\Omega_1(t)}{\Omega_2(t)}.
\end{equation}
When combined with the related photon numbers in the two radiation fields, the
eigenstates given by Eqs.~\eqref{Eq:eigenstates_trig} are called the ``dressed
states'' of the matter-field system. Although we do not keep track of the photon
numbers, we use this name here as well. The (time-dependent) dressed-state
eigenvalues are
\begin{equation}
    \omega^+ = +\half \sqrt{\Omega_1^2 + \Omega_2^2}, \qquad
    \omega^0 =  0, \qquad
    \omega^- = -\half \sqrt{\Omega_1^2 + \Omega_2^2}.
    \label{Eq:eigenvalues}
\end{equation}

\section{\label{Sec:adiabaticFollowing} The adiabatic following}

An alternative method for population transfer between two states is based on
sweeping the pulse frequency through a resonance. If the sweep is sufficiently
slow, then it is possible to produce complete population transfer between the
two states that are connected by the resonance~\cite{Shore:PRA45}. The adiabatic
process can be characterized by a steady state process~\cite{Morris:PR133}. That
is, the rates of change of the varying components of the incident laser fields
are assumed to be small enough that a quasi steady state is maintained
throughout the process~\cite{Hioe:PL83}. These processes have the advantage of
being insensitive to pulse area, pulse shape, and to the precise location of the
resonance. Then they are useful for producing population transfer in an ensemble
of atoms that have different Doppler shifts and different dipole moments.

\begin{figure}[h]
    \centering
    \includegraphics[scale=0.5]{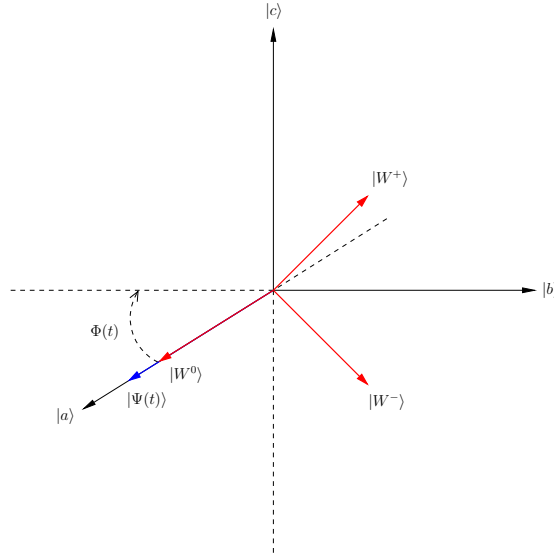}
    \caption{Graphic representation of the Hilbert space for the three-level
    system in the basis $\{\ket{a},\ket{b},\ket{c}\}$, and in the basis of the
    dressed states $\{\ket{W^+}$, $\ket{W^0}$, and $\ket{W^-}\}$. The dressed
    vectors are free to rotate in time following the evolution of the system.}
    \label{Fig:HilbertSpace}
\end{figure}

The condition for exact adiabatic following of a system from state $\ket{a}$ to
state $-\ket{b}$, without populating an excited state (which normally undergoes
spontaneous emission), can be seen from Fig.~\ref{Fig:HilbertSpace}.

In stimulated Raman scattering processes (STIRAP) the pulses are applied in
counterintuitive way, that is, the $\Omega_2(t)$ pulse precedes the pulse
$\Omega_1(t)$, though they partially overlap. In other words
\begin{equation}
    \lim_{t \to -\infty} \frac{\Omega_1(t)}{\Omega_2(t)} = 0, \quad
    \lim_{t \to +\infty} \frac{\Omega_1(t)}{\Omega_2(t)} = +\infty.
\end{equation}
Assuming that
\begin{equation}
    \tan\Phi(t) = \frac{\Omega_1(t)}{\Omega_2(t)},
\end{equation}
then we have
\begin{equation}
    \Phi(-\infty) = 0 \qquad \text{and} \qquad \Phi(+\infty) = \pi/2.
\end{equation}
Hence the adiabatic state $\ket{W^0}$ coincides with the state $\ket{a}$ before
the excitation and with state $-\ket{b}$ after it, so that initially only state
$\ket{W^0}$ among the adiabatic states is populated. If the excitation is
adiabatic, then the system will remain in this adiabatic state all time and the
population will eventually be completely transferred to state $\ket{b}$.
Moreover, no appreciable population will reside in the intermediate state
$\ket{c}$ at any time which makes the transfer efficiency insensitive to decay
from this state to other states~\cite{Vitanov:OC96}. Therefore in the adiabatic
process, the evolution of the wave vector $\ket{\Psi(t)}$ follows closely the
evolution of the ``dressed state'' $\ket{W^0}$, which goes from a direction
parallel to $\ket{a}$ to a direction antiparallel to $\ket{b}$.

We now consider the conditions under which the system evolves adiabatically.
Nonadiabatic coupling between the eigenstates is small when the rate of change
of the mixing angle $\Phi(t)$, is small compared with the separation of the
corresponding eigenvalues~\cite{Messiah}.
\begin{equation}
    \bigl| \braket{W^+}{\dot{W}^0} \bigr|
        \ll \bigl| \omega^{\pm} - \omega^0 \bigr|.
\end{equation}
For no detuning, this separation is given by
\begin{equation}
    |\omega^{\pm} - \omega^0| = \half \sqrt{\Omega_1^2 + \Omega_2^2}
                              = \Omega_{\text{eff}}.
\end{equation}
In addition, it is easy to find that $| \braket{W^+}{\dot{W}^0}| =
-\dot{\Phi}\sin\Theta$, and therefore the adiabaticity constraint, with
$\sin\Theta=1$, reads
\begin{equation}
    \bigl| \dot{\Phi} \bigr| \ll \Omega_{\text{eff}}.
\end{equation}
By differentiating the above expression with respect to time, we may get
\begin{equation}
    \Biggl| \frac{\dot{\Omega}_1\Omega_2 - \Omega_1\dot{\Omega}_2}
        {\Omega_1^2 + \Omega_2^2} \Biggr|
        \ll \Omega_{\text{eff}}.
\end{equation}

Finally, for a given counterintuitive sequence of pulses $\Omega_1$ and
$\Omega_2$, separated by some time delay $\Delta t /T$ (with $T$ being the pulse
length or interaction time), the adiabatic theorem leads to the condition
\begin{equation}
    \Omega_{\text{eff}}\,T \gg 1.
\end{equation}

\chapter{\label{Ch:02} Laser-induced population transfer in multilevel systems}

Cavity quantum electrodynamics (QED) is the part of physics that studies the
interaction of single atoms and photons coupled to an electromagnetic
resonator. Many interesting effects have been observed during the last twenty
years. They include the alteration of the atomic radiative rates by the presence
of a cavity around an atom, shifts in the atomic energy level due to the
coupling with the cavity, manipulation of photons by using the interaction with
individual atoms crossing the cavity, creation of non-classical field states,
among others~\cite{Oppo}.

In the recent few years, and due to the advancements in the technology of
high-$Q$ cavities and in atomic beam manipulation, many proposals in the area of
quantum information and quantum computation have been made. They include
applications to particle teleportation, quantum cryptography, spectroscopy and
conditional dynamics.

In this chapter we study numerically the transfer of population in a four-level
system by using $\pi$-pulse methods and adiabatic passage schemes. We present
some examples of the resonance-like features in the failure probability $p$. We
show that the features appear for Gaussian, hyperbolic secant, and Lorentzian
pulse profiles.

\section{The atom-cavity system}

\begin{figure}[h]
    \centering
    \includegraphics[scale=0.8]{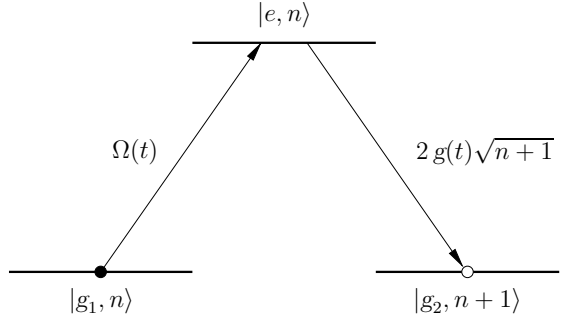}
    \caption{$\Lambda$ Three-level atom in a cavity.}
    \label{Fig:3levelResonance_02}
\end{figure}

Consider a three-level atom consisting of two ground states $\ket{\textsl{g}_1}$
and $\ket{\textsl{g}_2}$, and an excited state $\ket{e}$, interacting with a
coherent driving field $\Omega(t)$ of frequency $\omega_L$ and a cavity-mode
field $\textsl{g}(t)$ of frequency $\omega$, as indicated in
Fig.~\ref{Fig:3levelResonance_02}.

\noindent The state vector describing the three-level atom can be written in the
form
\begin{equation}
    \ket{\Psi(t)} = \sum_n \bigl[
          C_{\textsl{g}_1,n}(t) \ket{\textsl{g}_1,n} + C_{e,n}(t) \ket{e,n}
        + C_{\textsl{g}_2,n}(t) \ket{\textsl{g}_2,n} \bigr],
\end{equation}
where $C_{\textsl{g}_1,n}$, $C_{e,n}$, and $C_{\textsl{g}_2,n}$ are the
probability amplitudes of finding the atom in states $\ket{\textsl{g}_1,n}$,
$\ket{e,n}$, and $\ket{\textsl{g}_2,n}$, respectively. The ket
$\ket{\textsl{g}_1,n} \equiv \ket{\textsl{g}_1} \otimes \ket{n}$ represents the
state in which the atom is in the ground state $\ket{\textsl{g}_1}$ and the
cavity field has $n$ photons. Similar descriptions exist for the other two
states.

The time evolution of the system is described by the Schr\"{o}dinger equation
\begin{equation}
    i\hbar\,\ket{\dot{\Psi}(t)} = H(t) \ket{\Psi(t)}
\end{equation}
with the Hamiltonian operator $H(t)$ given by
\begin{align}
    H(t) & = \hbar\omega a^{\dagger} a
        + \omega_{e\textsl{g}} \ketbra{e}{e} - i \hbar \textsl{g}(t)
        \bigl(\ketbra{e}{\textsl{g}_2} a - \text{H.c.} \bigr) \notag \\
        & \qquad + i\hbar \Omega(t)
        \bigl( \ketbra{e}{\text{g}_1} e^{-i\omega_L t} - \text{H.c.} \bigr).
    \label{Eq:Hamiltonian_4.1}
\end{align}
Here $a$ represents the annihilator operator for the cavity mode. The time
dependence of $\Omega(t)$ and $\textsl{g}(t)$ may be provided simply by the
motion of the atom across the laser- and cavity-field profiles.

\subsection{Dark state}

It is interesting to observe that the interaction part of the
Hamiltonian~\eqref{Eq:Hamiltonian_4.1} can only cause transitions between states
within the family
\begin{equation}
    \bigl\{\ket{\textsl{g}_1,n}, \,\ket{e,n}, \,\ket{\textsl{g}_2,n+1} \bigr\},
\end{equation}
Therefore, in the rotating-wave approximation, the adiabatic energy eigenvalues
of the Hamiltonian associated with a particular family of states
are~\cite{Parkins:PRL71}
\begin{align}
    E_n       & = n\hbar\omega, \\
    E_n^{\pm} & = n\hbar\omega + \frac{\hbar}{2} \Bigl[ \Delta \pm
        \sqrt{\Delta^2 + 4\textsl{g}(t)^2 (n+1) + \Omega(t)^2} \Bigr],
\end{align}
where we have assumed that $\omega = \omega_L$, and
$\Delta = \omega_{e\textsl{g}} -\omega$ is the detuning. We are interested in
the eigenstate corresponding to $E_n=n\hbar\omega$, which is given by
\begin{equation}
    \ket{E_n (t)} = \frac{2 \textsl{g}(t)\sqrt{n+1}\ket{\textsl{g}_1,n} +
    \Omega(t)\ket{\textsl{g}_2,n+1}}
    {\sqrt{\Omega(t)^2 + 4\textsl{g}(t)^2 (n+1)}}.
\end{equation}
This eigenstate is, at all times, free of any contribution from the excited
state $\ket{e,n}$, and is independent of the detuning $\Delta$.

\subsection{Adiabatic following}

If we assume that only level $\ket{\textsl{g}_1,n}$ is initially populated,
complete population transfer occurs if~\cite{Kuklinski:PRA40}
\begin{equation}
    \frac{\Omega(t)}{\textsl{g}(t)} \Biggr|_{t \to -\infty}= 0
    \qquad \text{and} \qquad
    \frac{\textsl{g}(t)}{\Omega(t)} \Biggr|_{t \to +\infty}= 0,
\end{equation}
where $t \to -\infty$ and $t \to +\infty$ corresponds to times before and after
the interaction with the fields, respectively. Consequently, for the pulse
sequence in which the pulse $\Omega(t)$ is delayed with respect to
$\textsl{g}(t)$, the dark state is the appropriate vehicle for transferring
population from state $\ket{\textsl{g}_1,n}$ to state $\ket{\textsl{g}_2,n+1}$
without populating state $\ket{e,n}$.

If the condition for adiabaticity~\cite{Messiah}
\begin{equation}
    \Omega_0\,T,\:2\,\textsl{g}_0\sqrt{n+1}\,T \gg 1
\end{equation}
is satisfied (with $\Omega_0$ and $\textsl{g}_0$ representing peak intensities
for the respective fields) then the state vector of the system remains very
nearly an eigenvector of the time-dependent Hamiltonian at all times, that is
\begin{equation}
    \ket{\Psi(t)} \approx \ket{E_n (t)}.
\end{equation}
%

\subsection{Master equation}

We can generalize the previous ideas to more complicated and realistic
atomic-level structures. For example, this time we may consider Zeeman ground
state levels, and include in our analysis the spontaneous emission $\Gamma$, and
the cavity decay $\kappa$.

\begin{figure}[h]
    \centering
    \includegraphics[scale=0.8]{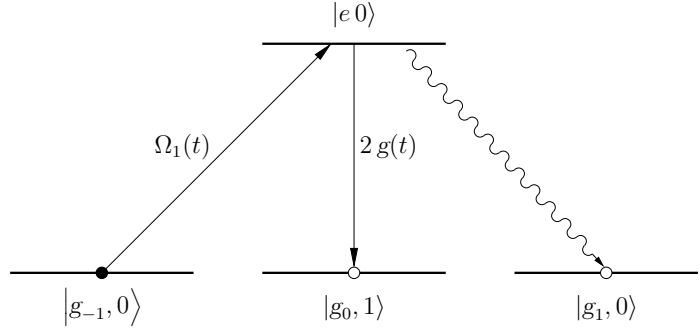}
    \caption{$\Lambda$ four-level atom in a cavity.}
    \label{Fig:4levelResonance}
\end{figure}

Consider the case of an atomic $J_{\textsl{g}}=1 \to J_e=0$ transition in a
four-level $\Lambda$ atom, as indicated in Fig.~\ref{Fig:4levelResonance}.
The transition $\ket{\textsl{g}_1,1} \to \ket{e,0}$ is strongly connected to a
$\pi$-polarized cavity-mode field with coupling strength $\textsl{g}(t)$. The
transition $\ket{\textsl{g}_{-1},0} \to \ket{e,0}$ is coupled to a coherent
$\sigma^+$-polarized laser field with frequency $\omega_L$ and Rabi frequency
$\Omega(t)$. The excited state is assumed to be able to decay to the three
ground states with the same decaying rate $\Gamma/3$.

In the short period of time compared to the natural decay times, the dynamical
evolution of the density matrix $\rho(t)$ of the atomic system is given by the
Liouville equation
\begin{equation}
    \frac{\partial\rho}{\partial t} = -i\bigl[
        H_{\text{eff}}\,\rho - \text{H.c.} \bigr]
        + \Gamma \sum_{\sigma=0,\pm1} A_{\sigma} \rho A_{\sigma}^{\dagger}
        + \kappa a \rho a^{\dagger},
    \label{Eq:masterEq1}
\end{equation}
where $\rho(t)$ is the reduced density operator of the system, and
\begin{align}
    H_{\text{eff}} = & \bigl(\Delta -i\Gamma/2 \bigr)
        \sum_{m_e}\ketbra{J_e m_e}{J_e m_e}
        - i \frac{\kappa}{2} a^{\dagger} a \notag \\
        & \quad - i \Omega(t) \bigl( A_{+1} - A_{+1}^{\dagger} \bigr)
                - i \textsl{g}(t) \bigl( a^{\dagger}A_0 - A_0^{\dagger}a \bigr),
    \label{Eq:Heff_1}
\end{align}
is the effective non-Hermitian Hamiltonian in the rotating-wave
approximation~\cite{Parkins:PRA51}. For the two-photon resonance problem
($\Delta=0$) this Hamiltonian reduces to
\begin{align}
    H_{\text{eff}} = & -i\frac{\Gamma}{2} \ketbra{e}{e}
        - i \frac{\kappa}{2} a^{\dagger} a \notag \\
        & \quad - i \Omega(t) \bigl( A_{+1} - A_{+1}^{\dagger} \bigr)
                - i \textsl{g}(t) \bigl( a^{\dagger}A_0 - A_0^{\dagger}a \bigr),
    \label{Eq:Heff_2}
\end{align}
where the atomic lowering operators $A_{\sigma}$ are given by
\begin{equation}
    A_{\sigma} = \sum_{m_e,m_{\textsl{g}}} \ket{J_{\textsl{g}} m_{\textsl{g}}}
        \braket{J_{\textsl{g}} m_{\textsl{g}}\,;1\sigma}{J_e m_e} \bra{J_e m_e},
\end{equation}
with $\braket{J_{\textsl{g}} m_{\textsl{g}}\,;1\sigma}{J_e m_e}$ the
Clebsch-Gordan coefficient for the dipole transition $\ket{e} \to
\ket{\textsl{g}}$ with polarization $\sigma=0,\pm1$. Working out these
coefficients, we may have
\begin{equation}
    A_{1}  =  \frac{1}{\sqrt{3}} \,\ketbra{\textsl{g}_{-1}}{e}, \qquad
    A_{0}  = -\frac{1}{\sqrt{3}} \,\ketbra{\textsl{g}_0}{e},    \qquad
    A_{-1} =  \frac{1}{\sqrt{3}} \,\ketbra{\textsl{g}_{1}}{e}.
    \label{Eq:A_operators}
\end{equation}
Upon substitution of Eqs.~\eqref{Eq:Heff_2} and~\eqref{Eq:A_operators} into
Eq.~\eqref{Eq:masterEq1}, we find that
\begin{align}
    \frac{\partial \rho}{\partial t} & = -\half \bigl( \Gamma\ketbra{e}{e}\rho
        + \kappa a^{\dagger} a \rho + \text{H.c.} \bigr)
        - \frac{\Omega(t)}{\sqrt{3}} \bigl( \ketbra{\textsl{g}_{-1}}{e}\rho
        - \ketbra{e}{\textsl{g}_{-1}}\rho + \text{H.c.} \bigr) \notag \\
        & \qquad - \frac{\textsl{g}(t)}{\sqrt{3}} \bigl(a^{\dagger}
          \ketbra{\textsl{g}_{0}}{e}\rho
        - \ketbra{e}{\textsl{g}_0} a \rho + \text{H.c.} \bigr)
        + \kappa a \rho a^{\dagger} \notag \\
        & \qquad + \frac{\Gamma}{3} \bigl( \ketbra{\textsl{g}_1}{e}\rho
          \ketbra{e}{\textsl{g}_1} + \ketbra{\textsl{g}_0}{e}\rho
          \ketbra{e}{\textsl{g}_0} + \ketbra{\textsl{g}_{-1}}{e}\rho
          \ketbra{e}{\textsl{g}_{-1}} \bigr).
    \label{Eq:masterEq2}
\end{align}
We can see from Eq.~\eqref{Eq:masterEq2} that only the following family
of eigenstates survive:
\begin{equation}
    \ket{\textsl{g}_{-1},0}, \quad
    \ket{e,0},      \quad
    \ket{\textsl{g}_0,1},    \quad
    \ket{\textsl{g}_0,0},    \quad \text{and} \quad
    \ket{\textsl{g}_1,0}.
\end{equation}
This means that in our density matrix approach, we have up to 25 matrix elements
of the form $\dot{\rho}_{ij} = \bra{i} \dot{\rho} \ket{j}$. However, we can get
rid of some of these matrix elements because they are electric dipole forbidden
or make no physical sense. In this way, the total number of matrix elements can
be reduced from 25 to only eight, which corresponds to the number of
differential equations describing the evolution of the system.

By introducing the following notation for the atomic levels:
\begin{equation}
    \ket{a} = \ket{\textsl{g}_{-1}}, \qquad
    \ket{b} = \ket{\textsl{g}_0},    \qquad
    \ket{c} = \ket{\textsl{g}_1};
\end{equation}
and calling $\rho_{ij} \equiv \text{Re}\bigl[ \bra{i} \rho \ket{j} \bigr]$, we
can write the equations of motion for the density matrix elements in the form
{\allowdisplaybreaks
\begin{subequations}
\label{Eq:fourLevelSystem}
\begin{align}
    \dot{\rho}_{a0a0} & = -2 \,\Omega(t) \,\rho_{a0e0}
        + \frac{\Gamma}{3} \,\rho_{e0e0}, \\
    \dot{\rho}_{b0b0} & =  \frac{\Gamma}{3} \,\rho_{e0e0}
        + \kappa \,\rho_{b1b1}, \\
    \dot{\rho}_{b1b1} & = -\kappa \,\rho_{b1b1}
        - 2 \,\textsl{g}(t) \,\rho_{b1e0}, \\
    \dot{\rho}_{c0c0} & =  \frac{\Gamma}{3} \,\rho_{e0e0}, \\
    \dot{\rho}_{e0e0} & = -\Gamma\,\rho_{e0e0} + 2\,\Omega(t)\,\rho_{a0e0}
        + 2 \,\textsl{g}(t) \,\rho_{b1e0}, \\
    \dot{\rho}_{a0e0} & = -\frac{\Gamma}{2} \,\rho_{a0e0} - \Omega(t) \bigl(
        \rho_{e0e0} - \rho_{a0a0} \bigr) + \textsl{g}(t) \,\rho_{a0b1}, \\
    \dot{\rho}_{a0b1} & = -\frac{\kappa}{2} \,\rho_{a0b1}
        - \Omega(t) \,\rho_{b1e0} - \textsl{g}(t) \,\rho_{a0e0}, \\
    \dot{\rho}_{b1e0} & = -\half\bigl( \Gamma + \kappa \bigr) \rho_{b1e0} +
        \Omega(t) \,\rho_{b1a0} - \textsl{g}(t)\bigl( \rho_{e0e0}
        - \rho_{b1b1} \bigr).
\end{align}
\end{subequations}}
%

\section{Numerical results}

In this section we have studied numerically two different methods for
transferring the population between two atomic levels: the $\pi$-pulses and the
adiabatic passage methods. The calculation of the transfer efficiencies was
done by numerically integrating the system of
equations~\eqref{Eq:fourLevelSystem} using the Runge-Kutta method of fifth order
with an adaptive mesh. By using Gaussian, hyperbolic secant, and Lorentzian
pulse shapes, we obtained the coherent transfer efficiencies for the four-level
system in a cavity, including radiative decay from the excited states. For
simplicity, we have considered a high-$Q$ cavity with no decay ($\kappa=0$).
Five different parameters were considered in our simulations: the Rabi frequency
and the width of the pulses, the time delay (for intuitive and counterintuitive
configurations), and the spontaneous emission of the excited states.

Because of the big volume of data manipulated during the simulations (necessary
to obtain results with some degree of accuracy) the computations were performed
in a cluster of four heterogeneous computers running Parallel Virtual Machine
(PVM). Surfaces of the logarithm of the failure probability were plotted for
some physically reasonable intervals for the Rabi frequencies and widths of the
pulses. For example, most of the time we studied widths in the interval
$[0,10]$. We also explored time delays in intervals between $[0,50]$ for
Gaussian pulses, and $[0,100]$ for Sech and Lorentzian pulses. The figures
obtained for all these parameters were given in ``absolute'' numbers (no units).
We did it in this way because we wanted to examine the ratios between quantities
with same dimensions rather than individual values taken by the parameters
themselves. This is why most of the graphs were plotted in terms of
$\Omega/\textsl{g}$, and $\Delta t/\sigma$. During the numerical simulations we
also fixed the value $\textsl{g}$ to unity for many of them. This allowed us to
have a point of comparison between the pulses' energies (Rabi frequencies) and
widths for the different failure probabilities obtained, and then to conclude
which set of parameters achieved the maximum transfer efficiency.

Our analysis starts with the $\pi$-pulses method. An intuitive sequence of these
pulses can, in principle, produce a complete population transfer between two
atomic levels~\cite{Shore:PRA45}. Here, we adjust the laser intensities and
pulses' duration so that the time integral of the Rabi frequencies (the pulse
area) have the value of $\pi$. The first pulse $\Omega$, takes the system from
the initially populated level $\ket{\textsl{g}_{-1}\:0}$ to the excited level
$\ket{e\:0}$; then the cavity-mode field $\textsl{g}$ takes the system to the
target level $\ket{\textsl{g}_0\:0}$. For this to be possible the excitation has
to be coherent. Results for the time delay and failure probability are shown in
Table~\ref{Table:pipulses} for different values of the Rabi frequency and the
width of the pulses.

\begin{longtable}{cccccccc}
\caption{$\pi$-pulses method.}\\
\label{Table:pipulses}\\
\hline
\endhead
\hline
\endfoot
Pulse profile & $\Gamma$   & $\Omega$ & $\sigma$   & $\textsl{g}$ &
$\sigma_{\textsl{g}}$ & $\Delta t$ & $\log_{10} p$ \\
\hline\\[-5pt]
Gaussian   & 0.01 & 2.14 & 0.29 & 1.00 & 0.63 & 1.26 & -2.05 \\
           & 0.02 & 2.15 & 0.29 & 1.00 & 0.63 & 1.16 & -1.78 \\
           & 0.05 & 2.03 & 0.31 & 1.00 & 0.63 & 1.01 & -1.44 \\
           & 0.10 & 2.11 & 0.30 & 1.00 & 0.63 & 0.89 & -1.19 \\
           & 0.20 & 2.18 & 0.29 & 1.00 & 0.63 & 0.75 & -0.95 \\
           &      &      &      &      &      &      &       \\
Sech       & 0.01 & 1.84 & 0.26 & 1.00 & 0.50 & 1.52 & -1.94 \\
           & 0.02 & 2.01 & 0.25 & 1.00 & 0.50 & 1.35 & -1.69 \\
           & 0.05 & 2.16 & 0.23 & 1.00 & 0.50 & 1.11 & -1.37 \\
           & 0.10 & 2.26 & 0.22 & 1.00 & 0.50 & 0.94 & -1.13 \\
           & 0.20 & 2.43 & 0.21 & 1.00 & 0.50 & 0.77 & -0.91 \\
           &      &      &      &      &      &      &       \\
Lorentzian & 0.01 & 5.71 & 0.09 & 1.00 & 0.50 & 2.28 & -1.63 \\
           & 0.02 & 5.53 & 0.09 & 1.00 & 0.50 & 1.80 & -1.43 \\
           & 0.05 & 5.56 & 0.09 & 1.00 & 0.50 & 1.32 & -1.16 \\
           & 0.10 & 5.80 & 0.09 & 1.00 & 0.50 & 1.04 & -0.96 \\
           & 0.20 & 6.28 & 0.08 & 1.00 & 0.50 & 0.82 & -0.77 \\[5pt]
\end{longtable}

\begin{figure}[htbp]
  \centering
  \mbox{\subfigure[$\Gamma=0.01$]{\includegraphics[scale=0.5]{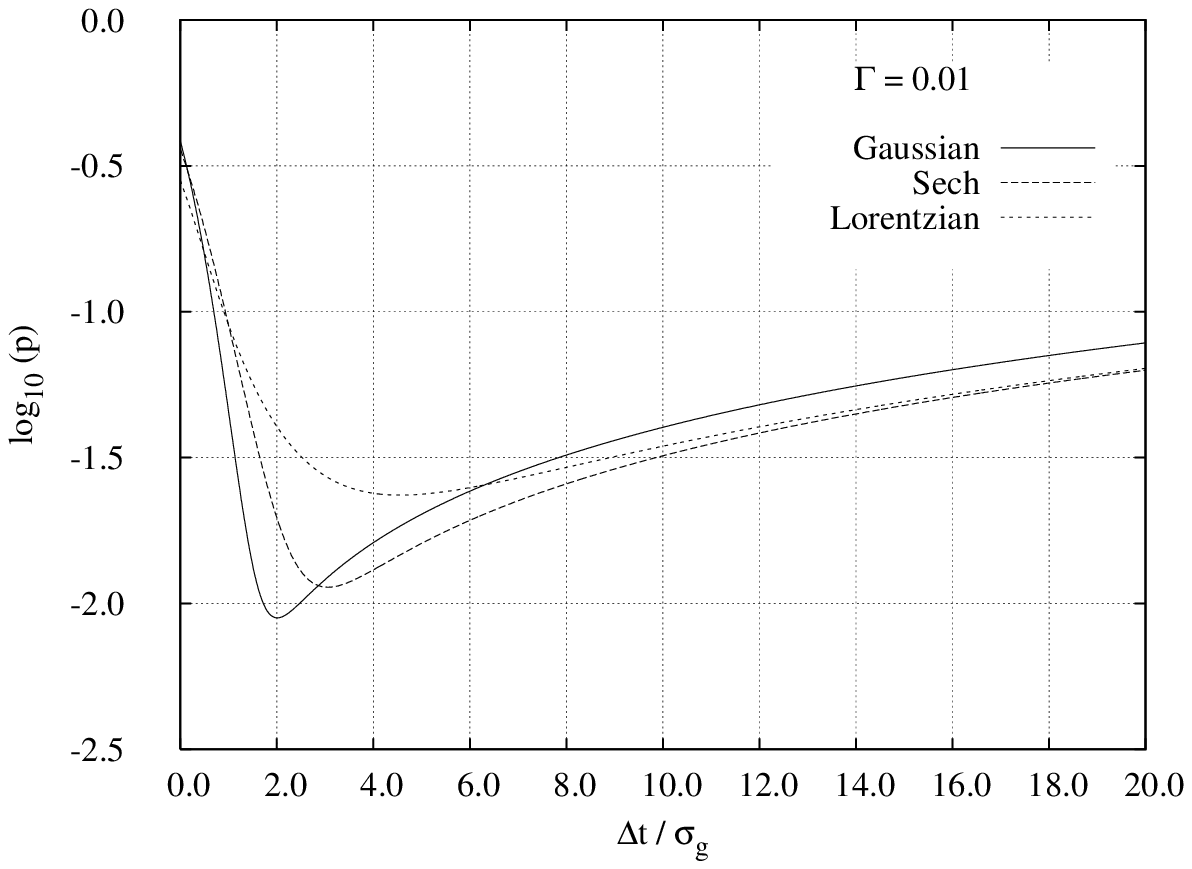}}}
  \mbox{\subfigure[$\Gamma=0.05$]{\includegraphics[scale=0.5]{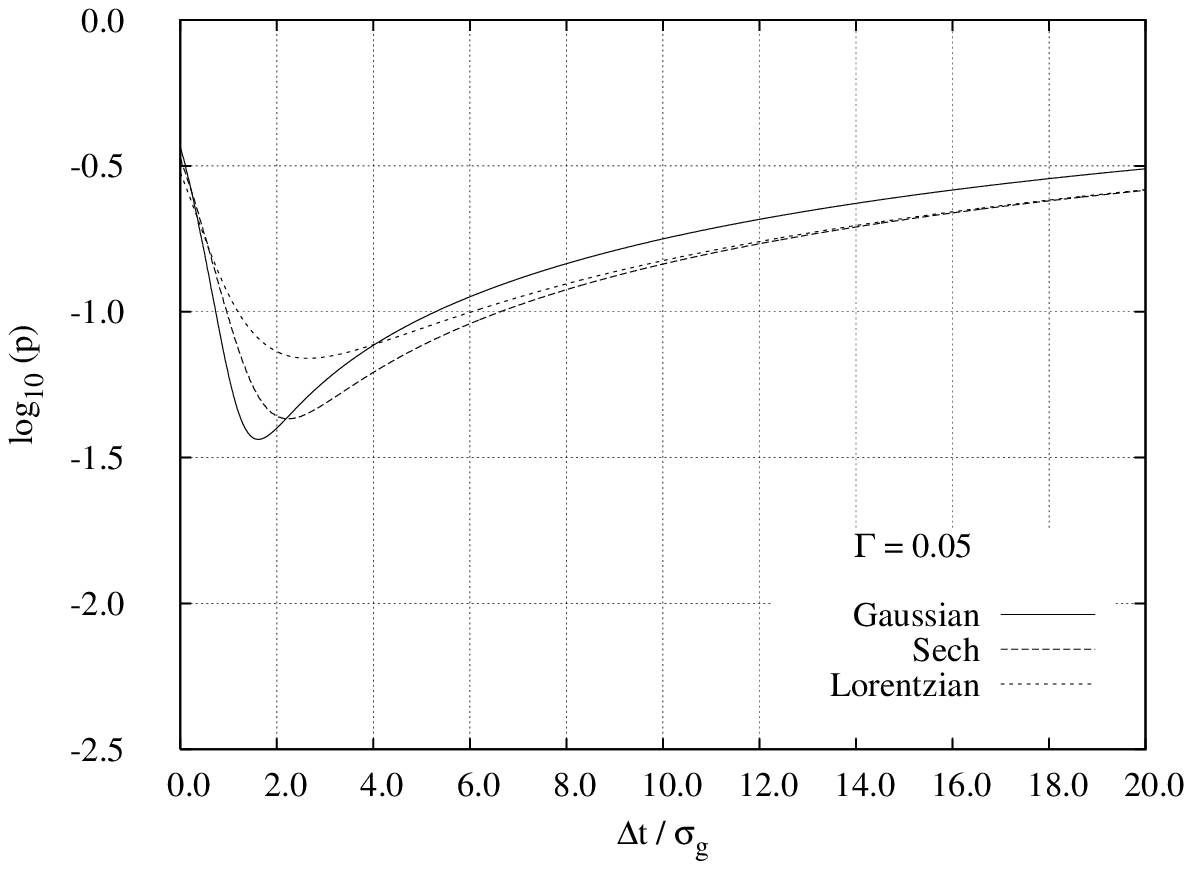}}}
  \mbox{\subfigure[$\Gamma=0.10$]{\includegraphics[scale=0.5]{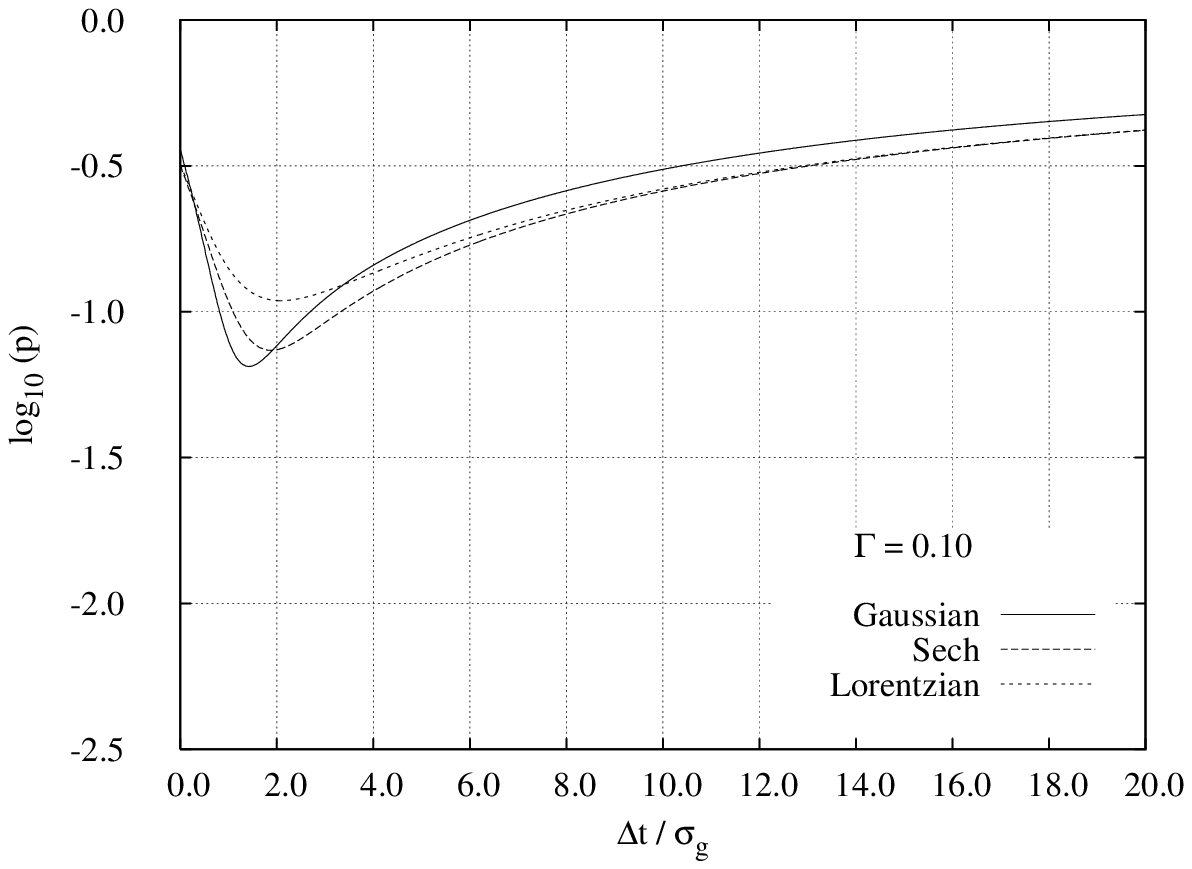}}}
  \mbox{\subfigure[$\Gamma=0.20$]{\includegraphics[scale=0.5]{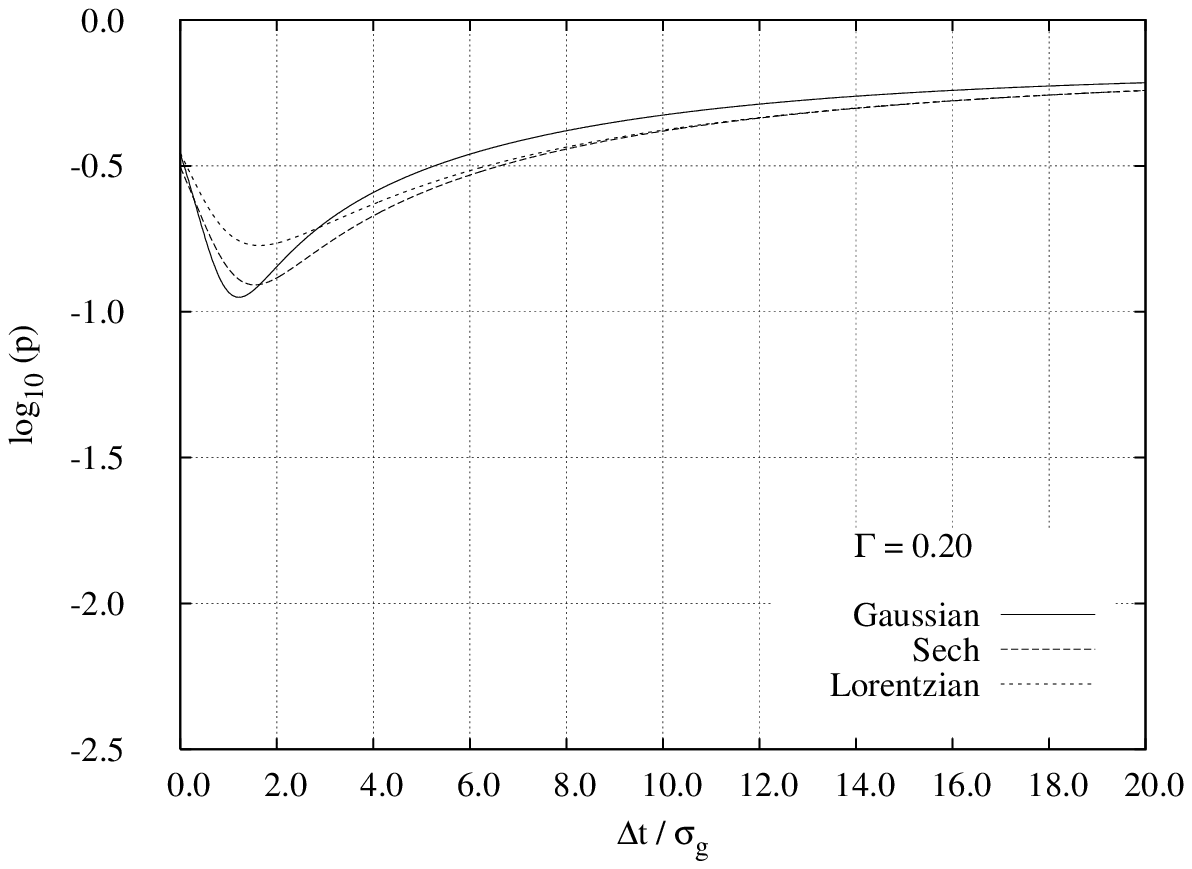}}}
  \caption{$\pi$-pulses failure probability for four different values of
  spontaneous emission. The values of the Rabi frequencies, widths, and time
  delay can be found in Table~\ref{Table:pipulses}}
  \label{Fig:pipulses01}
\end{figure}

\noindent As is well known, this method presents the most efficiency ($100$\%)
when the system undergoes no spontaneous emission. However, when the system
decays radiatively, the efficiency of the method decreases gradually as the
spontaneous emission rate increases. We observe that, when the spontaneous
emission is $\Gamma=0.01$, the Gaussian pulses have an efficiency of $99.1$\%,
while the secant and Lorentzian pulses presented maximum efficiencies of
$98.9$\% and $97.7$\%, respectively. These values were the best obtained for a
wide range of the examined pulse amplitudes and pulse widths.
Fig.~\ref{Fig:pipulses01} shows that for greater values of the spontaneous
emission, the transfer efficiency reduces in a considerable way.
The time delay between the pulses also reduces to shorter times for bigger
spontaneous emission values. This is understandable because more overlap between
the pulses is required to neutralize the decay rate of the excited states.

\begin{figure}[htbp]
    \centering
    \mbox{\subfigure[Gaussian]{\includegraphics[scale=0.6]{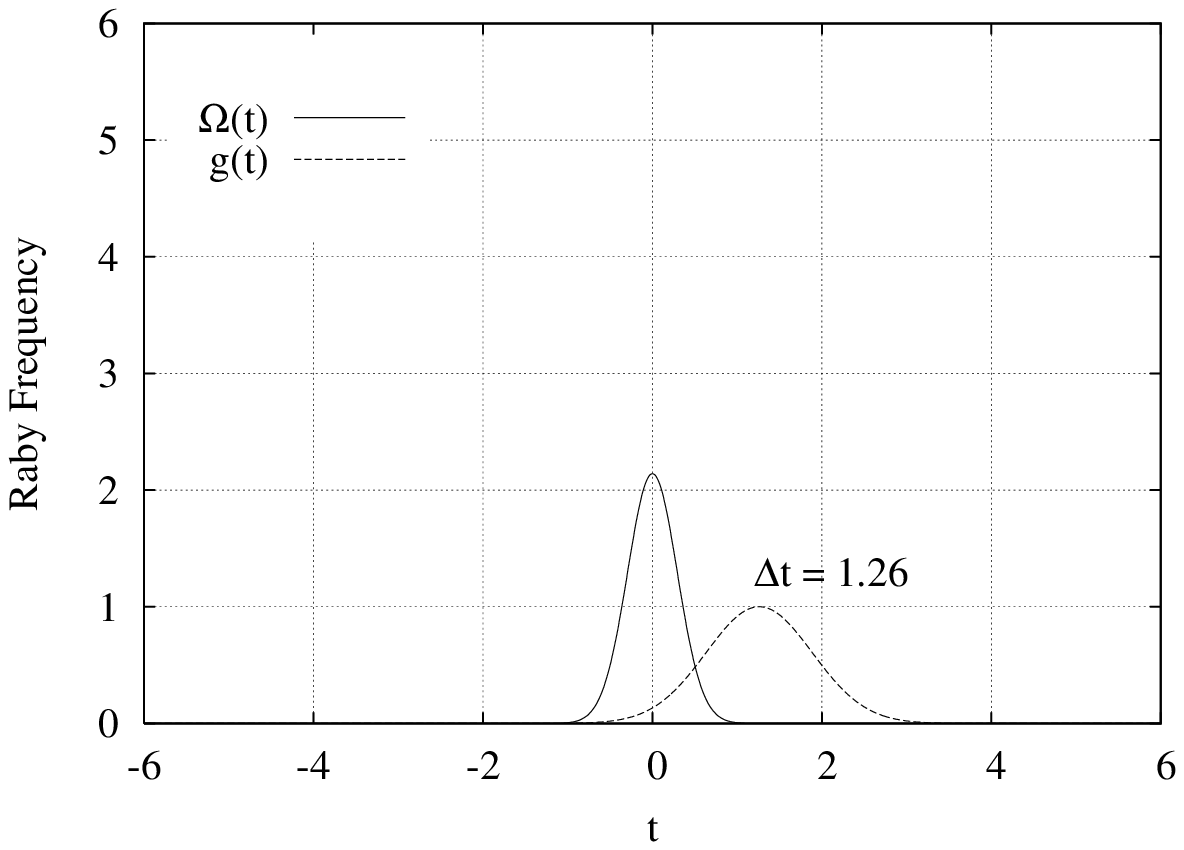}}}
    \mbox{\subfigure[Sech]{\includegraphics[scale=0.6]{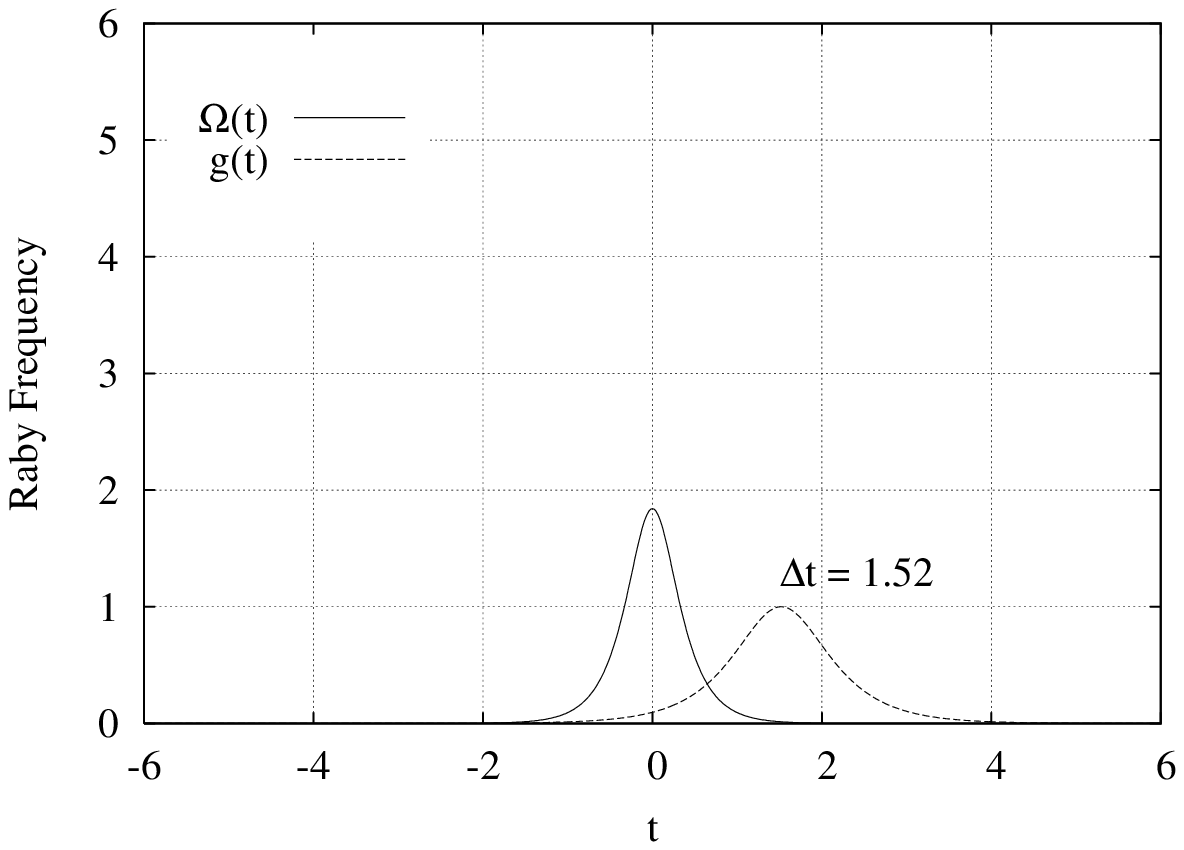}}}
    \mbox{\subfigure[Lorentzian]{\includegraphics[scale=0.6]{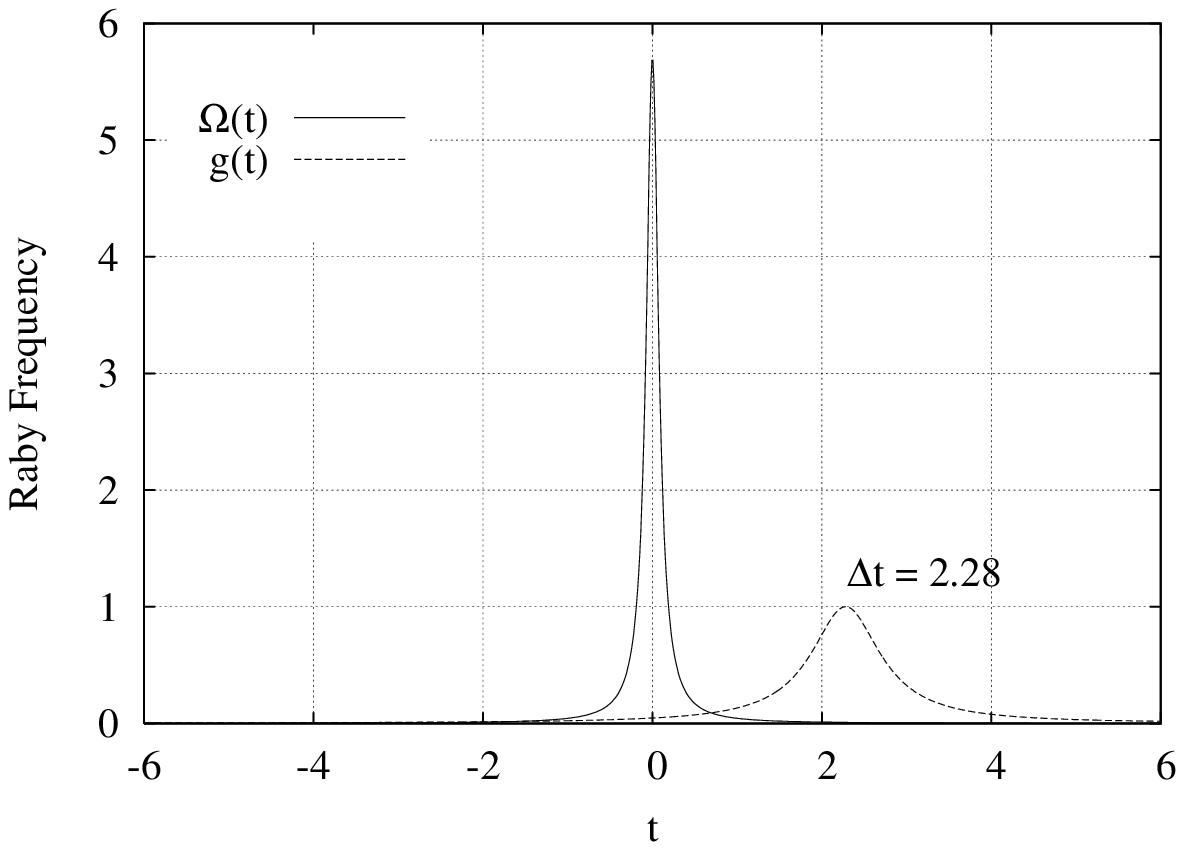}}}
    \caption{$\pi$-pulses' profiles for the best set of parameters when
    $\Gamma=0.01$.}
    \label{Fig:pipulses02}
\end{figure}

It is interesting to note that in this \textit{intuitive} sequence of pulses (as
it is the case for the $\pi$-pulses method), good transfer efficiencies were
achieved when the driving classical field (laser) was stronger than the
quantized cavity mode, as shown in Fig.~\ref{Fig:pipulses02}. Even though the
ratio of the Gaussian pulses ($\Omega/\textsl{g}$) was bigger than the ratio of
the Sech pulses, less interaction time ($\Delta t/\sigma_{g}$) was required for
the Gaussian pulses to obtain a bigger efficiency.
The Lorentzian pulses required the biggest energy (Rabi frequency) to achieve
even less efficiency than the other two types of pulses, as observed from the
ratio of the pulses shown in~\ref{Fig:pipulses02}.

The $\pi$-pulses method is a good and efficient technique to transfer the
population in atomic systems, but we will see that the adiabatic passage scheme
overtakes considerably the $\pi$-pulses performance for some very interesting
and particular situations.

The next method we studied numerically was the adiabatic passage scheme.
Table~\ref{Table:adiabatic} shows the values of time delay and failure
probability we obtained for different values of the spontaneous emission.

\begin{longtable}{cccccccc}
\caption{Adiabatic passage scheme}\\
\label{Table:adiabatic}\\
\hline
\endhead
\hline
\endfoot
Pulse profile & $\Gamma$   & $\Omega$ & $\sigma$   & $\textsl{g}$ &
$\sigma_{\textsl{g}}$ & $\Delta t$ & $\log_{10} p$ \\
\hline\\[-5pt]
Gaussian   & 0.00 &  2.00 & 1.00 &  2.00 & 1.00 &  1.31 & -4.88 \\
           &      &  4.00 & 1.00 & 19.20 & 1.00 &  1.90 & -4.53 \\
           &      &  6.00 & 1.00 &  5.70 & 1.00 &  1.50 & -6.83 \\
           &      &  6.70 & 1.50 &  2.00 & 1.00 &  2.72 & -5.67 \\
           &      &       &      &       &      &       &       \\
           & 0.10 &  3.39 & 3.23 &  1.00 & 2.45 &  5.85 & -1.99 \\
           &      &  2.75 & 3.09 &  1.00 & 2.48 &  5.29 & -2.00 \\
           &      &       &      &       &      &       &       \\
           & 0.20 &  3.30 & 3.30 &  1.00 & 2.50 &  5.90 & -1.73 \\
           &      &  2.40 & 3.30 &  1.00 & 3.00 &  5.38 & -1.80 \\
           &      &  2.30 & 4.20 &  1.00 & 4.48 &  6.62 & -1.96 \\
           &      &  2.10 & 4.60 &  1.00 & 5.00 &  7.09 & -2.01 \\
           &      &       &      &       &      &       &       \\
Sech       & 0.00 &  2.00 & 1.00 &  2.00 & 1.00 &  0.80 & -7.79 \\
           &      &       &      &       &      &       &       \\
           & 0.10 &  2.60 & 1.40 &  1.00 & 1.20 &  2.70 & -1.71 \\
           &      &  5.00 & 1.50 &  1.00 & 2.00 &  3.80 & -1.90 \\
           &      &  4.50 & 1.70 &  1.00 & 2.40 &  4.10 & -1.89 \\
           &      &  4.20 & 1.80 &  1.00 & 2.60 &  4.30 & -1.86 \\
           &      &  6.30 & 3.40 &  1.00 & 4.01 & 12.01 & -2.00 \\
           &      &       &      &       &      &       &       \\
           & 0.20 & 14.70 & 5.00 &  1.00 & 7.00 & 21.40 & -2.06 \\
           &      &       &      &       &      &       &       \\
Lorentzian & 0.00 &  2.00 & 1.00 &  2.00 & 1.00 &  0.32 & -4.46 \\
           &      &       &      &       &      &       &       \\
           & 0.10 &  9.20 & 0.60 &  1.00 & 2.00 &  2.29 & -1.05 \\[5pt]
\end{longtable}

\noindent Here, it is important to emphasize that the figures shown in the table
do not necessarily correspond to the best possible values of the transfer
efficiency achieved by the system. It happens that the efficiency can be
optimized as much as we want by increasing the values of the Rabi frequencies,
or equivalently, the widths of the pulses. Thus, for bigger Rabi frequencies,
bigger efficiencies. This is something that has no experimental worth. What we
want to examine here are those values of the Rabi frequency and time delay that
can be reproduced in a laboratory, and for which very good transfer efficiency
values can be obtained. So we have decided to consider a reasonable value of
$\log_{10}p=-2.00$ for our numerical exploration.

\begin{figure}[htbp]
    \centering
    \mbox{\subfigure[Surface]{\includegraphics[scale=1.0]{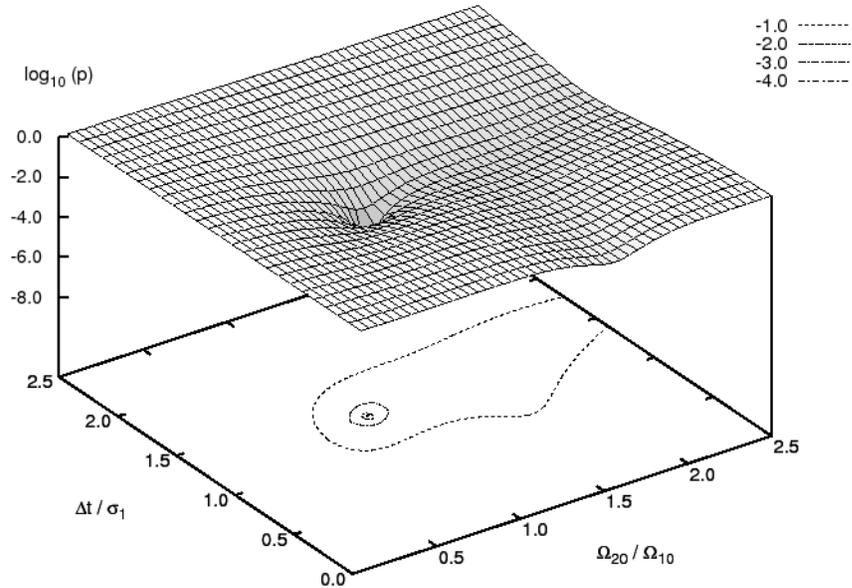}}}
    \mbox{\subfigure[Contour]{\includegraphics[scale=0.45]{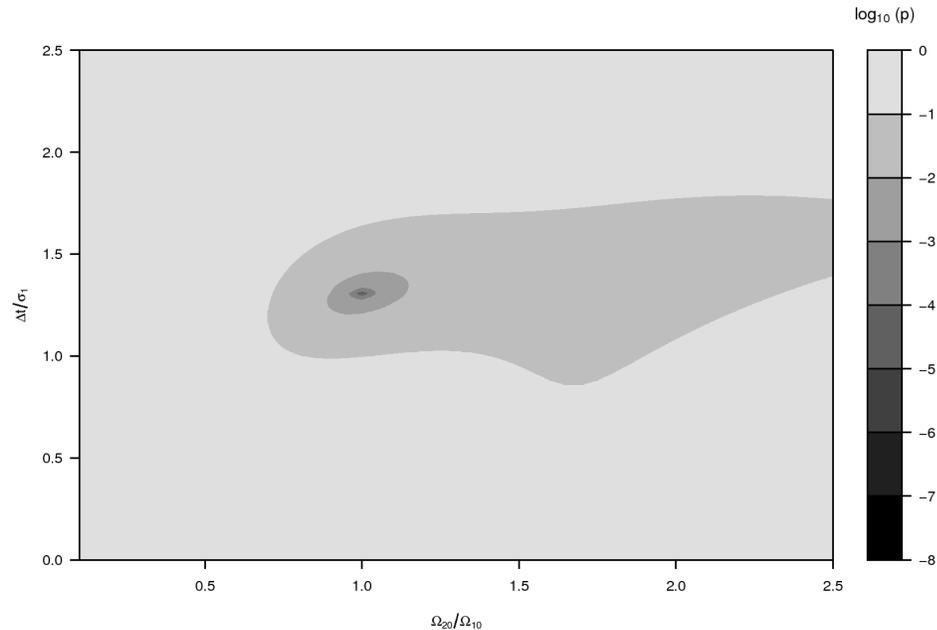}}}
    \caption{Gaussian surface and contour for $\Gamma=0.00$.
    Here $\Omega=2.00$, $\sigma=1.00$, $\textsl{g}=2.00$,
    $\sigma_{\textsl{g}}=1.00$, $\Delta t=1.31$, and $\log_{10}p=-4.88$.}
    \label{Fig:adiabatic01}
\end{figure}

\begin{figure}[htbp]
    \centering
    \mbox{\subfigure[Surface]{\includegraphics[scale=1.0]{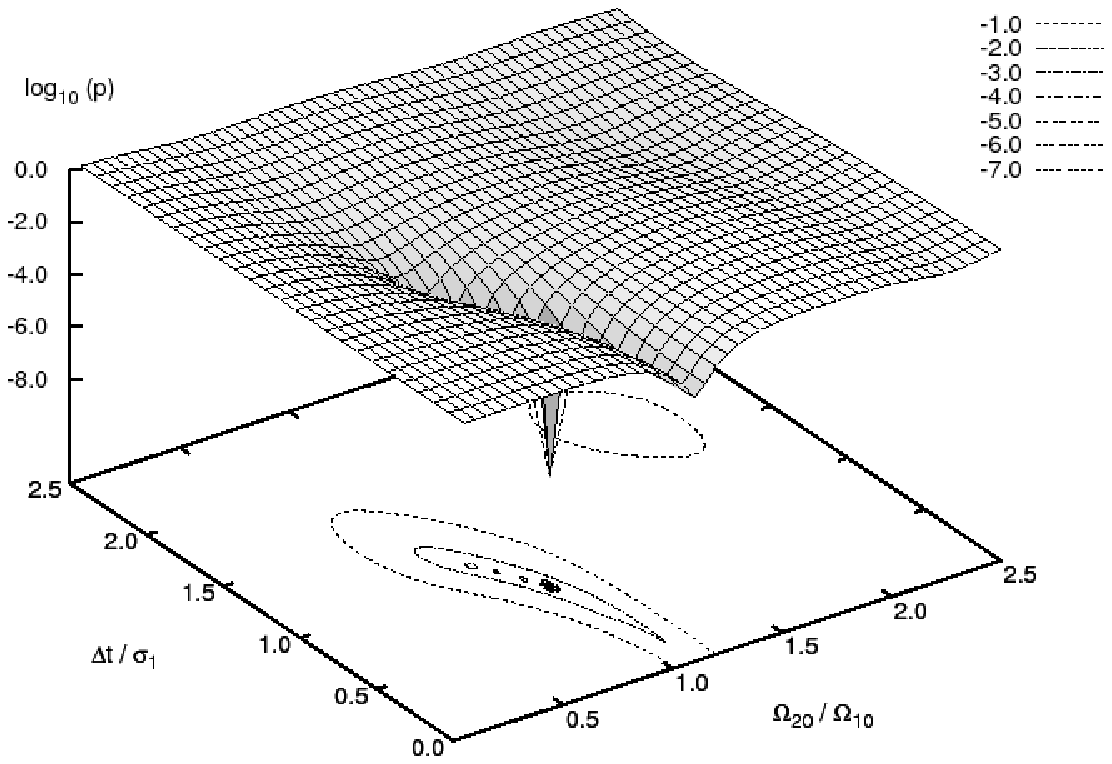}}}
    \mbox{\subfigure[Contour]{\includegraphics[scale=0.45]{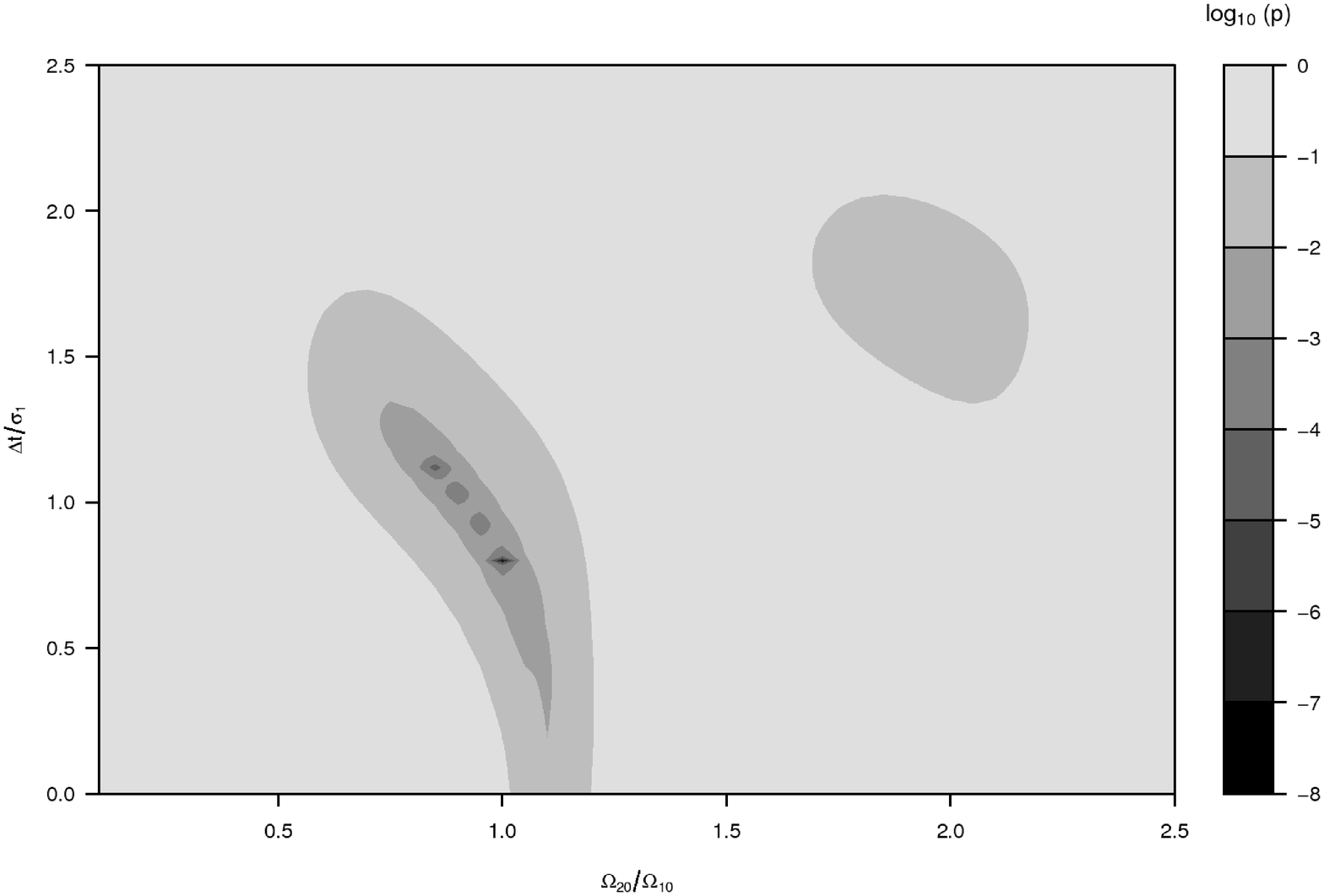}}}
    \caption{Hyperbolic secant surface and contour for $\Gamma=0.00$.
    Here $\Omega=2.00$, $\sigma=1.00$, $\textsl{g}=2.00$,
    $\sigma_{\textsl{g}}=1.00$, $\Delta t=0.80$, and $\log_{10}p=-7.79$.}
    \label{Fig:adiabatic02}
\end{figure}

\begin{figure}[htbp]
    \centering
    \mbox{\subfigure[Surface]{\includegraphics[scale=1.0]{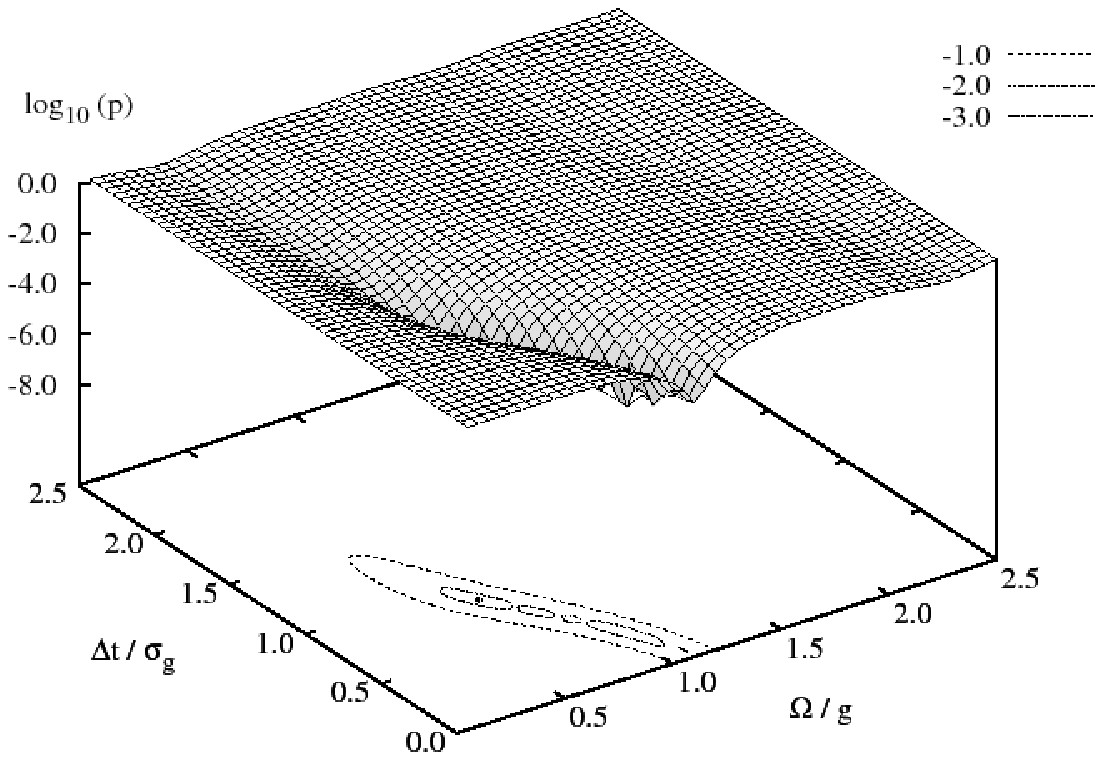}}}
    \mbox{\subfigure[Contour]{\includegraphics[scale=0.45]{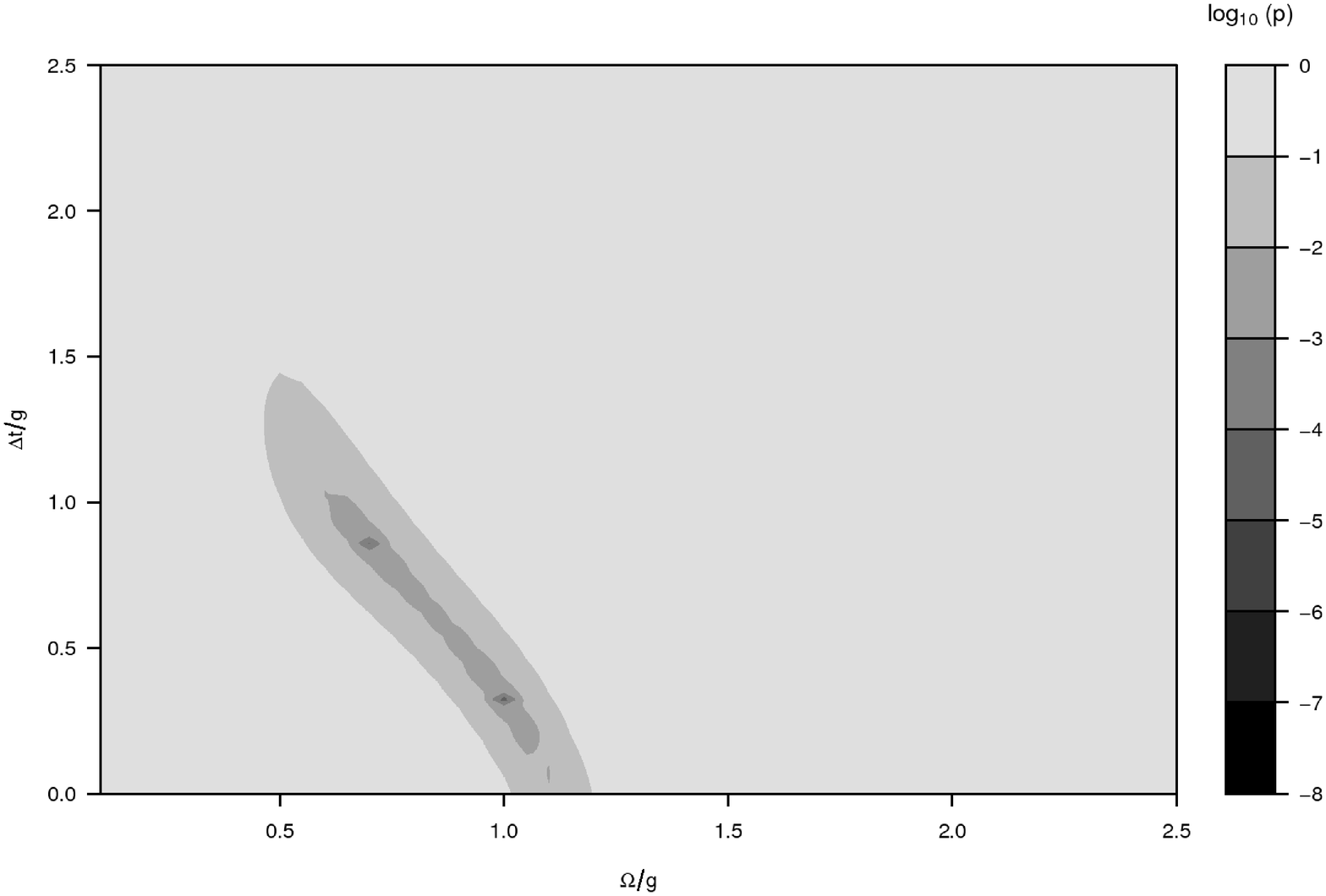}}}
    \caption{Lorentzian secant surface and contour for $\Gamma=0.00$.
    Here $\Omega=2.00$, $\sigma=1.00$, $\textsl{g}=2.00$,
    $\sigma_{\textsl{g}}=1.00$, $\Delta t=0.32$, and $\log_{10}p=-4.46$.}
    \label{Fig:adiabatic03}
\end{figure}

We analyze first the case of no spontaneous emission, when the system is driven
by Gaussian pulses. Fig.~\ref{Fig:adiabatic01} shows that for the particular set
of parameters $\{\Omega=2.00,\:\sigma=1.00,\:\textsl{g}=2.00,
\:\sigma_{\textsl{g}}=1.00,\:\Delta t=1.31\}$ a tremendous efficiency of
$99.9$\% or higher is obtained.
We observe that, while the failure probability remains almost the same over a
wide region of the pulses' amplitudes, there is one particular sector for which
a sudden fall in the failure probability occurs. A very sharp, deep, and
unexpected well just appears on the probability surface. Intriguingly, an even
more sharper and deeper well emerges when the driving pulses are hyperbolic
secants, and the set of parameter is $\{\Omega=2.00, \:\sigma=1.00,
\:\textsl{g}=2.00, \:\sigma_{\textsl{g}}=1.00, \:\Delta t=0.80\}$, as shown in
Fig.~\ref{Fig:adiabatic02}.
For this case we observe that the region nearby to the sharp dip is deformed a
little bit forming a valley of good values for the transfer efficiency.
Fig.~\ref{Fig:adiabatic03} shows another efficient set of parameters, this time
for Lorentzian pulses.

\begin{figure}[htbp]
    \centering
    \mbox{\subfigure[Failure
probability]{\includegraphics[scale=0.6]{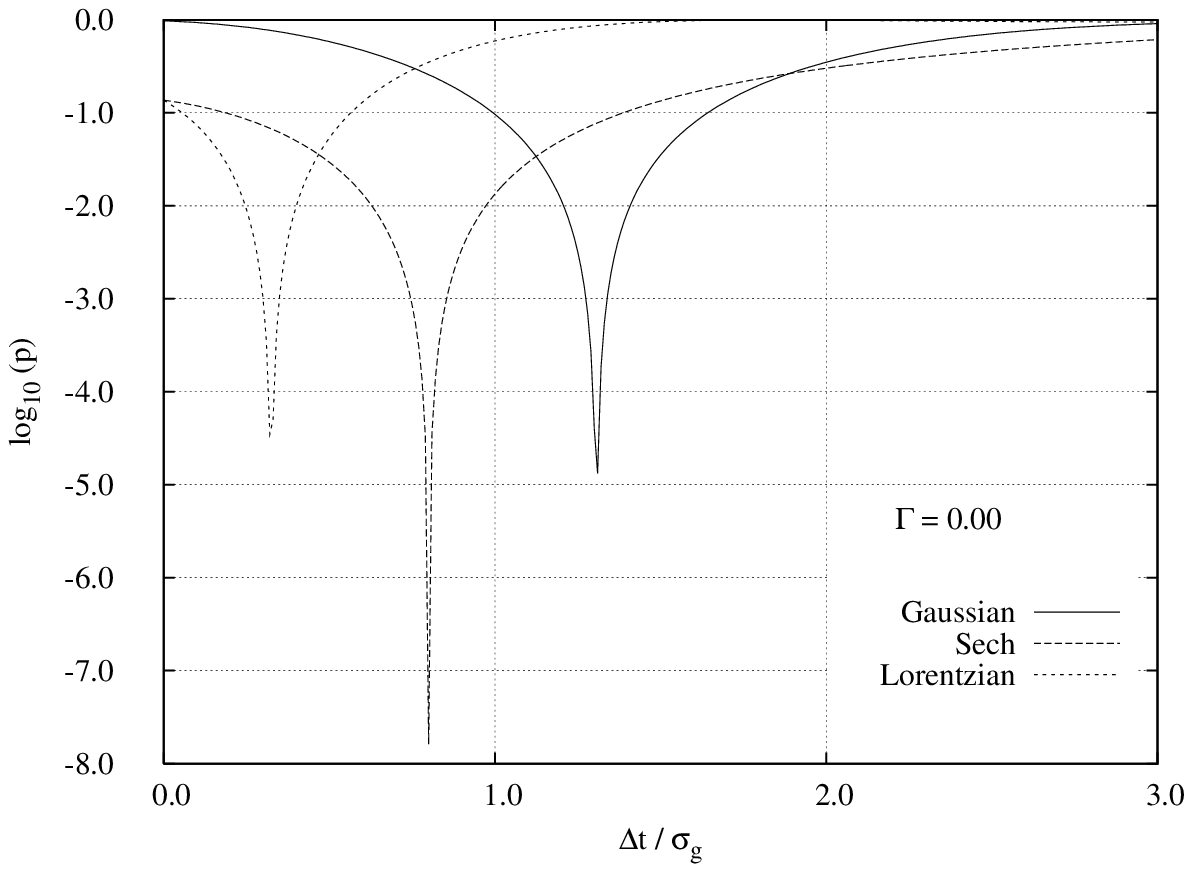}}}
    \mbox{\subfigure[Gaussian]{\includegraphics[scale=0.6]{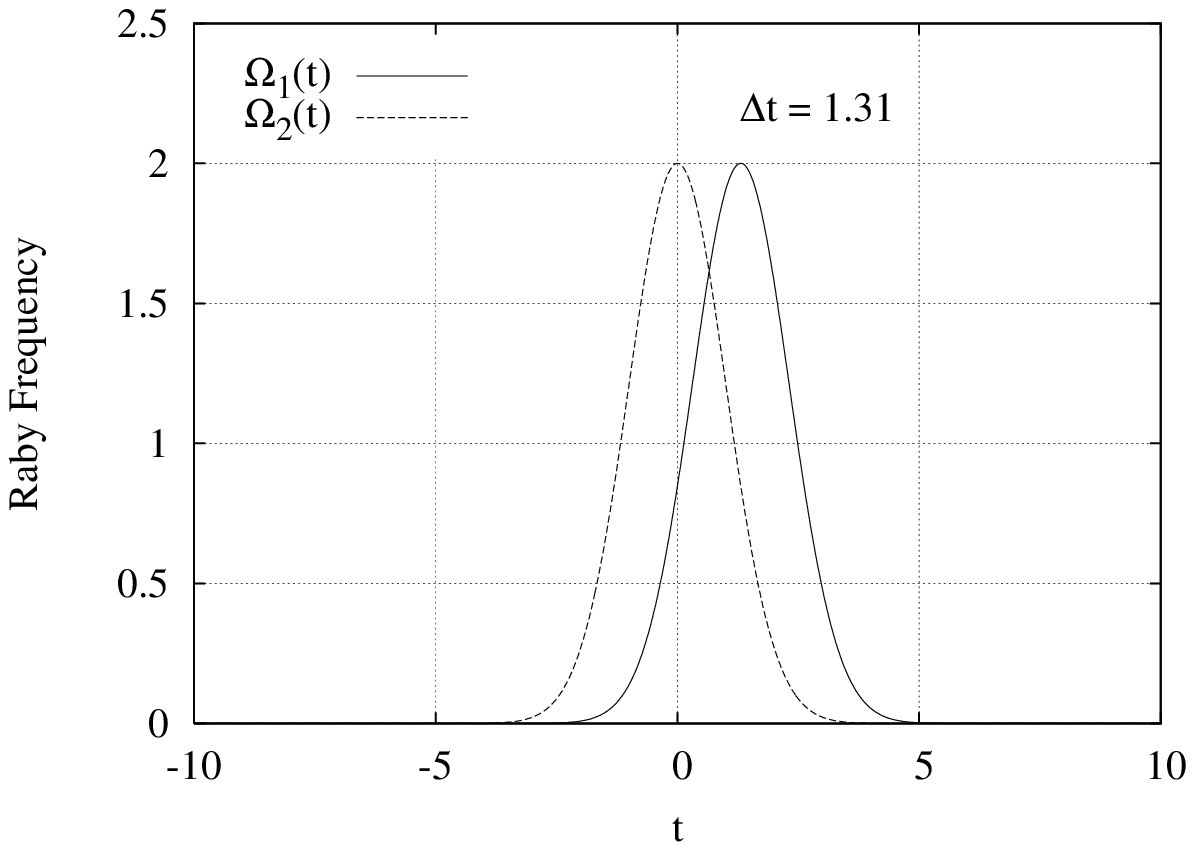}}}
    \mbox{\subfigure[Sech]{\includegraphics[scale=0.6]{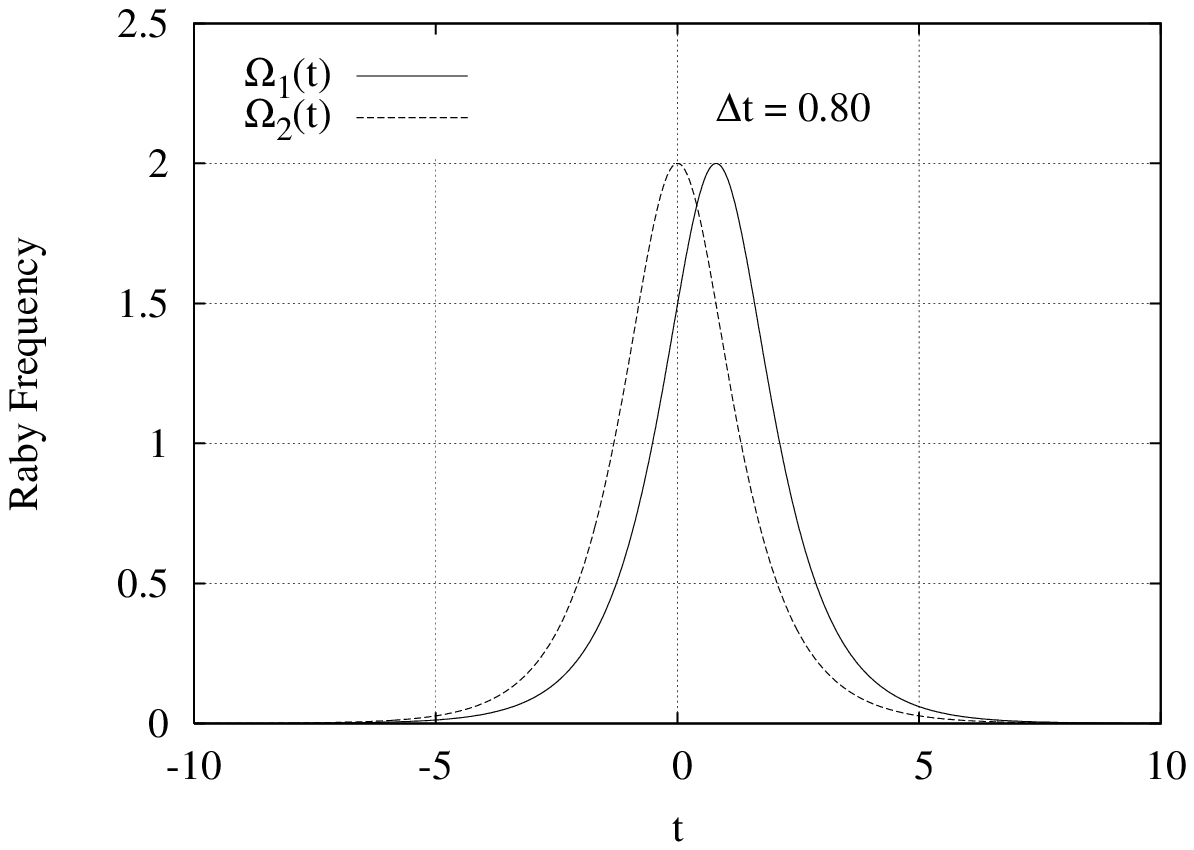}}}
    \mbox{\subfigure[Lorentzian]{\includegraphics[scale=0.6]{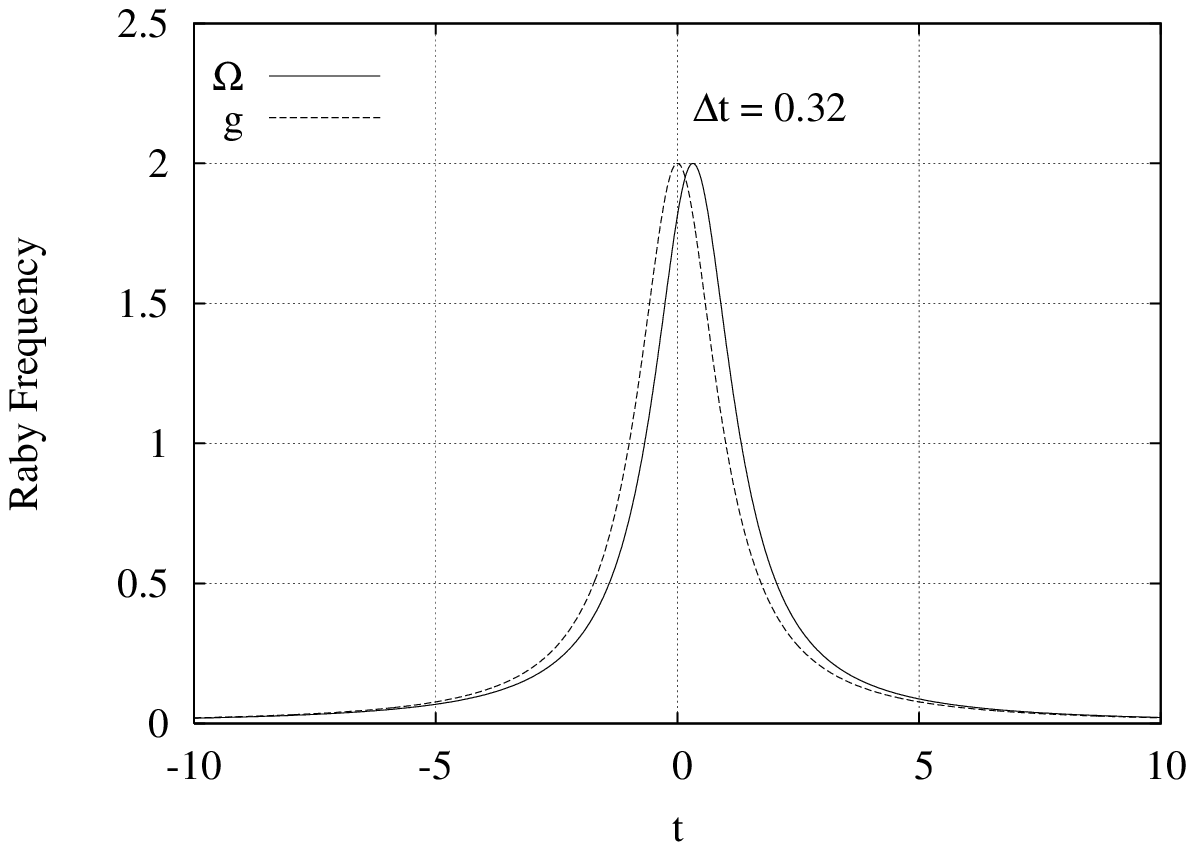}}}
    \caption{(a) Cross-sections of $\log_{10} p$ as a function of $\Delta
    t/\sigma_{\textsl{g}}$ for $\Gamma=0$. (b)-(d) Pulses' profiles.}
    \label{Fig:adiabatic04}
\end{figure}

We observe from Fig.~\ref{Fig:adiabatic04} that, with a shorter interaction time
($\Delta t/\sigma_{\textsl{g}}$) than the Gaussian pulses (but longer than
Lorentzian pulses), the Sech pulses achieved the best transfer efficiency.
Here, all the profiles have the same ratio for the pulses
($\Omega/\textsl{g}=1.0$).

\begin{figure}[htbp]
    \centering
    \mbox{\subfigure[Failure
probability]{\includegraphics[scale=1.0]{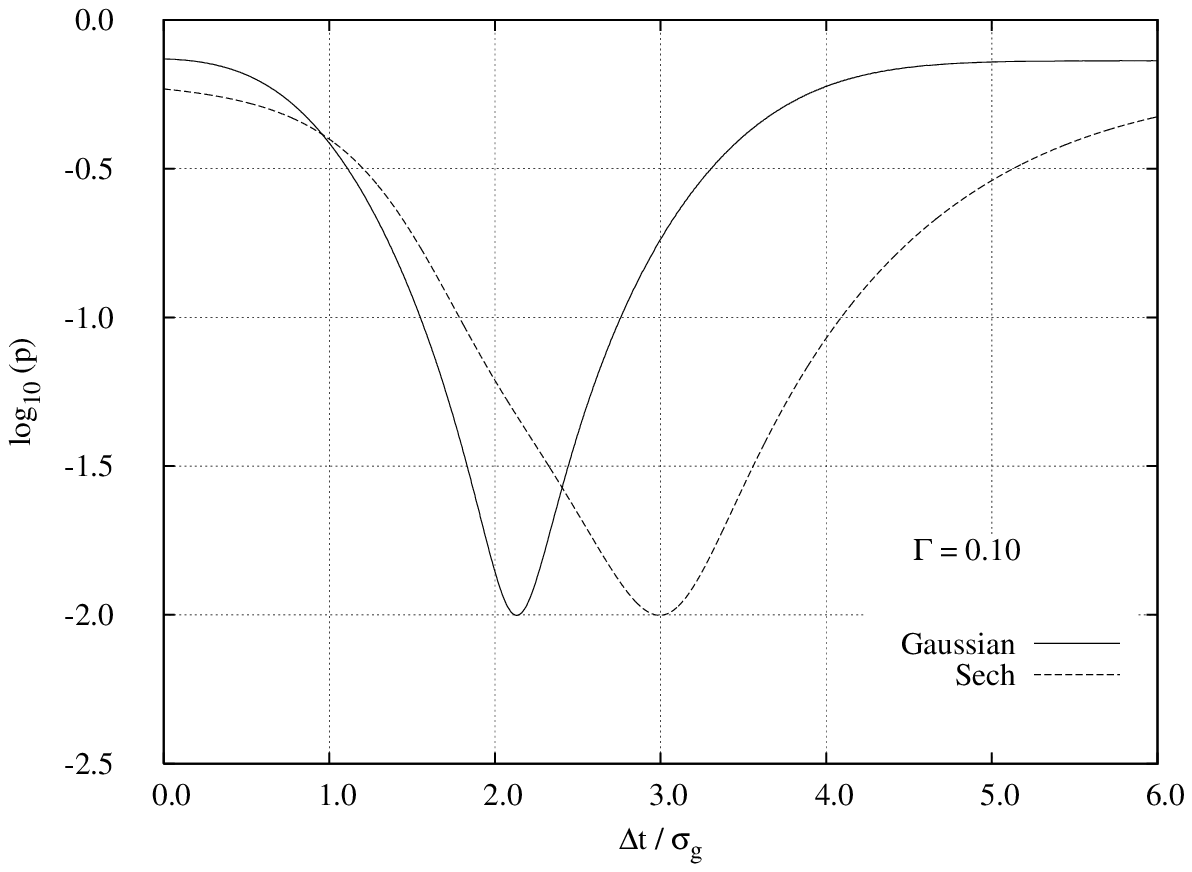}}}
    \mbox{\subfigure[Gaussian]{\includegraphics[scale=0.6]{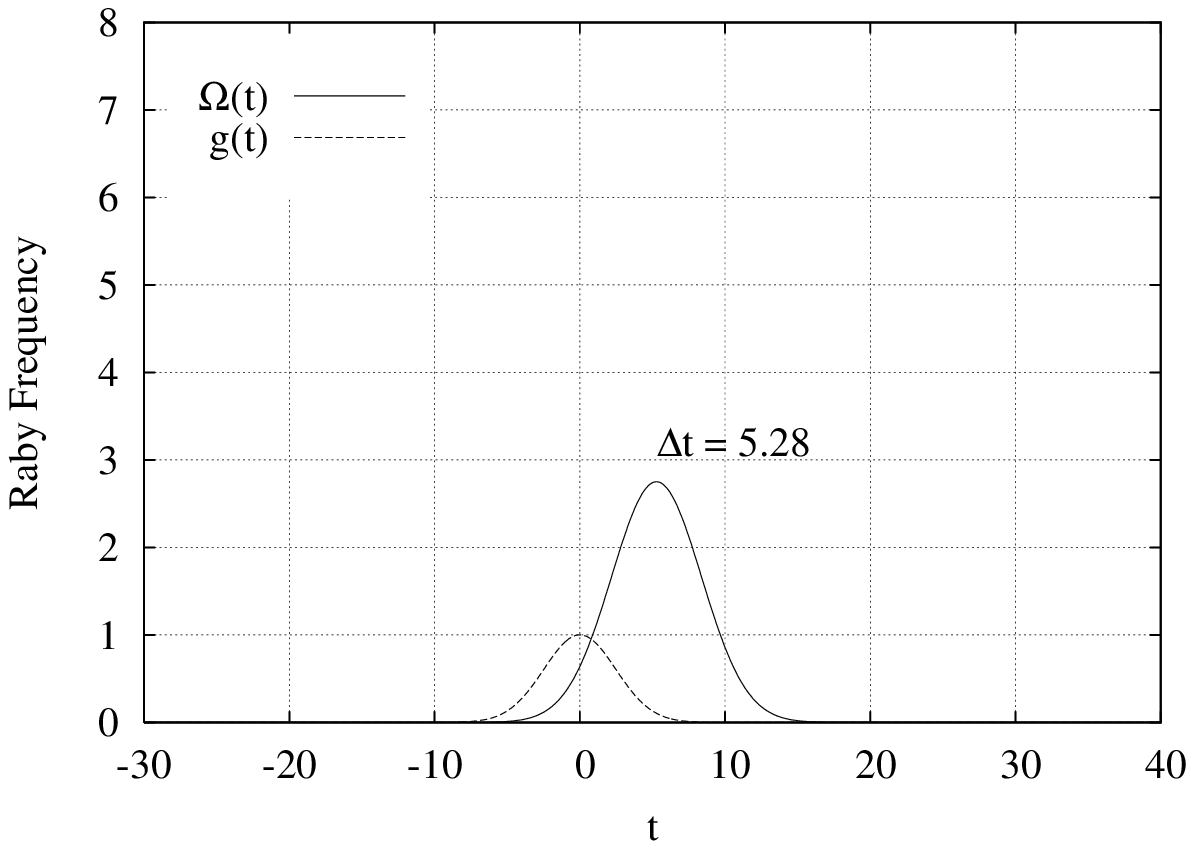}}}
    \mbox{\subfigure[Sech]{\includegraphics[scale=0.6]{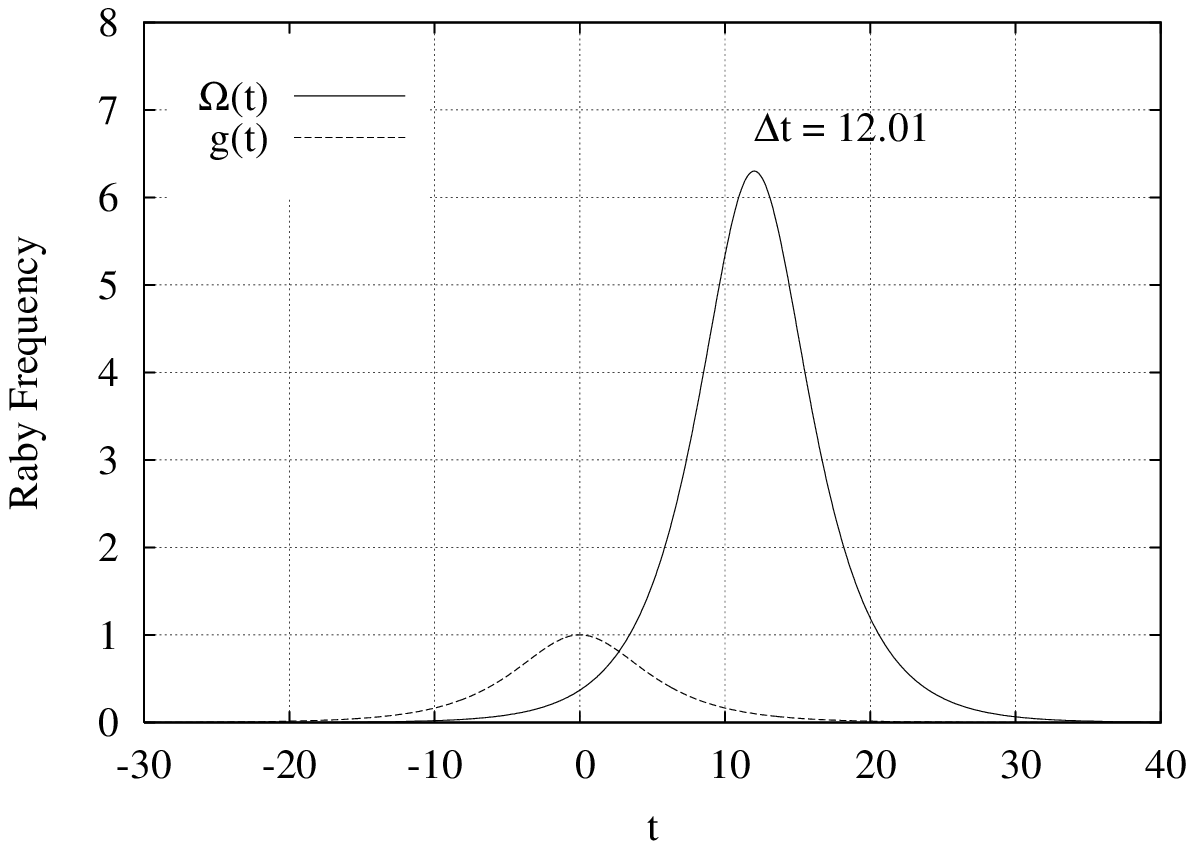}}}
    \caption{Adiabatic passage for $\Gamma=0.10$}
    \label{Fig:adiabatic05}
\end{figure}

\begin{figure}[htbp]
    \centering
    \mbox{\subfigure[Failure
probability]{\includegraphics[scale=1.0]{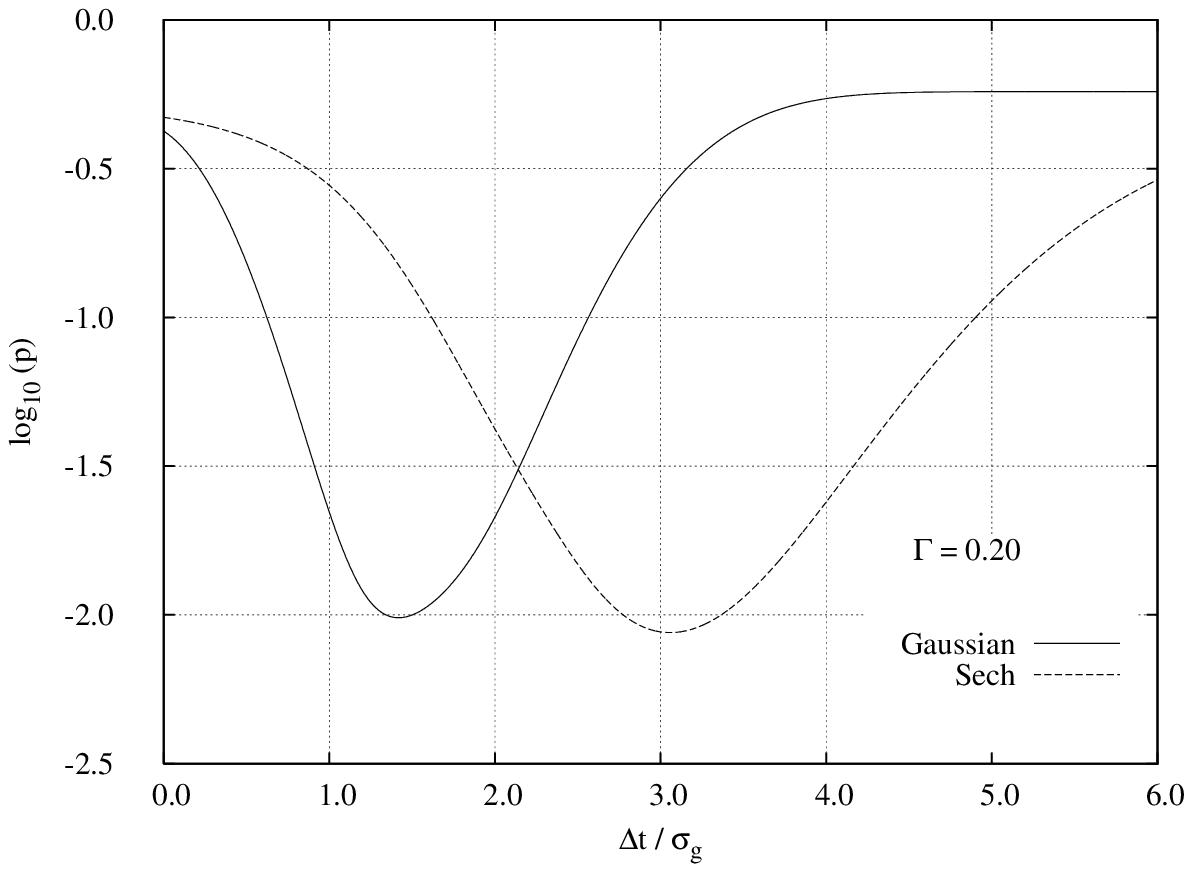}}}
    \mbox{\subfigure[Gaussian]{\includegraphics[scale=0.6]{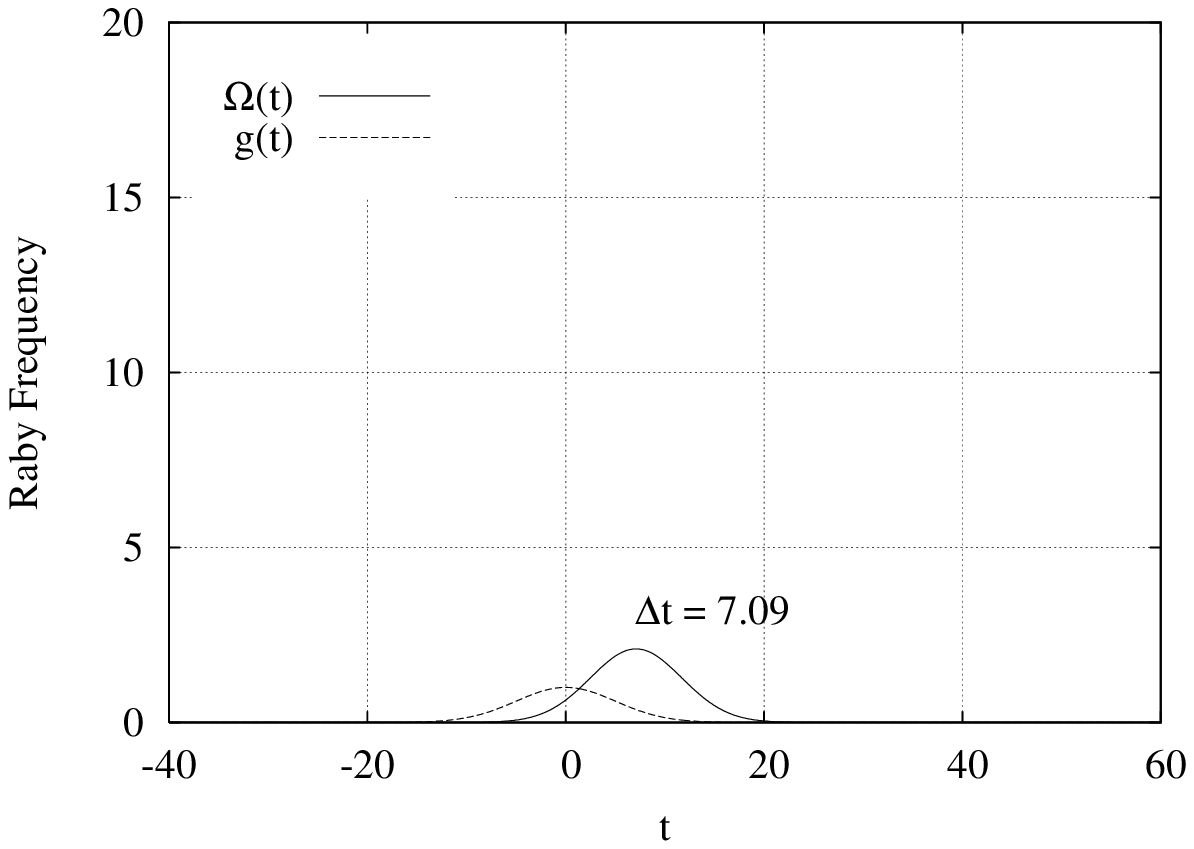}}}
    \mbox{\subfigure[Sech]{\includegraphics[scale=0.6]{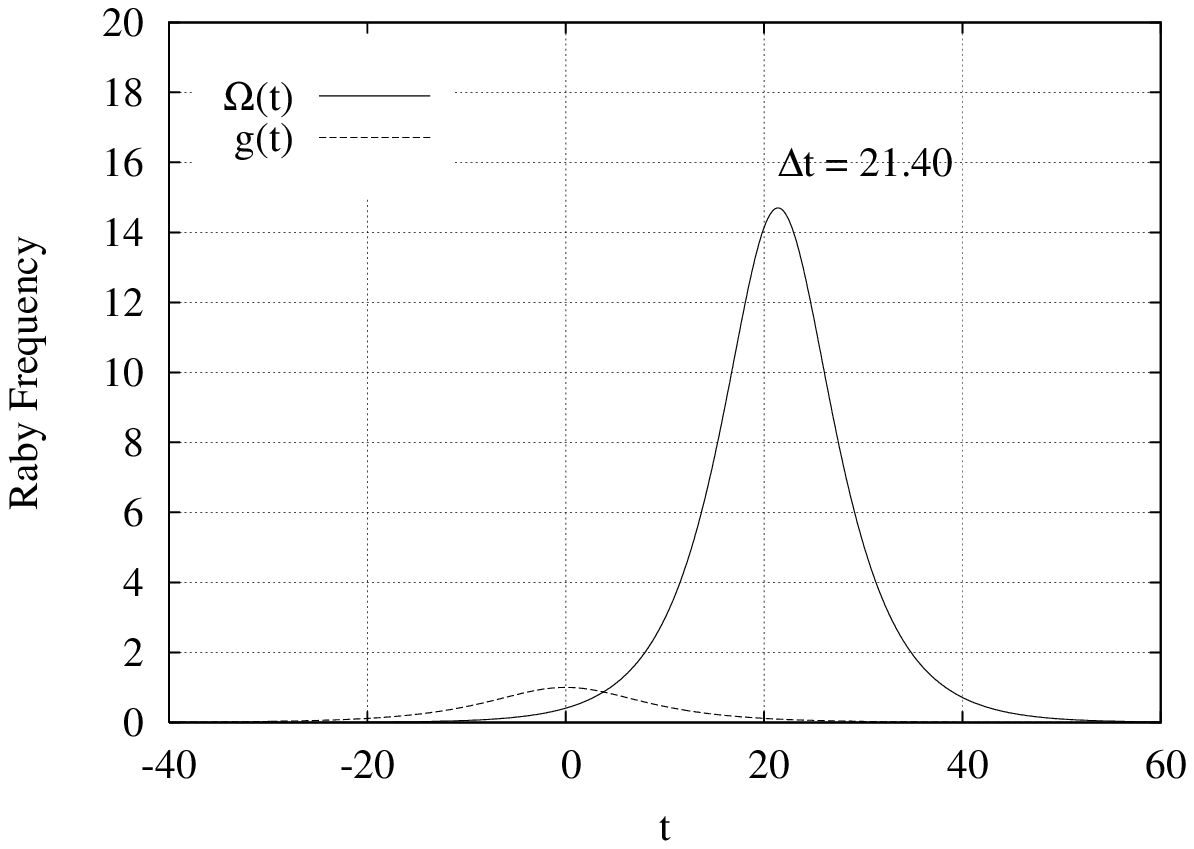}}}
    \caption{Adiabatic passage for $\Gamma=0.20$}
    \label{Fig:adiabatic06}
\end{figure}

When we consider spontaneous emission, the situation changes a little bit. We
are looking for good efficiencies of the order of $\log_{10}p=-2.00$.
Figs.~\ref{Fig:adiabatic05} and~\ref{Fig:adiabatic06} show the failure
probability for two different values of the spontaneous emission, when Gaussian
and secant pulses are used for driving the system.
In this case, we focus our attention in the Rabi frequencies, the pulses' widths
and the time delay. We immediately observe from the figures that a bigger energy
and a longer time delay is required by the Sech pulses to obtain the same
efficiency than that achieved by Gaussian pulses. We also see that a
considerable overlapping of the pulses is necessary to obtain good transfer
efficiencies. Gaussian pulses seem to be more effective this time, as the
comparison between the ratio of the pulses and the interaction time confirms.

Unfortunately, it was very difficult to obtain results for the Lorentzian pulses
for values of the spontaneous emission different from zero. These pulses are
very wide, and their numerical integration is a very complicated task. However,
it is possible to see from Table~\ref{Table:adiabatic}, that for bigger values
of the spontaneous emission, bigger values of the Rabi frequency are needed.

\chapter{\label{Ch:03} A simple analytical model}

In this chapter we introduce a simple analytical model that explains
qualitatively the origin of the nonadiabatic, resonance-like features observed
in adiabatic passage methods. We also present a mathematical description of the
three different pulse shapes used in our numerical simulations of chapters two
and three, including the nonadiabatic coupling terms. By using the analytical
model, we study the dependence of the failure probability $p$ on the system
parameters $\Omega\,T$, showing the exact solution of a simple problem. Although
the analytical results here reported have been obtained by others, the method is
somewhat different from those in the literature.

\section{\label{Sec:populationTransfer} Population transfer}

\begin{figure}[h]
    \centering
    \includegraphics[scale=0.8]{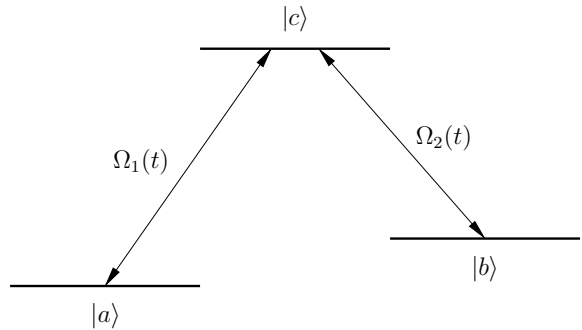}
    \caption{Three-level $\Lambda$ system driven by two lasers. The Rabi
    frequencies for the \textit{pump} and \textit{Stokes} laser are $\Omega_1$
    and $\Omega_2$, respectively.}
    \label{Fig:3levelResonance}
\end{figure}

Consider a three-level $\Lambda$ system driven by a two-mode classical coherent
field, as shown in Fig.~\ref{Fig:3levelResonance}.

\noindent We want to study the problem of coherent population transfer from the
initially populated state $\ket{a}$ to the final state $\ket{b}$ by using a pump
laser coupling states $\ket{a}$ and $\ket{c}$, and a Stokes laser coupling
states $\ket{c}$ and $\ket{b}$ (see~\cite{Kuklinski:PRA40}).

Under two-photon resonance condition and by using the rotating-wave
approximation, the evolution of the system is described by
\begin{subequations}
\label{Eq:C_resonance}
\begin{align}
    \dot{C}_a & = -\frac{\Omega_1(t)}{2} \,C_c, \label{Eq:Ca_resonance} \\
    \dot{C}_b & = -\frac{\Omega_2(t)}{2} \,C_c, \label{Eq:Cb_resonance} \\
    \dot{C}_c & =  \frac{\Omega_1(t)}{2} \,C_a
                 + \frac{\Omega_2(t)}{2} \,C_b, \label{Eq:Cc_resonance}
\end{align}
\end{subequations}
where $C_a$, $C_b$, and $C_c$ are the probability amplitudes of finding the
system in states $\ket{a}$, $\ket{b}$, and $\ket{c}$, respectively.

\begin{figure}[h]
    \centering
    \includegraphics[scale=0.8]{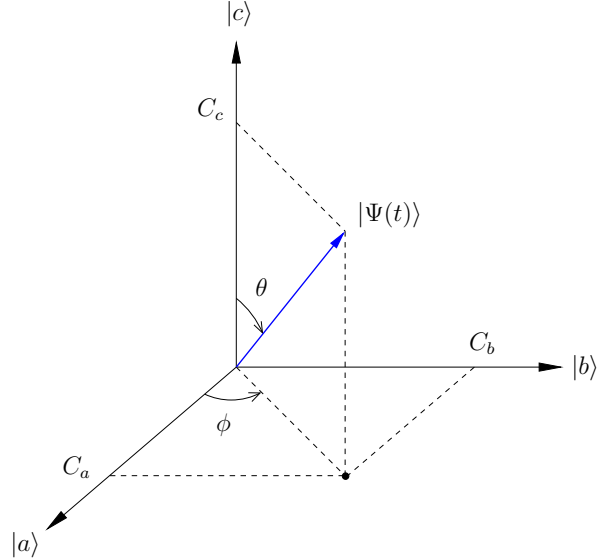}
    \caption{Graphic representation of the Hilbert space for the three-level
    system in the eigenbasis $\{ \ket{a}, \ket{b}, \ket{c} \}$. The state vector
    $\ket{\Psi(t)}$ is represented by its three Cartesian components $C_a$,
    $C_b$, and $C_c$, corresponding to the probability amplitudes.}
    \label{Fig:sphericalSystem}
\end{figure}

To obtain approximate solutions to this system, we may proceed as follows. Let
us first consider the normalized state vector $\ket{\Psi(t)}$ in spherical
coordinates, as indicated in Fig.~\ref{Fig:sphericalSystem}.
The state vector components $C_a$, $C_b$, and $C_c$ (probability amplitudes) can
be written in  the form
\begin{subequations}
\label{Eq:C_spherical}
\begin{align}
    C_a & = \sin\theta \cos\phi, \label{Eq:Ca}\\
    C_b & = \sin\theta \sin\phi, \label{Eq:Cb}\\
    C_c & = \cos\theta,          \label{Eq:Cc}
\end{align}
\end{subequations}
with time derivatives given by
\begin{subequations}
\label{Eq:C_spherical_dot}
\begin{align}
    \dot{C_a} & = \dot{\theta} \cos\theta \cos\phi
        - \dot{\phi}\sin\theta \sin\phi,   \label{Eq:Ca_dot} \\
    \dot{C_b} & = \dot{\theta} \cos\theta \sin\phi
        + \dot{\phi}\sin\theta \cos\phi,   \label{Eq:Cb_dot} \\
    \dot{C_c} & = -\dot{\theta}\sin\theta. \label{Eq:Cc_dot}
\end{align}
\end{subequations}
Now, plugging Eqs.~\eqref{Eq:C_spherical} and~\eqref{Eq:C_spherical_dot} into
Eqs.~\eqref{Eq:C_resonance}, and by using the \textit{mixing angle}
\begin{equation}
    \tan\Phi(t) = \frac{\Omega_1(t)}{\Omega_2(t)},
    \label{Eq:mixingAngle2}
\end{equation}
it is simple to verify that Eqs.~\eqref{Eq:C_resonance} reduce to a system of
two coupled nonlinear differential equations given by
\begin{subequations}
\label{Eq:NonlinearSystem}
\begin{align}
    \dot{\phi}(t)   & = \frac{\tan\bigl[\theta-\pi/2\bigr]}{2} \Omega(t)
        \cos\bigl[\phi+\Phi(t) \bigr],
    \label{Eq:phi_dot}\\
    \dot{\theta}(t) & = -\frac{\Omega(t)}{2} \sin\bigl[\phi + \Phi(t) \bigr].
    \label{Eq:theta_dot}
\end{align}
\end{subequations}
Here $\Omega(t) = \sqrt{\Omega_1^2(t)+\Omega_2^2(t)}$.

Taking a look at the asymptotic behavior of Eqs.~\eqref{Eq:C_resonance}, we see
that at very early times ($t \to -\infty$) when the state vector is parallel to
the initially populated state $\ket{a}$, the probability amplitude coefficients
take the values
\begin{equation}
\begin{split}
    C_a(-\infty) & = 1 = \sin\theta(-\infty)\cos\phi(-\infty), \\
    C_b(-\infty) & = 0 = \sin\theta(-\infty)\sin\phi(-\infty), \\
    C_c(-\infty) & = 0 = \cos\theta(-\infty).
\end{split}
\end{equation}
This implies that before the interaction with the lasers, the angular
coordinates of the state vector are
\begin{equation}
    \theta(-\infty) = \pi/2 \qquad \text{and} \qquad \phi(-\infty) = 0.
\end{equation}
Similarly, for very late times ($t \to +\infty$) when the state vector is
antiparallel to the final state $\ket{b}$ (see
Sec.~\ref{Sec:adiabaticFollowing}), the probability amplitudes are
\begin{equation}
\begin{split}
    C_a(\infty) & = 0  = \sin\theta(\infty)\cos\phi(\infty), \\
    C_b(\infty) & = -1 = \sin\theta(\infty)\sin\phi(\infty), \\
    C_c(\infty) & = 0  = \cos\theta(\infty).
\end{split}
\end{equation}
From these equations, we may conclude that after the interaction with the lasers
\begin{equation}
    \theta(\infty) =  \pi/2 \qquad \text{and} \qquad \phi(\infty)   = -\pi/2.
\end{equation}
Since we are interested in transferring population from the initial state
$\ket{a}$ to the final state $\ket{b}$ without populating the leaking excited
state $\ket{c}$ (which we assume undergoes radiative decay), then the state
vector $\ket{\Psi(t)}$ should evolve adiabatically by closely following the
``dressed state'' $\ket{W^0}$ of the Hamiltonian of
system~\eqref{Eq:C_resonance}. Thus the angular coordinates of the state vector
can be written as
\begin{equation}
    \theta_{ad}(t) = \pi/2 \qquad \text{and} \qquad \phi_{ad}(t) = -\Phi(t),
    \label{Eq:systemAdiabaticConditions}
\end{equation}
where the label ``\textit{ad}'' accounts for an \textit{adiabatic following} of
these angles to the corresponding angular coordinates of the dressed state.

If we assume small deviations from
conditions~\eqref{Eq:systemAdiabaticConditions}, the
system~\eqref{Eq:NonlinearSystem} can be linearized. By plugging the equations
\begin{subequations}
\label{Eq:changeVariables}
\begin{align}
    \theta & = \theta_{ad} + \delta\theta = \pi/2 + \delta\theta,
    \label{Eq:changeVariables_theta} \\
    \phi   & = \phi_{ad}   + \delta\phi   = -\Phi(t) + \delta\phi,
    \label{Eq:changeVariables_phi}
\end{align}
\end{subequations}
and their time derivatives
\begin{subequations}
\begin{align}
    \dot{\theta} & = \dot{\delta\theta},
    \label{Eq:theta_dotSmall} \\
    \dot{\phi}   & = -\dot{\Phi}(t) + \dot{\delta\phi},
    \label{Eq:phi_dotSmall}
\end{align}
\end{subequations}
into Eqs.~\eqref{Eq:NonlinearSystem}, we obtain
\begin{subequations}
\label{Eq:linearSystem}
\begin{align}
    \dot{\delta\phi}   & =  \frac{\Omega(t)}{2} \delta\theta + \dot{\Phi}(t),
    \label{Eq:smallphi_dot}\\
    \dot{\delta\theta} & = -\frac{\Omega(t)}{2} \delta\phi.
    \label{Eq:smalltheta_dot}
\end{align}
\end{subequations}

To integrate Eqs.~\eqref{Eq:linearSystem}, we need to make a change of variables
by defining a ``dimensionless phase'' $\tau$ in the form
\begin{equation}
    \tau(t) = \int_{-\infty}^{t} \frac{\Omega(t')}{2} \,dt'. \label{Eq:tau}
\end{equation}
This phase will play the role of a rescaled time in the following. Then, by
using the chain rule
\begin{equation}
    \frac{d}{dt} = \frac{d\tau}{dt}\frac{d}{d\tau} = \frac{\Omega(t)}{2}\,
    \frac{d}{d\tau},
    \label{Eq:chainRule}
\end{equation}
we can rewrite Eqs.~\eqref{Eq:linearSystem} in the form
\begin{subequations}
\begin{align}
    \frac{d}{d\tau}\delta\phi[\tau]   & = \delta\theta[\tau]
        + \frac{d}{d\tau}\Phi[\tau], \\
    \frac{d}{d\tau}\delta\theta[\tau] & = -\delta\phi[\tau].
\end{align}
\end{subequations}
Now, by differentiating the second equation and plugging it into the first,
we obtain the equation of an \textit{undamped driven harmonic oscillator}
\begin{equation}
    \frac{d\,^2}{d\tau^2}\delta\theta[\tau] + \delta\theta[\tau]
        = -\frac{d}{d\tau}\Phi[\tau].
    \label{Eq:FHO}
\end{equation}
The general solution to this equation is very well known and given by (see
Appendix~\ref{App:FHO})
\begin{equation}
    \delta\theta[\tau] = A \sin[\tau] + B \cos[\tau]
        - \int_0^{\tau}\sin[\,\tau-\tau'\,] \frac{d\Phi[\tau']}{d\tau'}\,d\tau'.
    \label{Eq:FHOsolution}
\end{equation}
The lower limit of the integral results when the time is $-\infty$ giving
\begin{equation}
    \tau[-\infty] = \int_{-\infty}^{-\infty} \frac{\Omega(t)}{2} \,dt = 0.
\end{equation}
In addition, we have that the difference between phases can be written as
\begin{equation}
    \tau-\tau' = \int_{t'}^{t} \frac{\Omega(t'')}{2} \,dt''.
\end{equation}
Now, returning to the original time scale in Eq.~\eqref{Eq:FHOsolution}, we get
\begin{align}
    \delta\theta(t) & =
       A  \sin\biggl[\int_{-\infty}^t \frac{\Omega(t')}{2}\,dt' \biggr]
     + B  \cos\biggl[\int_{-\infty}^t \frac{\Omega(t')}{2}\,dt'\biggr] \notag \\
        & \qquad - \int_{-\infty}^t\sin\biggl[
                   \int_{t'}^t \frac{\Omega(t'')}{2}\, dt''
        \biggr] \frac{d\Phi(t')}{dt'}\,dt';
    \label{Eq:FHOsolution01}
\end{align}
and by inserting Eq.~\eqref{Eq:FHOsolution01} into~\eqref{Eq:smalltheta_dot}, we
can solve for $\delta\phi$ obtaining
\begin{align}
    \delta\phi(t) & =
  - A \cos\biggl[\int_{-\infty}^t \frac{\Omega(t')}{2}\,dt' \biggr]
  + B \sin\biggl[\int_{-\infty}^t \frac{\Omega(t')}{2}\,dt'\biggr] \notag \\
    & \qquad+\int_{-\infty}^t\cos\biggl[\int_{t'}^t \frac{\Omega(t'')}{2}\,dt''
        \biggr] \frac{d\Phi(t')}{dt'}\,dt'.
    \label{Eq:FHOsolution02}
\end{align}
The initial conditions for the linearized system are
\begin{subequations}
\begin{align}
    \delta\theta(-\infty)       & = 0, \\
    \dot{\delta\theta}(-\infty) & = -\Omega(-\infty)\Phi(-\infty)/2, \\
    \delta\phi(-\infty)         & = \Phi(-\infty), \\
    \dot{\delta\phi}(-\infty)   & = \dot{\Phi}(-\infty).
\end{align}
\end{subequations}
And finally, by evaluating the constants of integration $A$ and $B$, we may
obtain
\begin{subequations}
\label{Eq:generalSolutions}
\begin{align}
    \delta\phi(t) & =
    \Phi(-\infty)\cos\biggl[\int_{-\infty}^t \frac{\Omega(t')}{2}\,dt'\biggr]
         +\int_{-\infty}^t\cos\biggl[\int_{t'}^t \frac{\Omega(t'')}{2}\,dt''
        \biggr] \frac{d\Phi(t')}{dt'}\,dt',
    \label{Eq:generalSolution01} \\
    \delta\theta(t) & =
   -\Phi(-\infty)\sin\biggl[\int_{-\infty}^t \frac{\Omega(t')}{2}\,dt'\biggr]
   -\int_{-\infty}^t\sin\biggl[\int_{t'}^t \frac{\Omega(t'')}{2}\,dt'' \biggr]
        \frac{d\Phi(t')}{dt'}\,dt'.
    \label{Eq:generalSolution02}
\end{align}
\end{subequations}

The two terms in each of the equations~\eqref{Eq:generalSolutions} represent
different things. The first term represents a coherent mixing of the adiabatic
states that occurs when the initial state of the system is not one of the
eigenstates of the interaction Hamiltonian. The second term, on the other hand,
represents corrections to the adiabatic approximation.

\section{Asymptotic behavior of the pulses}

\begin{figure}[h]
    \centering
    \scalebox{0.6}{\includegraphics{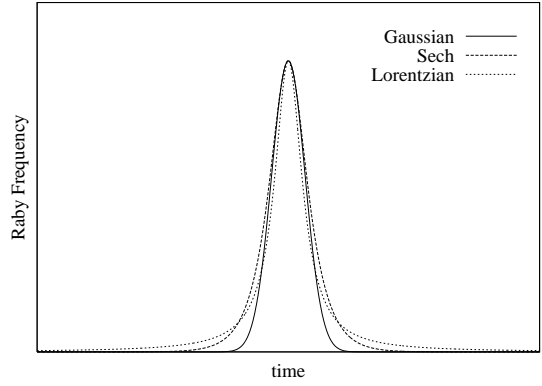}}
    \caption{Profiles of the three pulses used in this work.}
    \label{Fig:pulsesProfiles}
\end{figure}

In the present work, we used three different kinds of pulse shapes for the
numerical computations
\begin{subequations}
\label{Eq:pulses}
\begin{alignat}{2}
    & \text{Gaussian pulses:} & \qquad
    \Omega (t) & = \Omega_0\,\exp \bigl( -t^2/2\sigma^2 \bigr),   \\
    & \text{Hyperbolic Secant pulses:} & \qquad
    \Omega (t) & = \Omega_0\,\sech \bigl( t/\sigma \bigr), \\
    & \text{Lorentzian pulses:} & \qquad
    \Omega (t) & = \Omega_0\,\sigma^2/\bigl(t^2 + \sigma^2 \bigr).
\end{alignat}
\end{subequations}
Here $\Omega_0$ is the Rabi amplitude of the pulse, and $\sigma$ is a scale
parameter (width) describing the extent of the pulse. The different shapes are
illustrated in Fig.~\ref{Fig:pulsesProfiles}.

It is important to consider the asymptotic behavior of
functions~\eqref{Eq:pulses} because this determines if a sequence of pulses
satisfies the adiabatic conditions, and then the system can evolve
adiabatically. In addition, the \textit{nonadiabatic coupling} matrix element
\begin{equation}
    \dot{\Phi}(t) = \frac{\dot{\Omega}_1 \Omega_2 - \Omega_1 \dot{\Omega}_2}
                         {\Omega_1^2 + \Omega_2^2},
    \label{Eq:nonadiabatic_Coupling}
\end{equation}
can be used as a ``local'' adiabaticity criterion for a given shape of the laser
pulses.

\subsection{Gaussian pulses}

Consider two Gaussian pulses of the form
\begin{subequations}
\begin{align}
    \Omega_1(t) & = \Omega_{10}
        \exp \bigl[ -(t-\Delta t)^2/ 2\sigma_1^2 \bigr], \\
    \Omega_2(t) & = \Omega_{20}\,\exp \bigl( -t^2/2\sigma_2^2 \bigr).
\end{align}
\end{subequations}
These two pulses are in a \textit{counterintuitive} configuration, where the
pulse 2 precedes the pulse 1, and their separation is given by the \textit{time
delay} $\Delta t$.

\begin{figure}[h]
    \centering
    \scalebox{1.0}{\includegraphics{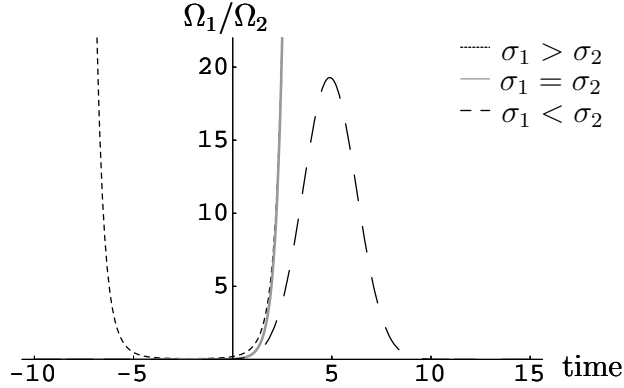}}
    \caption{Behavior of the Gaussian pulses ratio as a function of time, and
    for different pulse widths.}
    \label{Fig:asymptotic}
\end{figure}

The ratio of the pulses can be written as
\begin{equation}
    \frac{\Omega_1(t)}{\Omega_2(t)} = \frac{\Omega_{10}}{\Omega_{20}}
    \exp\biggl[ -\frac{(\Delta t)^2}{2\sigma_1^2}
                -\frac{\sigma_2^2 - \sigma_1^2}{2(\sigma_1\sigma_2)^2}\,t^2
                +\frac{\Delta t}{\sigma_1^2}\,t \biggr].
\end{equation}
We observe that at very early or very late times, the behavior of this ratio
depends on the pulses' widths as follows
\begin{subequations}
\begin{alignat}{2}
    &\sigma_1>\sigma_2: & \quad
    \Omega_1/\Omega_2 & \to +\infty \text{ when } t \to \pm\infty, \\
    &\sigma_1=\sigma_2: & \quad
    \Omega_1/\Omega_2 & \to
        \begin{cases}
            0 & \text{ when $t \to -\infty$}, \\
            +\infty& \text{ when $t \to +\infty$},
        \end{cases}
    \label{Eq:asymptotic_Gaussian} \\
    &\sigma_1<\sigma_2: & \quad
    \Omega_1/\Omega_2 & \to 0 \text{ when } t \to \pm\infty.
\end{alignat}
\end{subequations}
Equation~\eqref{Eq:asymptotic_Gaussian} shows that for pulses with equal widths
($\sigma_1 = \sigma_2$) the adiabatic conditions for the system are satisfied
(that is, the external field and the coupling field go to zero when the time
goes to $-\infty$ and $+\infty$, respectively) and the adiabatic theorem
supports the population inversion between the initial and the target levels. On
the other hand, for unequal widths ($\sigma_1 \neq \sigma_2$) the adiabatic
conditions are not \textit{fully} satisfied by the system, and the adiabatic
theorem \textit{alone} cannot validate the transferring process. In this case we
need additional arguments to help us to understand why, in some situations like
these, large transfer probabilities are still achieved.
Fig.~\ref{Fig:asymptotic} shows the asymptotic response of the pulses' ratio.

The nonadiabatic coupling for these Gaussian pulses is given by
\begin{equation}
    \dot{\Phi}(t) = \frac{
    \bigl[\Omega_{10}\Omega_{20}/(\sigma_1 \sigma_2)^2\bigr]
    \bigl[\sigma_2 \Delta t - (\sigma_2^2 - \sigma_1^2) t\bigr]}
         {\Omega_{10}^2\,\exp(t^2/\sigma_2^2) +
          \Omega_{20}^2\,\exp\bigl[(t-\Delta t)^2/\sigma_2^2\bigr]}\,
    \exp\biggl[ \frac{(t-\Delta t)^2}{2\sigma_1^2} +
                \frac{t^2}{2\sigma_2^2}\biggr].
\end{equation}
%

\subsection{Hyperbolic secant pulses:}

Consider two Sech pulses of the form
\begin{subequations}
\begin{align}
    \Omega_1(t) & = \Omega_{10}\,\sech \bigl[ (t-\Delta t) / \sigma_1 \bigr], \\
    \Omega_2(t) & = \Omega_{20}\,\sech \bigl( t/\sigma_2 \bigr).
\end{align}
\end{subequations}
The ratio of the two pulses is given by
\begin{equation}
    \frac{\Omega_1(t)}{\Omega_2(t)} =
    \biggl( \frac{\Omega_{10}}{\Omega_{20}} \biggr)
    \frac{\exp\bigl(t/\sigma_2\bigr) + \exp\bigl(-t/\sigma_2\bigr)}
         {\exp\bigl[(t-\Delta t)/\sigma_1\bigr]
        + \exp\bigl[-(t-\Delta t)/\sigma_1\bigr]},
\end{equation}
and the asymptotic behavior of the pulses can be expressed as:
\begin{subequations}
\begin{alignat}{2}
    &\sigma_1>\sigma_2: & \quad
    \Omega_1/\Omega_2 & \to +\infty \text{ when } t \to \pm\infty,
    \label{Eq:asymptotic_Sech_a} \\
    &\sigma_1=\sigma_2: & \quad
    \Omega_1/\Omega_2 & \to
        \begin{cases}
            \bigl(\Omega_{10}/\Omega_{20}\bigr)
            \exp\bigl(-\Delta t/\sigma \bigr) &
            \text{ when $t \to -\infty$}, \\
            \bigl(\Omega_{10}/\Omega_{20}\bigr)
            \exp\bigl(\Delta t/\sigma \bigr) &
            \text{ when $t \to +\infty$},
        \end{cases}
    \label{Eq:asymptotic_Sech_b} \\
    &\sigma_1<\sigma_2: & \quad
    \Omega_1/\Omega_2 & \to 0 \text{ when } t \to \pm\infty.
    \label{Eq:asymptotic_Sech_c}
\end{alignat}
\end{subequations}
From these relations we observe that hyperbolic secant pulses behave like
Gaussian pulses when the widths are unequal ($\sigma_1 \neq \sigma_2$). In
contrast, for equal widths ($\sigma_1 = \sigma_2$), the ratio of the pulses
reaches constant values different to zero or infinity. In both situations the
adiabatic conditions are not satisfied by the system. However, we will see in
Ch.~\ref{Ch:02} that Sech pulses still allow us to get very good transfer
efficiencies.

For Sech pulses, the nonadiabatic coupling is given by the function
\begin{align}
    \dot{\Phi}(t) & = \frac{
    \bigl[\Omega_{10}\Omega_{20}/\sigma_1 \sigma_2 \bigr]
    \sech\bigl[(t-\Delta t)/\sigma_1\bigr] \sech\bigl[t/\sigma_2\bigr]}
    {\Omega_{10}^2 \sech^2\bigl[(t-\Delta t)/\sigma_1\bigr] +
     \Omega_{20}^2 \sech^2\bigl[t/\sigma_2\bigr]} \notag \\
    & \quad\qquad \times \Biggl[ \sigma_1\tanh\biggl(\frac{t}{\sigma_2}\biggr)-
      \sigma_2\tanh\biggl(\frac{t-\Delta t}{\sigma_1}\biggr) \Biggr].
\end{align}
%

\subsection{Lorentzian pulses:}

Finally, let us consider two Lorentzian pulses of the form
\begin{subequations}
\begin{align}
    \Omega_1(t) & = \Omega_{10}\,
        \sigma_1^2/\bigl[(t-\Delta t)^2 + \sigma_1^2 \bigr], \\
    \Omega_2(t) & = \Omega_{20}\,
    \sigma_2^2/\bigl(t^2 + \sigma_2^2 \bigr).
\end{align}
\end{subequations}
The ratio of the two pulses is given by
\begin{equation}
    \frac{\Omega_1(t)}{\Omega_2(t)} =
        \biggl(\frac{\Omega_{10}\,\sigma_1^2}{\Omega_{20}\,\sigma_2^2}\biggr)
        \frac{t^2 + \sigma_2^2}
             {\bigl(t-\Delta t \bigr)^2 + \sigma_1^2},
\end{equation}
and the asymptotic behavior expressed as
\begin{subequations}
\label{Eq:Lorentzian}
\begin{alignat}{3}
    &\sigma_1>\sigma_2: & \quad
    \Omega_1/\Omega_2 & \to  \Omega_{10}\,\sigma_1^2 / \Omega_{20}\,\sigma_2^2
    &\qquad\text{ when } t \to \pm\infty, \\
    &\sigma_1=\sigma_2: & \quad
    \Omega_1/\Omega_2 & \to \Omega_{10} / \Omega_{20}
    &\qquad\text{ when $t \to \pm\infty$}, \\
    &\sigma_1<\sigma_2: & \quad
    \Omega_1/\Omega_2 & \to \Omega_{10}\,\sigma_1^2 / \Omega_{20}\,\sigma_2^2
    &\qquad\text{when } t \to \pm\infty.
\end{alignat}
\end{subequations}
These equations show a very interesting property: whatever the time delay of the
pulses is, their ratio is the same constant in both directions ($\pm t$).
Lorentzian pulses behave in a very strange and sometimes unpredictable way.
Ch.~\ref{Ch:02} will show some numerical results obtained when Lorentzian pulses
were used to drive a quantum system.

The nonadiabatic coupling of the Lorentzian pulses is given by
\begin{equation}
    \dot{\Phi}(t) =
    \frac{2\,\Omega_{10}\,\Omega_{20}\,\sigma_1^2\,\sigma_2^2
    \bigl\{ t\bigl[ (t-\Delta t)^2 + \sigma_1^2 \bigr]
         - \bigl(t-\Delta t\bigr)\bigl(t^2 + \sigma_2^2\bigr)\bigr\}}
         {\Omega_{10}^2 \sigma_1^4 \bigl(t^2 + \sigma_2^2\bigr)^2 +
          \Omega_{20}^2 \sigma_2^4 \bigl[(t-\Delta t)^2 + \sigma_1^2\bigr]^2}.
\end{equation}
%

\section{The failure probability}

To determine the degree of efficiency in the process of transferring atomic
population from the initial state $\ket{a}$ to the final state $\ket{b}$, we
define the ``failure probability'' as
\begin{equation}
    p = 1 - \bigl| C_b(\infty) \bigr|^2.
    \label{Eq:failure01}
\end{equation}
This quantity gives us information about how much the process fails to get the
population transferred to the target state. If we substitute Eqs.~\eqref{Eq:Cb}
and~\eqref{Eq:changeVariables} into Eq.~\eqref{Eq:failure01}, and consider only
the lowest order in $\delta\theta$ and $\delta\phi$, the probability failure can
be written in the form
\begin{equation}
    p = \bigl[(\delta\phi)^2 + (\delta\theta)^2 \bigr]_{t \to \infty}.
    \label{Eq:failure02}
\end{equation}

Let us first consider the case of two Gaussian pulses driving the three-level
system described by equations~\eqref{Eq:C_resonance}. If the pulses' widths are
considered equal ($\sigma_1=\sigma_2$) then the system satisfies the adiabatic
conditions and the mixing angle takes the values $\Phi(-\infty)=0$ and
$\Phi(\infty)=\pi/2$. Solutions for the small angle
deviations~\eqref{Eq:generalSolutions} can now be written as
\begin{subequations}
\label{Eq:SolutionsGaussian}
\begin{align}
    \delta\phi(t) & =
         \int_{-\infty}^t\cos\biggl[\int_{t'}^t \frac{\Omega(t'')}{2}\,dt''
        \biggr] \frac{d\Phi(t')}{dt'}\,dt',
    \label{Eq:SolutionGaussian01} \\
    \delta\theta(t) & =
   -\int_{-\infty}^t\sin\biggl[\int_{t'}^t \frac{\Omega(t'')}{2}\,dt'' \biggr]
        \frac{d\Phi(t')}{dt'}\,dt'.
    \label{Eq:SolutionGaussian02}
\end{align}
\end{subequations}
Upon substitution of these equations into Eq.~\eqref{Eq:failure02}, and by
using the relation
\begin{equation}
    \bigl| \delta\phi \pm i\,\delta\theta \bigr|^2
        = (\delta\phi)^2 + (\delta\theta)^2,
\end{equation}
we find the following closed-form expression for the failure probability
\begin{equation}
      p = \biggl| \int_{-\infty}^{\infty}
          \exp \biggl( i \int_t^{\infty} \frac{\Omega(t')}{2}\,dt' \biggr)
          \frac{d\Phi}{dt}\,dt \biggr|^2.
    \label{Eq:failure03}
\end{equation}

It can be readily seen that integral~\eqref{Eq:failure03} is small enough to
explain the large transfer probabilities observed in adiabatic passage
processes. As long as the adiabatic conditions are satisfied ($\Omega\sigma \gg
1$), the exponential term in the integrand can oscillate rapidly over the
interval of integration, making the integral to take small values. The local
adiabaticity criterion also suggests that the nonadiabatic coupling $d\Phi/dt$
remains small over that interval, contributing with a small value to the
integrand. Similarly, we note that the very sharp and profound dips found in the
Ch.~\ref{Ch:02} simulations for some particular choices of the pulses'
parameters (see Figs.~\ref{Fig:adiabatic01} and~\ref{Fig:adiabatic05})
correspond to zeros, or near-zeros of Eq.~\eqref{Eq:failure03}.

By making the change of variable
\begin{equation}
    \tau (t) = \int_{-\infty}^t \frac{\Omega(t')}{2} \,dt',
\end{equation}
the probability failure~\eqref{Eq:failure03} can be written in the form
\begin{equation}
    p = \biggl|\int_{0}^{\tau(\infty)} e^{-i\tau}\frac{d\Phi}{dt}\,dt\biggr|^2.
    \label{Eq:failure04}
\end{equation}
This expression can be interpreted as the Fourier transform of a function equal
to $d\Phi/d\tau$ between $\tau=0$ and $\tau(\infty)$, and zero elsewhere. This
accounts for the oscillatory character of integral~\eqref{Eq:failure03}, and
explains the existence of zeros (or near-zeros) for certain values of the
parameters.

By way of example, let us consider a simple case for which
integral~\eqref{Eq:failure03} can be evaluated analytically. If the pulses are
given by the equations
\begin{align}
    \Omega_1(t) & = \Omega_0 \sin(t/T), \notag \\
    \Omega_2(t) & =
        \begin{cases}
            \Omega_0 \cos(t/T) & \text{if $-T\pi/2 \leq t \leq T\pi/2$}, \\
                   0           & \text{elsewhere,}
        \end{cases} \notag
\end{align}
we may have that
\begin{equation}
    \Phi(t) = \frac{t}{T}, \qquad \text{and} \qquad
    \Omega (t) = \Omega_0. \notag
\end{equation}
Plugging these values into Eq.~\eqref{Eq:failure03}, we find that
\begin{equation}
    p = \biggl| \lim_{a \to \infty} \frac{1}{T} \int_{-\infty}^{\infty}
        e^{-i \Omega_0 (t-a)/2} \,dt \biggr|^2
      = \frac{16\,\sin^2 (\pi\,\Omega_0\,T/4)}{\Omega_0^2\,T^2}. \notag
\end{equation}
This result reveals a power-law decay of the failure probability in the form
$1/(\Omega_0\,T)^2$, and also shows zeros for this function when
$\Omega_0\,T=4n$.

Although it is well known that adiabatic passage produces very large transfer
efficiencies provided the adiabatic conditions are satisfied, the dependence on
$\Omega\,T$ of the failure probability is complicated. A ``perfect adiabatic
transfer'' is characterized by a probability failure that decreases
exponentially as a function of $\Omega\,T$ (this is called the DDP
result~\cite{Davis:JCP64}). Recent results have shown that this exponential
behavior is no longer satisfied for sufficiently large values of $\Omega\,T$,
giving instead a power-law decay of the probability failure with respect to
these parameters~\cite{Laine:PRA96,Vitanov:OC96}. In addition to this power-law,
oscillations of the failure probability have been also
observed coming from a nonadiabatic component of the total transition
amplitude~\cite{Drese:EPJD3}.

Now, let us explore a bit what happens when the conditions for adiabaticity are
not fully satisfied, for example, the case of hyperbolic secant pulses. As
before, we want to transfer population from the initial level to the target
level by closely following the evolution of the dressed (adiabatic) state of the
system. If the pulses' widths are in such a way that the ratio
$\Omega_1/\Omega_2$ goes to constant values at very early or very late times
(see Eq.~\eqref{Eq:asymptotic_Sech_b}), the adiabatic conditions are obviously
not satisfied by the system. Then an adiabatic following of the dressed state by
the state vector is not possible, at least under those \textit{initial}
conditions, because these to vectors are not parallel to each other before the
interaction with the lasers takes place. In other words, the state vector is
parallel to the initially populated level, while the dressed vector is
somewhere else (see Fig.~\ref{Fig:HilbertSpace} for a representation of these
vectors in the Hilbert space). However, it happens that at some moment during
the evolution of the system, the dressed state approaches the initial state and
the target state close enough to take sufficient population from the initial
level, evolve parallel to the state vector for a moment, leave the population in
the target state, and finally continue with its own evolution to some other
place in the Hilbert state. Although the system does not evolve in an adiabatic
manner, strictly speaking, there is a region in time (not too early, not too
late) for which this evolution resembles that of an adiabatic process. In this
``near-adiabatic'' process, large transfer efficiencies are still achieved.

\begin{figure}[htbp]
    \centering
    \mbox{\subfigure[Gaussian]{\includegraphics[scale=1.0]{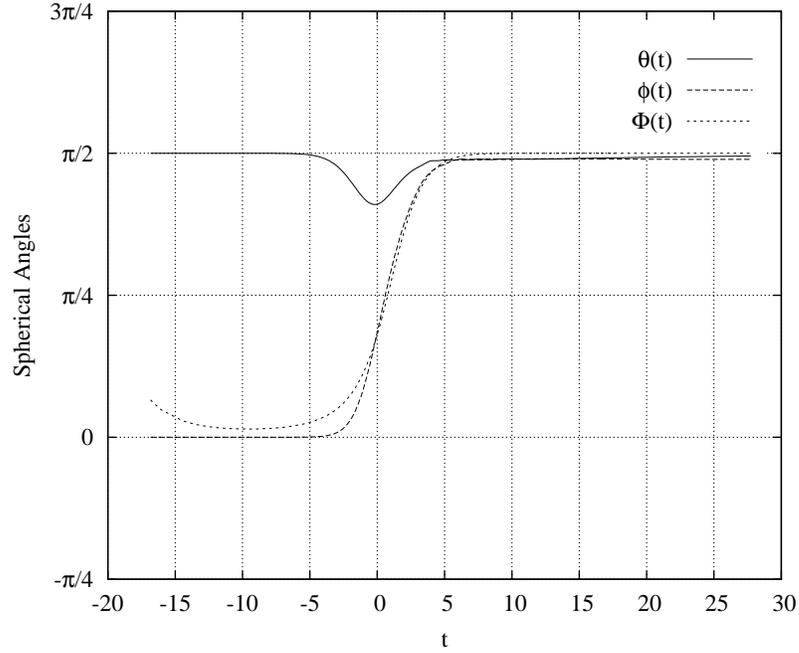}}}
    \mbox{\subfigure[Sech]{\includegraphics[scale=1.0]{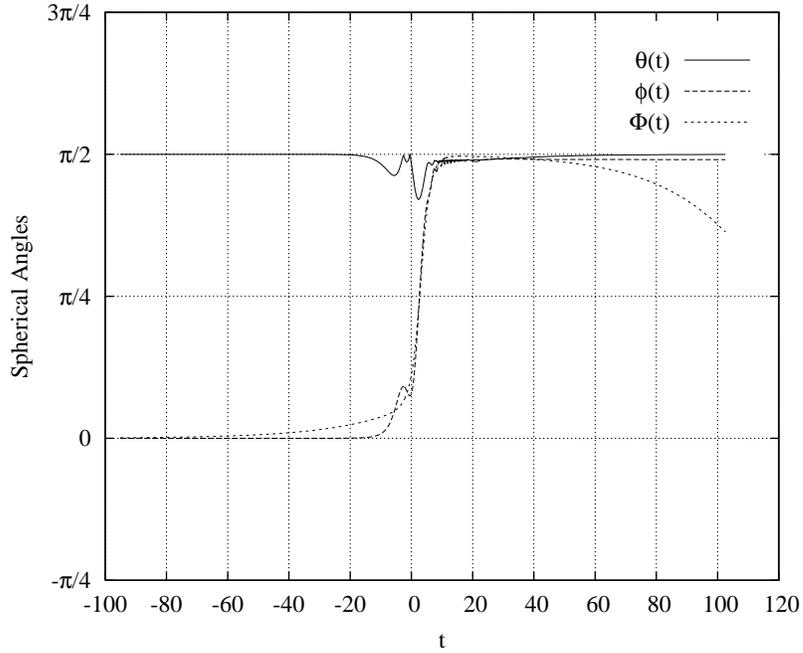}}}
    \caption{Spherical angles. For the Gaussian pulses we have $\Omega=2.75$,
    $\sigma=3.09$, $\textsl{g}=1.00$, $\sigma_{\textsl{g}}=2.48$, $\Delta
    t=5.28$, and $\log_{10}p=-2.00$. For the sech pulses we have $\Omega=6.30$,
    $\sigma=3.40$, $\textsl{g}=1.00$, $\sigma_{\textsl{g}}=4.01$, $\Delta
    t=12.01$, and $\log_{10}p=-2.00$. In both cases the spontaneous emission was
    $\Gamma=0.10$}
    \label{Fig:sphericalAngles}
\end{figure}

Fig.~\ref{Fig:sphericalAngles} shows the evolution of the system for two
particular set of parameters corresponding to Gaussian and Sech pulses.
We observe from these figures that the mixing angle is close to zero for a short
period of time before the interaction, and nearly $\pi/2$ for a short period of
time after the interaction. When $t \to -\infty$, the dressed state $\ket{W^0}$
is parallel to the empty bare state $\ket{b}$ and carries no population.
However, the dressed ``unstable'' states $\ket{W^+}$ and $\ket{W^-}$ (see
Sec.~\ref{Sec:adiabaticFollowing} have components along the bare state $\ket{a}$
(which has the initial entire population) and each one carry half of the total
population. These two dressed states also have components along the excited
state $\ket{c}$ which decays radiatively. The adiabatic frame of reference
rotates faster than the frequency of the lasers (and perhaps than the decay rate
itself) and the coupling classical field $\Omega(t)$ has no populated enough the
decaying level. Then the system is not going to lose too much population at the
beginning of the interaction with the cavity-mode $\textsl{g}$. By the time in
which $\Omega(t)$ starts to act upon the levels, the adiabatic state $\ket{W^0}$
is almost parallel to the initial state $\ket{a}$. At this point, this adiabatic
state will carry mostly the entire population of the initial state (not all of
it because the other unstable adiabatic states have already contributed with
population to the target state) and the wave vector will follow adiabatically
the evolution of this adiabatic state.

Our simple analytical model can also explain the behavior of the system for
those near-adiabatic conditions. Since the goal is to have
\begin{equation}
    \phi(\infty)   = -\frac{\pi}{2}, \qquad \text{and} \qquad
    \theta(\infty) =  \frac{\pi}{2},
\end{equation}
we see from equations~\eqref{Eq:generalSolutions} that the condition
\begin{equation}
    \int_{-\infty}^{\infty} \frac{\Omega(t)}{2}\,dt = n\pi,
    \qquad \text{with $n$ odd}
\end{equation}
must be satisfied. This ensures that the sine in the first term of the
$\delta\theta$ equation cancels out. The second terms of
equations~\eqref{Eq:generalSolutions}, which are proportional to $\dot{\Phi}$,
can be ignored for a moment (we will retake these components later on).
Therefore $\delta\phi \simeq -\Phi(-\infty)$. By plugging this result into
Eq.~\eqref{Eq:changeVariables_phi}, and considering the simple case in which
$\Omega_{10}=\Omega_{20}$, we may obtain
\begin{align}
    \phi(\infty) & \simeq -\Phi(\infty) - \Phi(-\infty) \notag \\
                 & \simeq -[\Phi(\infty) + \Phi(-\infty)] \notag \\
                 & \simeq -[\arctan(e^{\Delta t/\sigma}) +
                       \arctan(e^{-\Delta t/\sigma})] \notag \\
                 & \simeq -\pi/2.
\end{align}

Now, we have to examine what happens with the second terms
in~\eqref{Eq:generalSolutions}. Since those terms are typically small, because
of the near-adiabatic condition (that is, they are inversely proportional to
$\Omega\,\sigma$), a small deviation from the parameter values that make the
first terms in~\eqref{Eq:generalSolutions} vanish exactly may be enough to,
instead, make the first and second terms (nearly) cancel each other, and this
is, in fact, what we observe in our numerical calculations.

\chapter{\label{Ch:04} Transfer of atomic coherence}

The concepts of atomic coherence and interference has been extensively studied
and applied to many areas of atomic physics and quantum optics for the last
fifteen years. Applications of these ideas include electromagnetically induced
transparency (EIT), electromagnetically induced absorption (EIA), lasing without
inversion (LWI), sensitive spectroscopy in coherent media, among other
things~\cite{Fleischhauer:OC179}.

Recent developments in quantum information and quantum computation have
stimulated interest in processes that can prepare entangled states, as well as
perform conditional quantum dynamics and logic gates. The coherent manipulation
of these entangled states are fundamental to realizing a quantum computer, and
promises a novel atomic spectroscopy with resolution better than the standard
quantum limit~\cite{Pellizzari:PRL95}.

In this chapter we present a numerical analysis of the Zeeman coherence transfer
between two atoms by using adiabatic passage methods. We also present a
simplified model for the two-atom + cavity system that help us to better
understand the process of coherence transfer between two atoms coupled with a
cavity.

\section{The two-atom + cavity system}

\begin{figure}[h]
    \centering
    \includegraphics[scale=0.8]{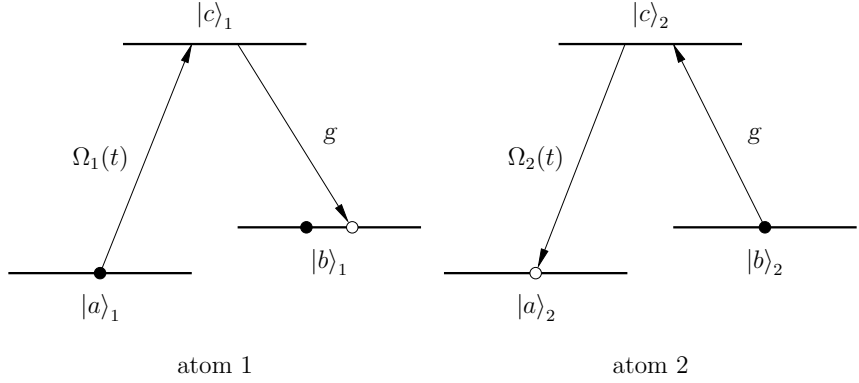}
    \caption{Transfer of atomic coherence between two $\Lambda$ systems.}
    \label{Fig:2atomCoherence}
\end{figure}

We consider a system consisting of a single-mode cavity containing two
three-level atoms in the $\Lambda$ configuration, as shown in
Fig.~\ref{Fig:2atomCoherence}. The atoms are fixed inside the cavity at
distances apart much larger than the wavelength of the cavity mode and
interacting individually with laser beams.
The transitions $\ket{a}_j \to \ket{c}_j$ of both atoms ($j=1,2$) are coupled to
separate classical coherent driving fields with frequency $\omega_L$ and Rabi
frequencies $\Omega_j$. The transitions $\ket{b}_j \to \ket{c}_j$ are strongly
coupled to the \textit{same} quantized cavity-mode field with coupling strength
\textsl{g} and frequency $\omega$. The dynamics of this system is described by
the interaction Hamiltonian:
\begin{equation}
    H_I = \frac{\hbar}{2} \sum_{j=1,2}
        \bigl[ \Omega_j(t)e^{-i \omega_L t}\ket{c}_{jj}\bra{a}
        + \textsl{g}\,\ket{c}_{jj}\bra{b} a \bigr] + \text{H.c.},
    \label{Eq:HI_coherence}
\end{equation}
where $a$ is the annihilation operator for the cavity mode and \textsl{g}
gives the coupling strength between the atoms and the field mode (vacuum
Rabi-frequency).

\subsection{Probability amplitude method}

We are interested in transferring the atomic coherence according to
\begin{equation}
    (A\ket{a}_1 + B\ket{b}_1)\,\ket{b}_2\ket{0}_c \longrightarrow
    \ket{b}_1\,(A\ket{a}_2 + B\ket{b}_2)\ket{0}_c,
\end{equation}
where $A$ and $B$ are arbitrary coefficients. If we consider the particular case
in which the system evolves without populating the dark state
$\ket{b}_1\ket{b}_2\ket{0}_c$ (that is, for $B=0$) then the evolution of the
quantum system takes place as follows:
\begin{equation}
\begin{array}{ccccc}
    \ket{a}_1\ket{b}_2\ket{0}_c & & & & \ket{b}_1\ket{a}_2\ket{0}_c \\[0.2cm]
    \Big\downarrow \vcenter{\rlap{$\scriptstyle{\Omega_1(t)}$}} & & & &
    \Big\uparrow \vcenter{\rlap{$\scriptstyle{\Omega_2(t)}$}}       \\[0.2cm]
    \ket{c}_1\ket{b}_2\ket{0}_c & \stackrel{\textsl{g}}{\longrightarrow} &
    \ket{b}_1\ket{b}_2\ket{1}_c & \stackrel{\textsl{g}}{\longrightarrow} &
    \ket{b}_1\ket{c}_2\ket{0}_c
\end{array}
\label{Eq:CoherentBasis}
\end{equation}
The state vector of the system can be constructed as a linear superposition of
these \textit{basis} eigenvectors in the form
\begin{align}
    \ket{\Psi(t)} & = C_{ab}(t)\,\ket{a\,b\,0}
                    + C_{cb}(t)\,\ket{c\,b\,0}
                    + C_{bb}(t)\,\ket{b\,b\,1} \notag \\
           & \qquad + C_{bc}(t)\,\ket{b\,c\,0}
                    + C_{ba}(t)\,\ket{b\,a\,0},
    \label{Eq:psi_coherence}
\end{align}
with the probability amplitudes given by
\begin{equation}
    \begin{array}{ccc}
        C_{ab}(t) = \braket{a\,b\,0}{\Psi(t)}, & \quad
        C_{cb}(t) = \braket{c\,b\,0}{\Psi(t)}, & \quad
        C_{bb}(t) = \braket{b\,b\,1}{\Psi(t)},   \\
        C_{bc}(t) = \braket{b\,c\,0}{\Psi(t)}, & \quad
        C_{ba}(t) = \braket{b\,a\,0}{\Psi(t)}. &
    \end{array}
    \label{Eq:ProbabilityAmplitudes}
\end{equation}
The corresponding time-dependent Schr\"{o}dinger equation is
\begin{equation}
    i \hbar \ket{\dot{\Psi}(t)} = H_I \ket{\Psi(t)},
    \label{Eq:Schrodinger_coherence}
\end{equation}
with
\begin{align}
    H_I & = \frac{\Omega_1(t)}{2} \,\ket{c}_{11}\bra{a}
        + \frac{\textsl{g}}{2} \,\ket{c}_{11}\bra{b} a
        + \frac{\Omega_2(t)}{2} \,\ket{c}_{22}\bra{a}
        + \frac{\textsl{g}}{2} \,\ket{c}_{22}\bra{b} a \notag \\
        & \qquad + \frac{\Omega_1^*(t)}{2} \,\ket{a}_{11}\bra{c}
        + \frac{\textsl{g}}{2}^* \, a^{\dagger} \,\ket{b}_{11}\bra{c}
        + \frac{\Omega_2^*(t)}{2} \,\ket{a}_{22}\bra{c}
        + \frac{\textsl{g}}{2}^* \, a^{\dagger} \,\ket{b}_{22}\bra{c}.
    \label{Eq:HI_operator}
\end{align}

By introducing Eqs.~\eqref{Eq:psi_coherence} and~\eqref{Eq:HI_operator} into
Eq.~\eqref{Eq:Schrodinger_coherence}, and multiplying the resulting equation by
every bra vector of the basis~\eqref{Eq:CoherentBasis}, we find that
\begin{subequations}
\label{Eq:Eqs_motion}
\begin{align}
    \dot{C}_{ab} & = -\frac{\Omega_1^*(t)}{2} \:\widetilde{C}_{cb},
    \label{Eq:Eqs_motion_ab} \\
    \dot{\widetilde{C}}_{cb} & = -\frac{\Gamma}{2} \:\widetilde{C}_{cb}
        + \frac{\Omega_1(t)}{2} \:C_{ab} + \frac{\textsl{g}}{2} \:C_{bb},
    \label{Eq:Eqs_motion_cb} \\
    \dot{C}_{bb} & = -\frac{\textsl{g}}{2}^* \bigl( \widetilde{C}_{cb}
        + \widetilde{C}_{bc} \bigr),
    \label{Eq:Eqs_motion_bb} \\
    \dot{\widetilde{C}}_{bc} & = -\frac{\Gamma}{2} \:\widetilde{C}_{bc}
        + \frac{\Omega_2(t)}{2} \:C_{ba} + \frac{\textsl{g}}{2} \:C_{bb},
        \label{Eq:Eqs_motion_bc} \\
    \dot{C}_{ba} & = -\frac{\Omega_2^*(t)}{2} \:\widetilde{C}_{bc}.
    \label{Eq:Eqs_motion_ba}
\end{align}
\end{subequations}
These are the equations of motion for the probability amplitudes of the two-atom
+ cavity system. Here we have considered the spontaneous emission of the excited
levels by adding phenomenological decay terms $\Gamma$ to the equations. We have
also avoided the complex character resulting in some of the equations by
introducing the notation
\begin{equation}
    \widetilde{C}_{bc} = i \,C_{bc} \qquad \text{and} \qquad
    \dot{\widetilde{C}}_{bc} = i \,\dot{C}_{bc}.
\end{equation}
%

\subsection{Dark state}

Under two-photon resonance condition, and by using the rotating-wave
approximation, the evolution of system~\eqref{Eq:Eqs_motion} is governed by the
non-Hermitian Hamiltonian
\begin{align}
    H' = \frac{\hbar}{2}
        \begin{bmatrix}
                0    & -\Omega_1^*     &      0     &       0         &     0 \\
            \Omega_1 & -\Gamma         & \textsl{g} &       0         &     0 \\
                0    & -\textsl{g}\,^* &      0     & -\textsl{g}\,^* &     0 \\
                0    &       0         & \textsl{g} & -\Gamma         & \Omega_2
\\
                0    &       0         &      0     & -\Omega_2^*     &     0
        \end{bmatrix}.
\end{align}
This operator has a \textit{dark} state with eigenvalue zero given by
\begin{equation}
    \ket{D} = \frac{\Omega_2 \,\textsl{g} \,\ket{a\:b\:0}
                  - \Omega_1 \,\Omega_2   \,\ket{b\:b\:1}
                  + \Omega_1 \,\textsl{g} \,\ket{b\:a\:0}}
                  {\sqrt{\Omega_0^2\,\textsl{g}^2 + \Omega_1^2 \Omega_2^2}},
    \label{Eq:Dark4.1}
\end{equation}
where $\Omega_0 = \sqrt{\Omega_1^2 + \Omega_2^2}$.

We observe that states of the type given by Eq.~\eqref{Eq:Dark4.1} are immune
against decay from the excited atomic levels, since they have no contribution of
such states. Therefore, the dark state $\ket{D}$ becomes the right mechanism to
transfer, from one atom to the other, the Zeeman coherence of the ground state
levels by using adiabatic passage~\cite{Pellizzari:PRL95}.

\subsection{Conditions for adiabatic following}

If the pulses are applied in a counterintuitive sequence, that is the pulse on
atom 2 preceding the pulse on atom 1, an adiabatic transfer of the dark state
between the atomic levels $\ket{a\:b\:0}$ and $\ket{b\:a\:0}$ may be achieved.
To see clearly how this happens, let define the mixing angle $\Phi$ by the
relationship
\begin{equation}
    \tan\Phi(t) = \frac{\Omega_1(t)}{\Omega_2(t)},
    \label{Eq:MixingAngle4}
\end{equation}
and express the dark state~\eqref{Eq:Dark4.1} in the form
\begin{equation}
    \ket{D} = K \biggl[ \cos\Phi\,\ket{a\:b\:0}
        - \frac{\Omega_0}{\textsl{g}} \sin\Phi\cos\Phi\,\ket{b\:b\:1}
        + \sin \Phi\,\ket{b\:a\:0} \biggr],
\end{equation}
where $K = \Omega_0\,\textsl{g}/\sqrt{\Omega_0^2\,\textsl{g}^2 +
\Omega_1^2\Omega_2^2}$. At very early times, just before the interaction with
the pulses, the condition $\Omega_1/\Omega_2 \to 0$ is satisfied, and the dark
state $\ket{D} \to \ket{a\:b\:0}$, as $\Phi(-\infty) \to 0$ and $K \to 1$. Then,
at very late times, the pulses satisfy the condition $\Omega_2/\Omega_1 \to 0$,
and the dark state $\ket{D} \to \ket{b\:a\:0}$, as $\Phi(+\infty) \to \pi/2$ and
$K \to 1$.

It is important to note that the strength $\textsl{g}$ of the quantized cavity
mode (which is assumed to be constant for simplicity) plays an important role
here. If the condition $\textsl{g} \gg \Omega_0$ is satisfied, then $K \to 1$
for all times, and the intermediate level $\ket{b\:b\:1}$ is never populated
($\Omega_0/\textsl{g} \to 0$). This reduces the probability of finding a photon
roaming inside the cavity, decreasing the chances for cavity decay. In addition,
if the conditions for adiabatic evolution are
fulfilled~\cite{Thompson:PRL68,Morin:PRL73}, that is
\begin{equation}
    \textsl{g}\,T, \:\Omega_0\,T \gg 1 \qquad \text{and} \qquad
    \textsl{g}\,\Omega_0 \gg \Gamma, \kappa,
\end{equation}
with $T$ the laser pulse duration, $\Gamma$ the radiative decay, and $\kappa$
the cavity decay; then the state vector of the system $\ket{\Psi(t)}$ may evolve
adiabatically, following closely the dark state $\ket{D}$.

\subsection{Density matrix and equations of motion}

For a realistic description of the two-atom + cavity system, in which
dissipative channels are accounted for, we must employ a master equation
description. The time evolution of the system is described by the following
Liouville equation:
\begin{equation}
    \dot{\rho} = -i \bigl[ H_{\text{eff}}\,\rho
        - \rho\,H_{\text{eff}}^{\dagger} \bigr] + \sum_{j=1}^2 J_{\Gamma_j}\rho
        + J_{\kappa}\rho,
    \label{Eq:Master4}
\end{equation}
where, in the interaction picture and on resonance
\begin{equation}
    H_{\text{eff}} = -i\kappa a^{\dagger} a
        - i \frac{\Gamma}{2} \sum_{j=1}^2 \ket{c}_{jj}\bra{c}
        + \sum_{j=1}^2 \biggl(\frac{\Omega_j(t)}{2}\,\ket{c}_{jj}\bra{a}
        + \frac{\textsl{g}}{2} \,\ket{c}_{jj}\bra{b} a + \text{H.c.} \biggr)
\end{equation}
is a non-Hermitian Hamiltonian including decay terms from spontaneous emission
and cavity decay. The superoperator $J_{\Gamma_j}$ describes the return of the
electron to the atomic ground states after a spontaneous emission, and
$J_{\kappa}$ the corresponding term for the cavity
decay~\cite{Pellizzari:PRL95}.

The decay $\Gamma$ terms can be obtained by doing a simple analysis of the
spontaneous emission rates of the excited levels. For example, the initially
populated state $\ket{a\:b\:0} \equiv \ket{a}_1\ket{b}_2\ket{0}_c$ can only have
contributions from the radiative decay of states $\ket{c\:b\:0}$ and
$\ket{a\:c\:0}$. However, the state $\ket{c\:b\:0}$ is the only level which
really contributes to the decay, since the state $\ket{a\:c\:0}$ never happens
in this scheme. In addition, this state can also spontaneously emit to other two
different states: the state $\ket{b\:b\:1}$ and some other arbitrary state which
is not taken into account in our system evolution. Therefore, the contribution
made by state $\ket{c\:b\:0}$ is $\Gamma/3$. A similar analysis applies to the
other possibilities.

By constructing a density operator with the elements of
basis~\eqref{Eq:CoherentBasis}, plugging it into Eq.~\eqref{Eq:Master4}, and
left- and right-multiplying the resulting equation by the elements of the basis,
we find
{\allowdisplaybreaks
\begin{subequations}
\label{Eq:CoherenceSystem}
\begin{align}
\intertext{Diagonal matrix elements:}
    \dot{\rho}_{abab} & = \frac{\Omega_1(t)}{2}
        \bigl(\widetilde{\rho}_{abcb} - \widetilde{\rho}_{cbab} \bigr)
        + \frac{\Gamma}{3}\,\rho_{cbcb}, \\
    \dot{\rho}_{cbcb} & = \frac{\Omega_1(t)}{2}
        \bigl(\widetilde{\rho}_{cbab} - \widetilde{\rho}_{abcb} \bigr)
        + \frac{\textsl{g}}{2}
        \bigl(\widetilde{\rho}_{cbbb} - \widetilde{\rho}_{bbcb} \bigr)
        - \Gamma\,\rho_{cbcb}, \\
    \dot{\rho}_{bbbb} & = \frac{\textsl{g}}{2}
        \bigl(\widetilde{\rho}_{bbcb} + \widetilde{\rho}_{bbbc}
            - \widetilde{\rho}_{cbbb} - \widetilde{\rho}_{bcbb} \bigr)
        + \frac{\Gamma}{3} \bigl( \rho_{cbcb} + \rho_{bcbc} \bigr), \\
    \dot{\rho}_{bcbc} & = \frac{\Omega_2(t)}{2}
        \bigl(\widetilde{\rho}_{bcba} - \widetilde{\rho}_{babc} \bigr)
        + \frac{\textsl{g}}{2}
        \bigl(\widetilde{\rho}_{bcbb} - \widetilde{\rho}_{bbbc} \bigr)
        - \Gamma\,\rho_{bcbc}, \\
    \dot{\rho}_{baba} & = \frac{\Omega_2(t)}{2}
        \bigl(\widetilde{\rho}_{babc} - \widetilde{\rho}_{bcba} \bigr)
        + \frac{\Gamma}{3}\,\rho_{bcbc}, \\
\intertext{Off-diagonal matrix elements:}
    \dot{\widetilde{\rho}}_{abcb} & = \frac{\Omega_1(t)}{2}
        \bigl(\rho_{cbcb} - \rho_{abab} \bigr)
        - \frac{\textsl{g}}{2}\,\rho_{abbb}
        - \frac{\Gamma}{2}\,\widetilde{\rho}_{abcb}, \\
    \dot{\rho}_{abbb} & = -\frac{\Omega_1(t)}{2}\,\widetilde{\rho}_{cbbb}
        + \frac{\textsl{g}}{2}
        \bigl(\widetilde{\rho}_{abcb} + \widetilde{\rho}_{abbc} \bigr), \\
    \dot{\widetilde{\rho}}_{abbc} & = \frac{\Omega_1(t)}{2}\,\rho_{cbbc}
        - \frac{\Omega_2(t)}{2}\,\rho_{abba}
        - \frac{\textsl{g}}{2}\,\rho_{abbb}
        - \frac{\Gamma}{2}\,\widetilde{\rho}_{abbc}, \\
    \dot{\rho}_{abba} & = -\frac{\Omega_1(t)}{2}\,\widetilde{\rho}_{cbba}
        + \frac{\Omega_2(t)}{2}\,\widetilde{\rho}_{abbc},\\
    \dot{\widetilde{\rho}}_{cbab} & = \frac{\Omega_1(t)}{2}
        \bigl(\rho_{abab} - \rho_{cbcb} \bigr)
        + \frac{\textsl{g}}{2}\,\rho_{bbab}
        - \frac{\Gamma}{2}\,\widetilde{\rho}_{cbab}, \\
    \dot{\widetilde{\rho}}_{cbbb} & = \frac{\Omega_1(t)}{2}\,\rho_{abbb}
        + \frac{\textsl{g}}{2}
        \bigl(\rho_{bbbb} - \rho_{cbcb} - \rho_{cbbc} \bigr)
        - \frac{\Gamma}{2}\,\widetilde{\rho}_{cbbb}, \\
    \dot{\rho}_{cbbc} & = -\frac{\Omega_1(t)}{2}\,\widetilde{\rho}_{abbc}
        + \frac{\Omega_2(t)}{2}\,\widetilde{\rho}_{cbba}
        + \frac{\textsl{g}}{2}
        \bigl(\widetilde{\rho}_{cbbb} - \rho_{bbbc} \bigr)
        - \Gamma\,\rho_{cbbc}, \\
    \dot{\widetilde{\rho}}_{cbba} & = \frac{\Omega_1(t)}{2}\,\rho_{abba}
        - \frac{\Omega_2(t)}{2}\,\rho_{cbbc}
        + \frac{\textsl{g}}{2}\,\rho_{bbba}
        - \frac{\Gamma}{2}\,\widetilde{\rho}_{cbba}, \\
    \dot{\rho}_{bbab} & = \frac{\Omega_1(t)}{2}\,\rho_{bbcb}
        - \frac{\textsl{g}}{2}
        \bigl(\widetilde{\rho}_{cbab} + \widetilde{\rho}_{bcab} \bigr), \\
    \dot{\widetilde{\rho}}_{bbcb} & = -\frac{\Omega_1(t)}{2}\,\rho_{bbab}
        + \frac{\textsl{g}}{2}
        \bigl(\rho_{cbcb} + \rho_{bccb} - \rho_{bbbb} \bigr)
        - \frac{\Gamma}{2}\,\widetilde{\rho}_{bbcb}, \\
    \dot{\widetilde{\rho}}_{bbbc} & = -\frac{\Omega_2(t)}{2}\,\rho_{bbba}
        + \frac{\textsl{g}}{2}
        \bigl(\rho_{cbbc} + \rho_{bcbc} - \rho_{bbbb} \bigr)
        - \frac{\Gamma}{2}\,\widetilde{\rho}_{bbbc}, \\
    \dot{\rho}_{bbba} & = \frac{\Omega_2(t)}{2}\,\widetilde{\rho}_{bbbc}
        - \frac{\textsl{g}}{2}
        \bigl(\widetilde{\rho}_{cbba} + \widetilde{\rho}_{bcba} \bigr), \\
    \dot{\widetilde{\rho}}_{bcab} & = -\frac{\Omega_1(t)}{2}\,\rho_{bccb}
        + \frac{\Omega_2(t)}{2}\,\rho_{baab}
        + \frac{\textsl{g}}{2}\,\rho_{bbab}
        - \frac{\Gamma}{2}\,\widetilde{\rho}_{bcab}, \\
    \dot{\rho}_{bccb} & = \frac{\Omega_1(t)}{2}\,\widetilde{\rho}_{bcab}
        - \frac{\Omega_2(t)}{2}\,\widetilde{\rho}_{bacb}
        + \frac{\textsl{g}}{2}
        \bigl(\widetilde{\rho}_{bcbb} - \widetilde{\rho}_{bbcb} \bigr)
        - \Gamma\,\rho_{bccb}, \\
    \dot{\widetilde{\rho}}_{bcbb} & = \frac{\Omega_2(t)}{2}\,\rho_{babb}
        + \frac{\textsl{g}}{2}
        \bigl(\rho_{bbbb} -\rho_{bccb} - \rho_{bcbc} \bigr)
        - \frac{\Gamma}{2}\,\widetilde{\rho}_{bcbb}, \\
    \dot{\widetilde{\rho}}_{bcba} & = \frac{\Omega_2(t)}{2}
        \bigl(\rho_{baba} - \rho_{bcbc} \bigr)
        + \frac{\textsl{g}}{2}\,\rho_{bbba}
        - \frac{\Gamma}{2}\,\widetilde{\rho}_{bcba}, \\
    \dot{\rho}_{baab} & = \frac{\Omega_1(t)}{2}\,\widetilde{\rho}_{bacb}
        - \frac{\Omega_2(t)}{2}\,\widetilde{\rho}_{bcab},\\
    \dot{\widetilde{\rho}}_{bacb} & = -\frac{\Omega_1(t)}{2}\,\rho_{baab}
        + \frac{\Omega_2(t)}{2}\,\rho_{bccb}
        - \frac{\textsl{g}}{2}\,\rho_{babb}
        - \frac{\Gamma}{2}\,\widetilde{\rho}_{bacb}, \\
    \dot{\rho}_{babb} & = -\frac{\Omega_2(t)}{2}\,\widetilde{\rho}_{bcbb}
        + \frac{\textsl{g}}{2}
        \bigl(\widetilde{\rho}_{bacb} + \widetilde{\rho}_{babc} \bigr), \\
    \dot{\widetilde{\rho}}_{babc} & = \frac{\Omega_2(t)}{2}
        \bigl(\rho_{bcbc} - \rho_{baba} \bigr)
        - \frac{\textsl{g}}{2}\,\rho_{babb}
        - \frac{\Gamma}{2}\,\widetilde{\rho}_{babc}.
\end{align}
\end{subequations}}
These are the density matrix equations of motion of the two-atom + cavity
system, where $\rho_{pqrs} \equiv \bra{p\,q} \rho \ket{r\,s}$. As before, we
have avoided the complex nature of some of these equations by introducing the
notation
\begin{equation}
    \widetilde{\rho}_{pqrs} = i \rho_{pqrs}, \qquad
    \dot{\widetilde{\rho}}_{pqrs} = i \dot{\rho}_{pqrs}.
\end{equation}
%

\section{Numerical results}

This section presents numerical results obtained by integration of the system of
differential equations~\eqref{Eq:CoherenceSystem} using the Runge-Kutta method.
Basically, we studied the adiabatic passage method for transferring interatomic
coherence, when Gaussian and Sech pulses were used to drive a two-atom + cavity
system. For simplicity, we considered a high-$Q$ cavity with constant coupling
strength \textsl{g}, and no cavity decay ($\kappa=0$). For this problem, six
different parameters were considered in our simulations: the Rabi frequency and
the width of the pulses, the time delay, the coupling strength of the cavity
mode, and the spontaneous emission of the excited states.

Numerical results corresponding to different values of the Rabi frequency, the
pulse width, the coupling constant, and the time delay are shown in
Table~\ref{Table:coherence}.
\begin{longtable}{ccccccccc}
\caption{Transfer of coherence. Adiabatic passage} \\
\label{Table:coherence}\\
\hline
\endhead
\hline
\endfoot
Pulse profile & $\Gamma$     & $\Omega_{10}$ & $\sigma_1$   & $\Omega_{20}$ &
$\sigma_2$ &
$\textsl{g}$ & $\Delta t$    & $\log_{10} p$ \\
\hline\\[-5pt]
Gaussian & 0.00 &  2.60 & 10.00 & 1.00 &  1.00 & 1.00 &  6.50 & -0.68 \\
         &      & 14.40 & 10.10 & 1.00 &  4.00 & 1.00 & 25.29 & -1.54 \\
         &      &  0.70 &  7.00 & 1.00 &  2.00 & 2.00 &  3.48 & -2.19 \\
         &      &       &       &      &       &      &       &       \\
         & 0.01 &  1.10 &  4.50 & 1.00 &  1.00 & 1.00 &  0.00 & -0.61 \\
         &      & 15.00 & 10.20 & 1.00 &  4.00 & 1.00 & 25.70 & -1.40 \\
         &      & 14.80 & 13.60 & 1.00 &  7.00 & 1.00 & 36.10 & -2.13 \\
         &      &       &       &      &       &      &       &       \\
         & 0.10 &  2.50 &  7.30 & 1.00 &  4.00 & 2.00 & 11.69 & -1.09 \\
         &      &  2.00 &  6.80 & 1.00 &  4.00 & 4.00 &  9.80 & -1.10 \\
         &      &  1.90 &  6.70 & 1.00 &  4.00 & 6.00 &  9.40 & -1.10 \\
         &      & 25.00 & 30.00 & 1.00 & 30.00 & 1.00 & 80.50 & -1.76 \\
         &      &       &       &      &       &      &       &       \\
Sech     & 0.00 &  0.90 &  2.30 & 1.00 &  2.00 & 1.00 &  0.00 & -1.14 \\
         &      &  3.80 &  4.40 & 1.00 &  5.00 & 1.00 & 11.51 & -3.71 \\
         &      &       &       &      &       &      &       &       \\
         & 0.01 &  3.80 &  4.40 & 1.00 &  5.00 & 1.00 & 11.48 & -2.06 \\
         &      &       &       &      &       &      &       &       \\
         & 0.02 &  5.20 &  4.10 & 1.00 &  5.00 & 2.00 & 11.00 & -1.85 \\
         &      &  6.90 &  4.10 & 1.00 &  5.00 & 3.00 & 11.80 & -1.85
\end{longtable}
As in Ch.~\ref{Ch:02}, we were looking for values of the failure probability of
$\log_{10}p=-2.00$ or better. During the simulations we fixed the value of
$\Omega_{20}$ to unity. In this way, we could compare the pulses' energies (Rabi
frequencies) and widths for the different failure probabilities obtained, and
then to conclude for which set of parameters we achieved the maximum transfer
efficiency. The other parameters of the system were allowed to take values from
physically reasonable intervals. For example, we studied widths in the
interval $[0,10]$ most of the time, going a little bit further in some special
cases. We also explored time delays in intervals between $[0,50]$ for
Gaussian pulses, and $[0,100]$ for Sech and Lorentzian pulses.

\begin{figure}[htbp]
    \centering
    \mbox{\subfigure[Surface]{\includegraphics[scale=1.0]{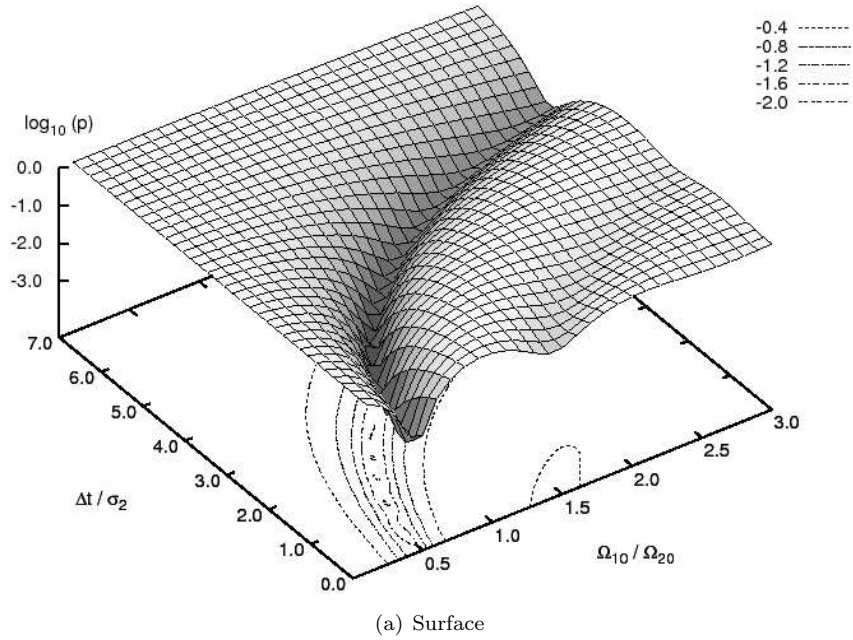}}}
    \mbox{\subfigure[Contour]{\includegraphics[scale=0.45]{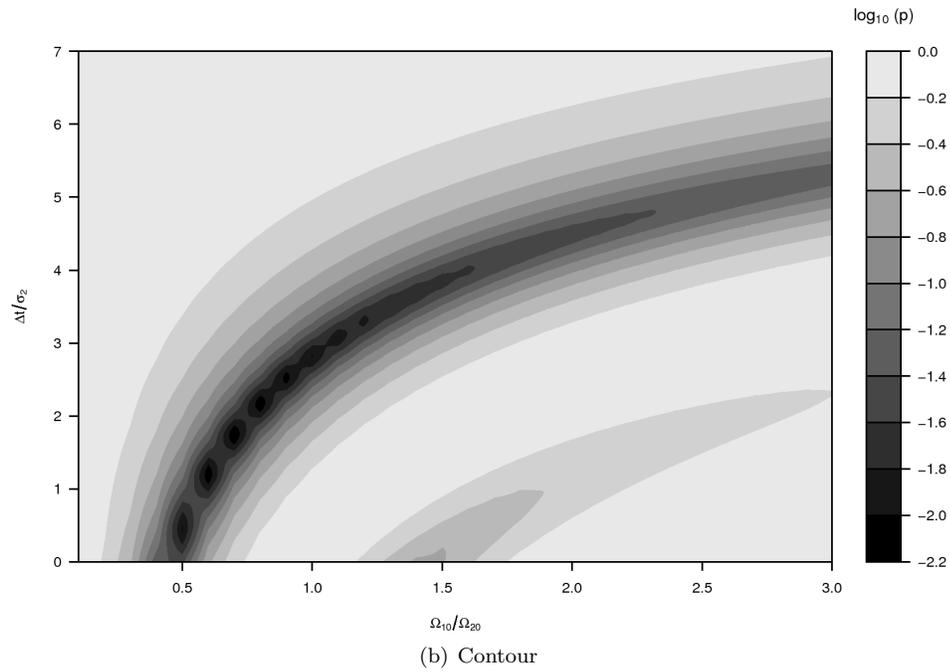}}}
    \caption{Gaussian surface and contour for $\Gamma=0.00$. Here we have
    $\Omega_{10}=0.70$, $\sigma_1=7.00$, $\Omega_{20}=1.00$, $\sigma_2=2.00$,
    $\textsl{g}=2.00$, $\Delta t=3.48$, and $\log_{10}p=-2.19$.}
    \label{Fig:coherence01}
\end{figure}

\begin{figure}[htbp]
\centering
\mbox{\subfigure[Surface]{\includegraphics[scale=1.0]{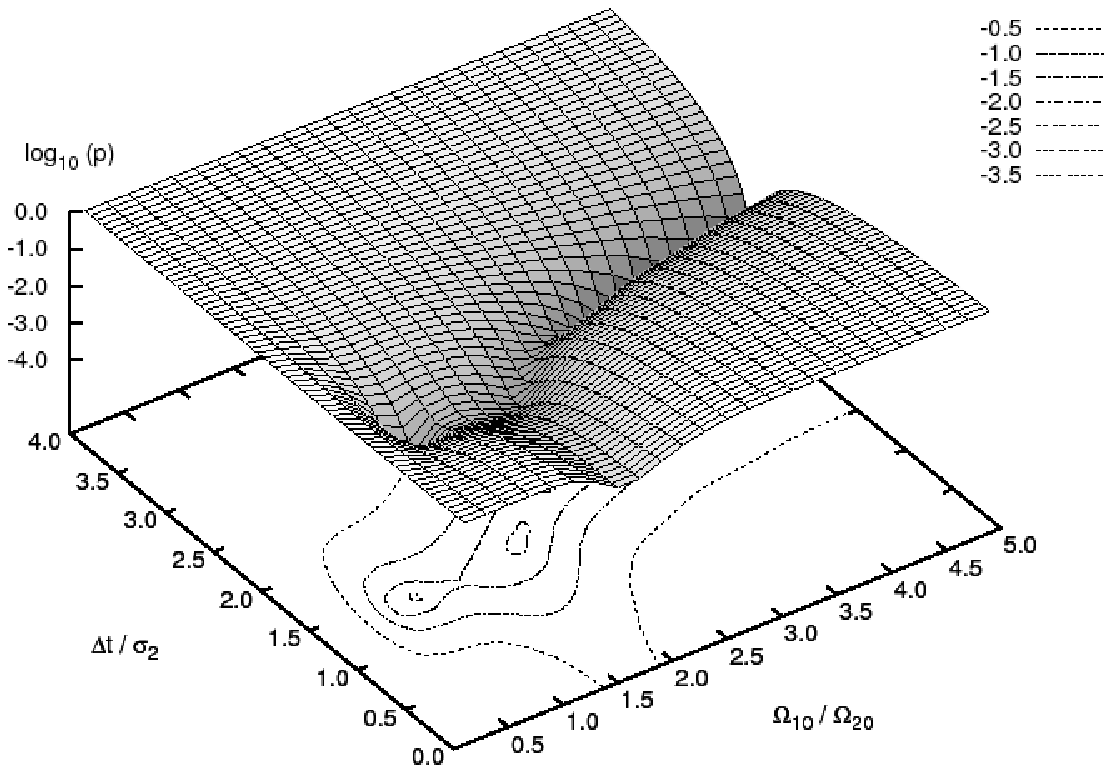}}}
\mbox{\subfigure[Contour]{
\includegraphics[scale=0.45,angle=0]{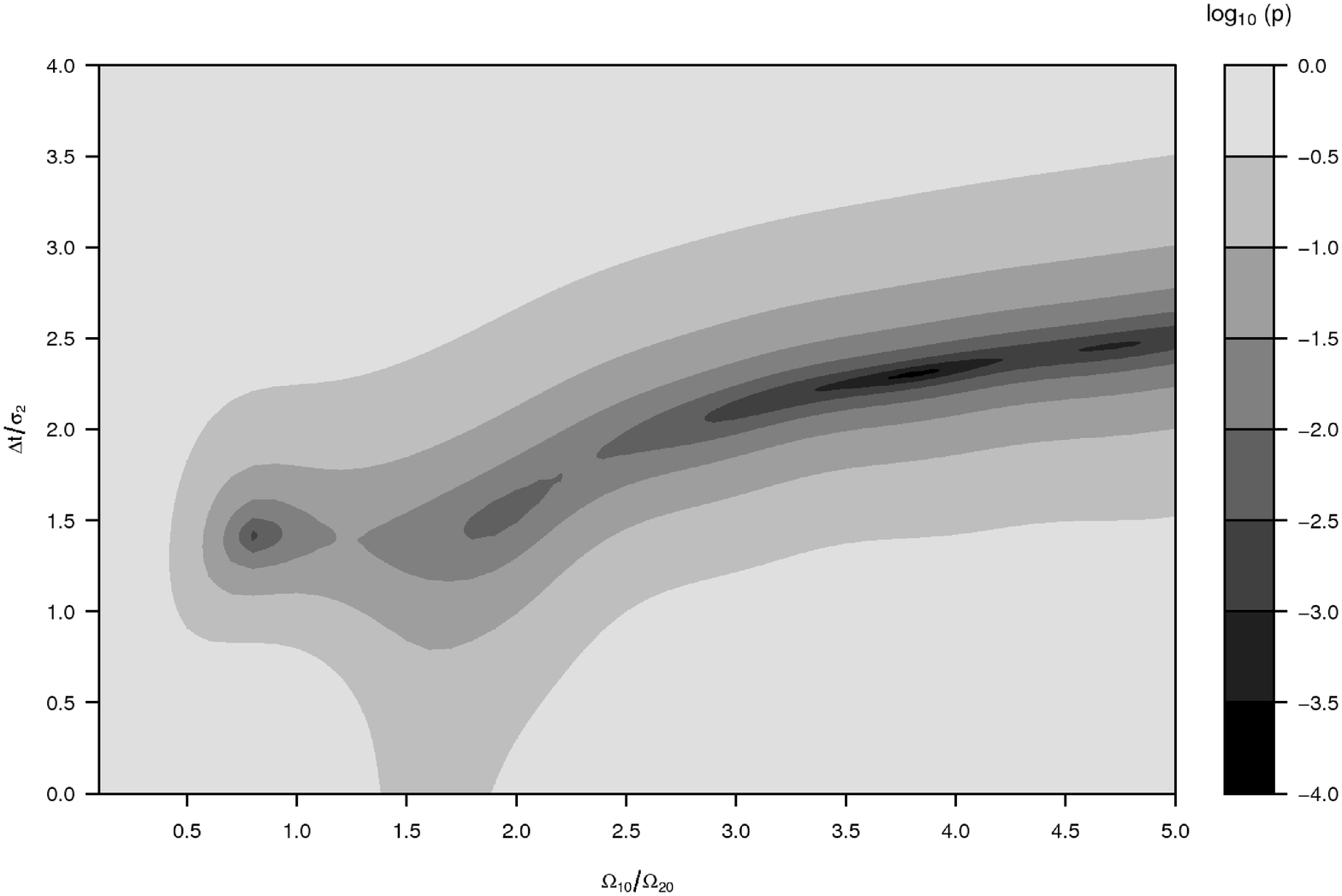}}
}
\caption{Sech surface and contour for $\Gamma=0.00$. Here we have
$\Omega_{10}=3.80$, $\sigma_1=4.40$, $\Omega_{20}=1.00$, $\sigma_2=5.00$,
$\textsl{g}=1.00$, $\Delta t=11.51$, and $\log_{10}p=-3.71$.}
\label{Fig:coherence02}
\end{figure}

We first studied the case of no spontaneous emission. We observed from
Fig.~\ref{Fig:coherence01} that this time there are no sharp dips, but instead
we found a long and profound valley consisting of very good transfer of
efficiency parameters.
It is interesting to note that for Gaussian pulses and no spontaneous emission,
a good transfer efficiency was achieved for a small value of the pulse amplitude
compared with its width (see Table~\ref{Table:coherence}).
Fig.~\ref{Fig:coherence02} shows again that Sech pulses are very efficient when
there is no spontaneous emission.
However, bigger pulse amplitudes and longer time delays were required, making
them less efficient that Gaussians. We tried to extrapolate numerically the
apparent \textit{exponential} behavior of these valleys formed by the maximum
probability surface, but the results we obtained showed no easy connections with
the parameters of the system.

We did not consider Lorentzian pulses in this analysis because they were very
computationally demanding.

\begin{figure}[htbp]
    \centering
    \mbox{\subfigure[Surface]{\includegraphics[scale=1.0]{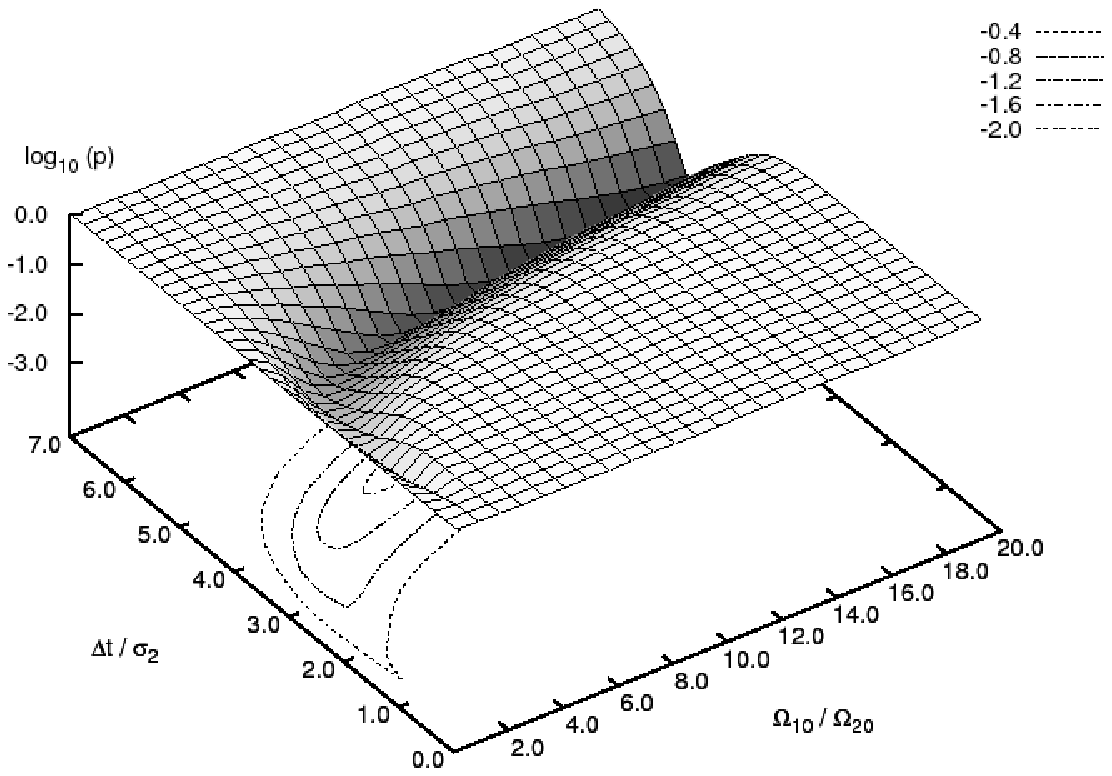}}}
    \mbox{\subfigure[Contour]{\includegraphics[scale=0.45]{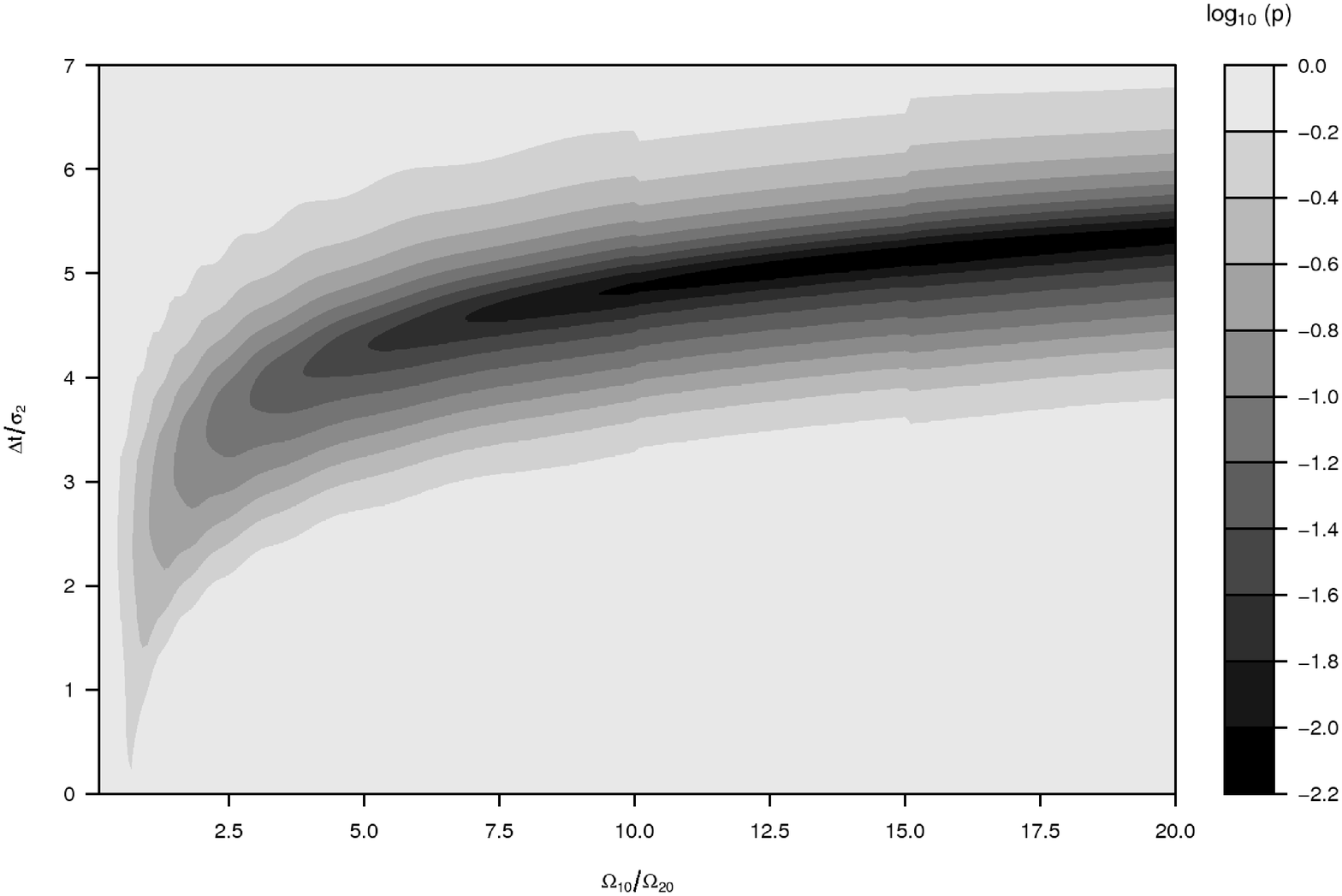}}}
    \caption{Gaussian surface and contour for $\Gamma=0.01$. Here we have
    $\Omega_{10}=15.10$, $\sigma_1=13.60$, $\Omega_{20}=1.00$, $\sigma_2=7.00$,
    $\textsl{g}=1.00$, $\Delta t=36.20$, and $\log_{10}p=-2.14$.}
    \label{Fig:coherence03}
\end{figure}

\begin{figure}[htbp]
    \centering
    \mbox{\subfigure[Surface]{\includegraphics[scale=1.0]{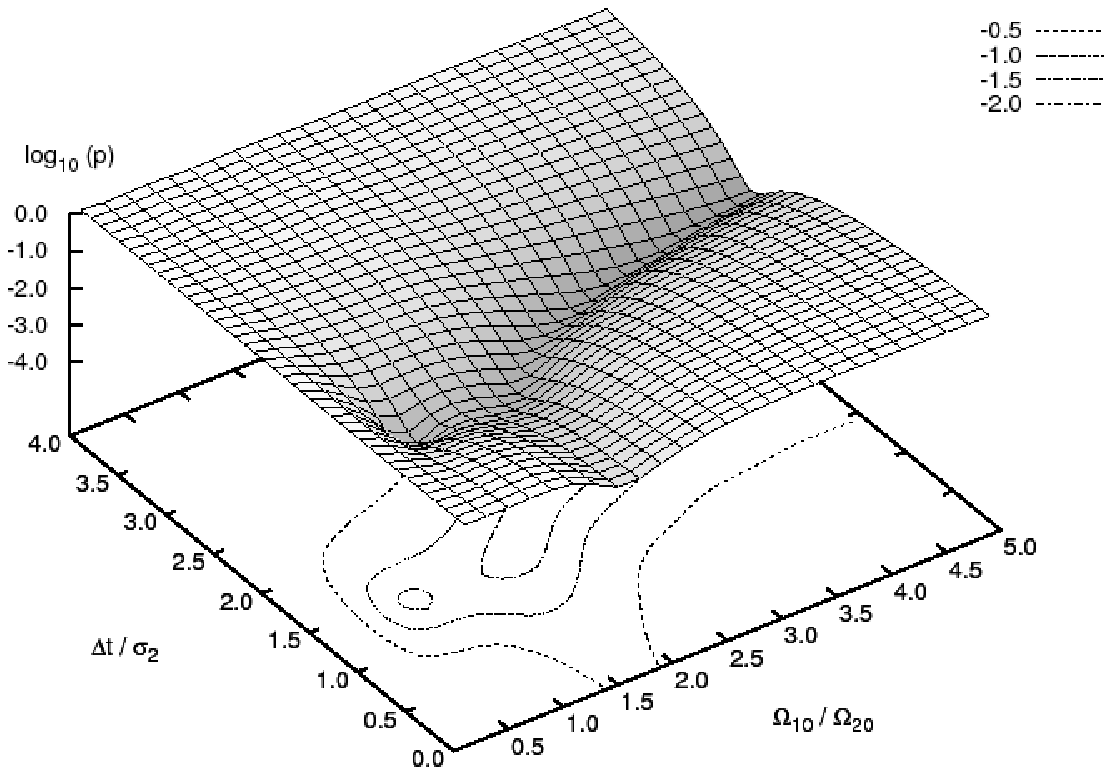}}}
    \mbox{\subfigure[Contour]{\includegraphics[scale=0.45]{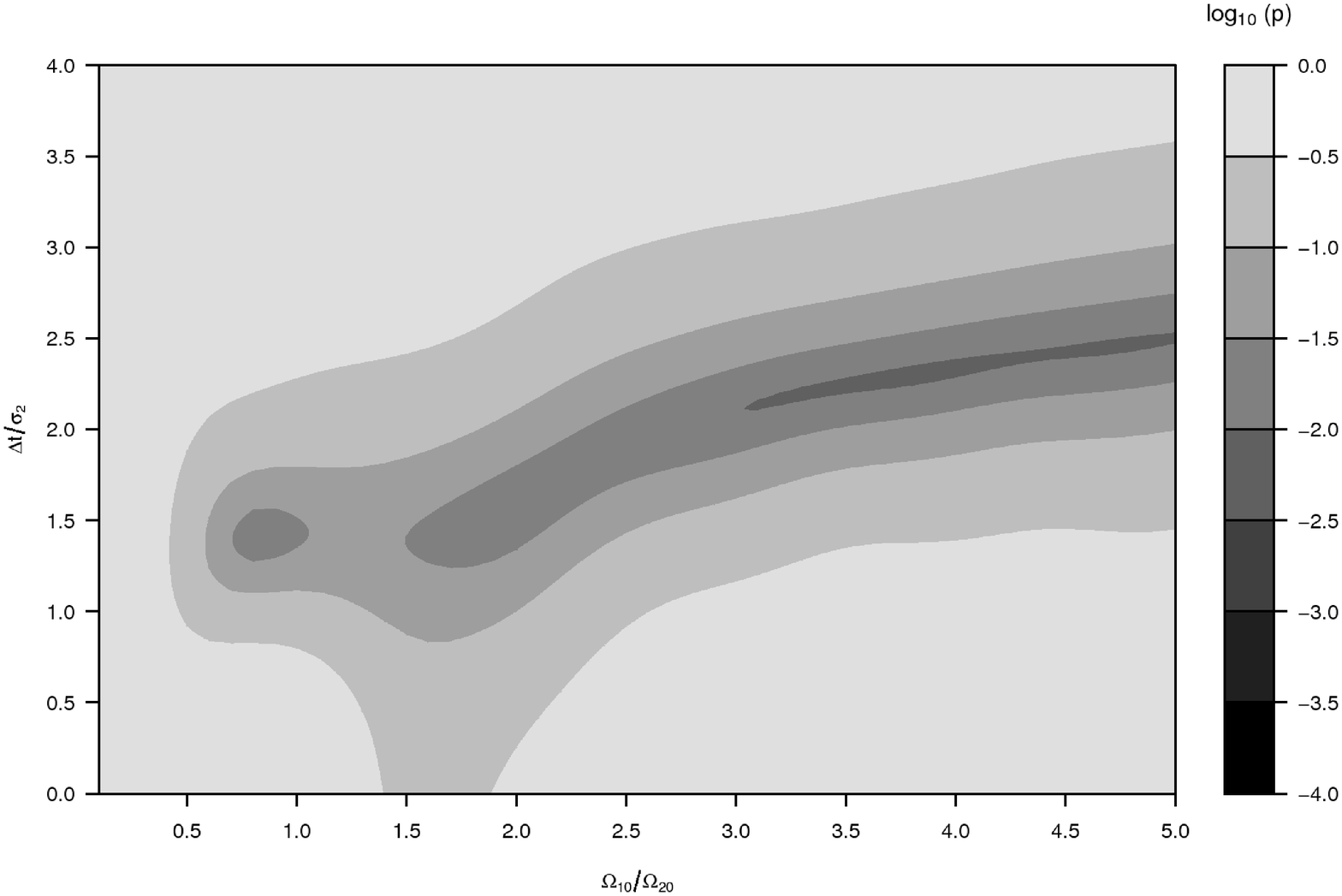}}}
    \caption{Sech surface and contour for $\Gamma=0.01$. Here we have
    $\Omega_{10}=3.80$, $\sigma_1=4.40$, $\Omega_{20}=1.00$, $\sigma_2=5.00$,
    $\textsl{g}=1.00$, $\Delta t=11.48$, and $\log_{10}p=-2.06$.}
    \label{Fig:coherence04}
\end{figure}

\begin{figure}[htbp]
    \centering
    \mbox{\subfigure[Surface]{\includegraphics[scale=1.0]{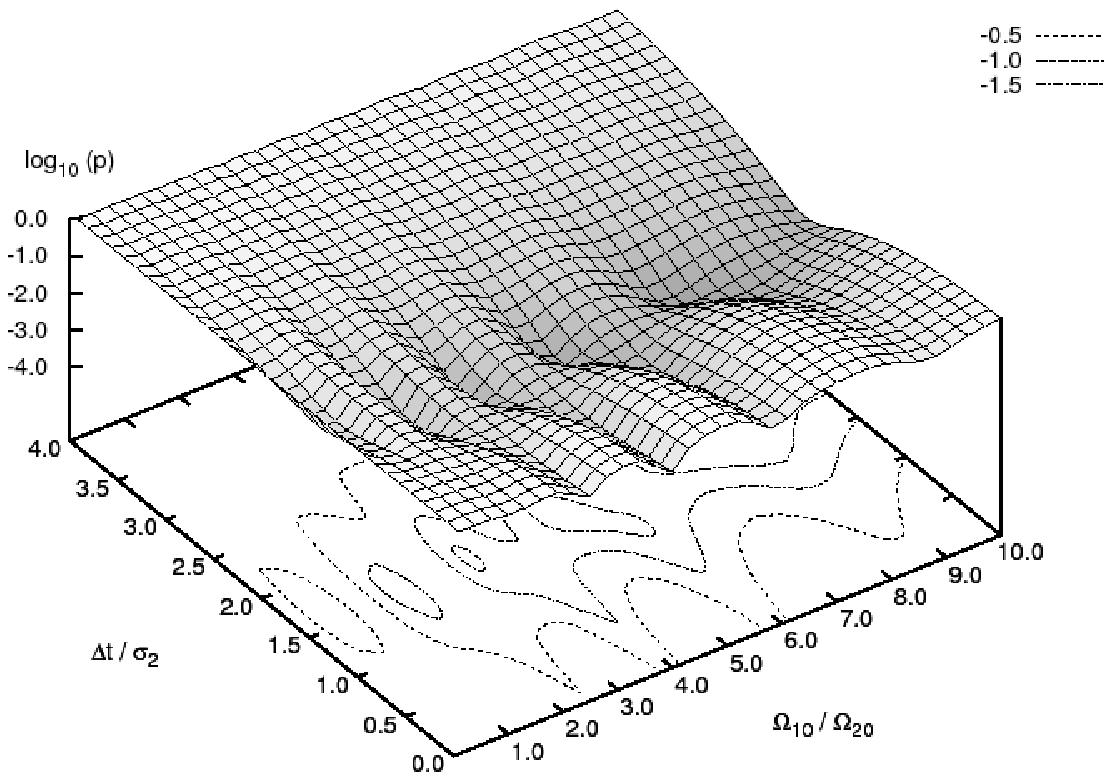}}}
    \mbox{\subfigure[Contour]{\includegraphics[scale=0.45]{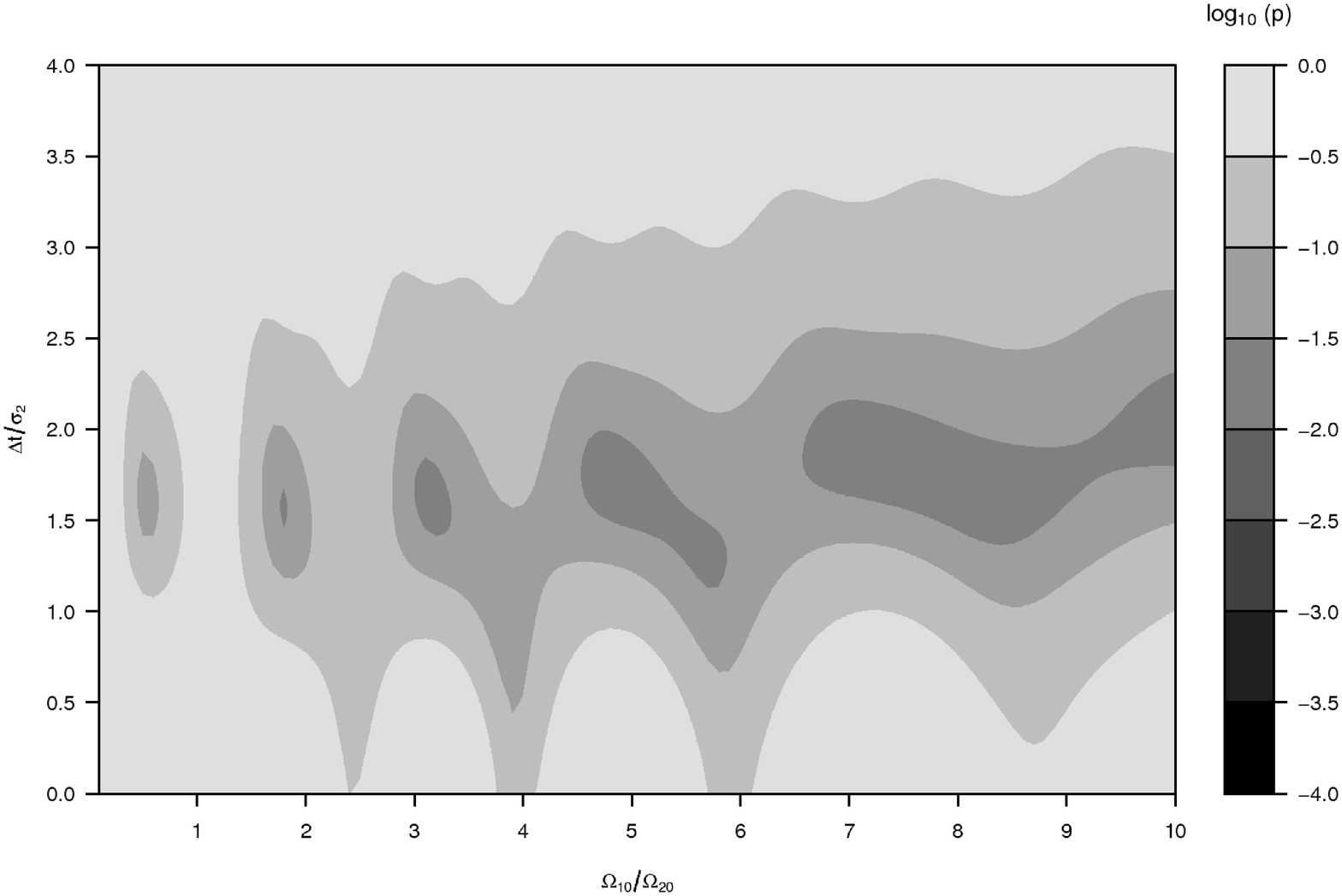}}}
    \caption{Sech surface and contour for $\Gamma=0.02$. Here we have
    $\Omega_{10}=7.30$, $\sigma_1=5.00$, $\Omega_{20}=1.00$, $\sigma_2=8.00$,
    $\textsl{g}=3.00$, $\Delta t=14.70$, and $\log_{10}p=-1.96$.}
    \label{Fig:coherence05}
\end{figure}

For spontaneous emission values of $\Gamma=0.01$ and $0.02$,
Figs.~\ref{Fig:coherence03}, \ref{Fig:coherence04}, and~\ref{Fig:coherence05}
showed that Sech pulses performed this time better than Gaussian pulses. This
was also checked by comparison with the ratio of the pulses and the interaction
times.
This time, the valleys were shallower than those for no decay.

It is very important to note that bigger values of the coupling strength
\textsl{g} increased the coherence transfer efficiency for no spontaneous
emission. However, when the system underwent radiative decay, an increment in
the magnitude of the cavity mode showed no effect in the failure probability.
Instead, if the width of the pulses was increased, so was the transfer
efficiency.

In the case of Gaussian pulses, for bigger values of the spontaneous emission,
we needed really big values of the pulse amplitudes and time delays to get a
good transfer efficiency. However, we still did not reach the minimum value of
the failure probability we were looking for, as shown in
Table~\ref{Table:coherence}.

\newpage
\section{An alternative model}

\begin{figure}[h]
    \centering
    \includegraphics[scale=0.8]{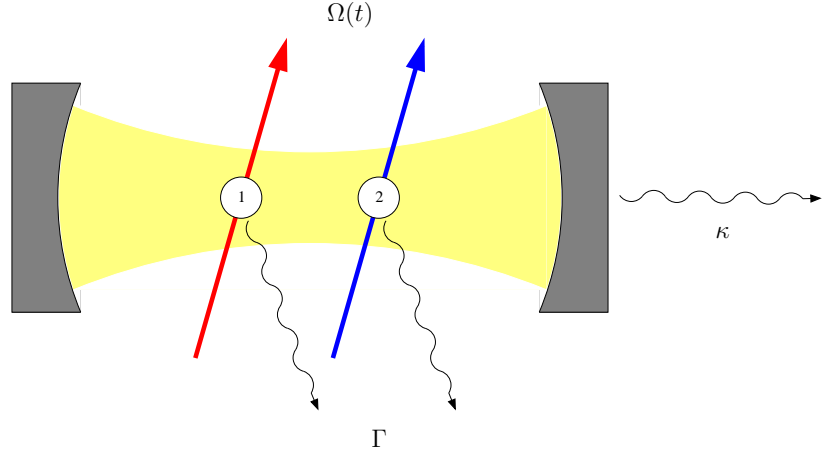}
    \caption{Schematic representation of two qubits interacting with a classical
    coherent field $\Omega(t)$ and a microcavity mode. The qubits and the cavity
    can decay with rates $\Gamma$ and $\kappa$, respectively.}
    \label{Fig:QC}
\end{figure}

Consider for simplicity the case in which the two Rabi-frequencies are the same,
that is $\Omega_1(t) = \Omega_2(t) = \Omega(t)$. We assume that the interatomic
distance is much less than the wavelength of the coherent classical fields, so
that the intracavity registration of a photon cannot be used to identify which
atom is the source of this radiation. The system now has \textit{some} kind of
symmetry and can be consider as one consisting of two \textit{identical}
particles interacting with a laser field with Rabi-frequency $\Omega(t)$ and a
quantized cavity mode with coupling strength $\textsl{g}$. The situation is
illustrated in Fig.~\ref{Fig:QC}.

It is interesting to observe that in view of the symmetry of the couplings, the
dynamics of the two-atom + cavity system can be examined as the evolution of two
different systems (one symmetric and the other antisymmetric) interacting
individually with the classical coherent field and the quantum cavity mode, but
reciprocally exchanging the interatomic coherence.

\subsection{The associated three-level system: EIT}

\begin{figure}[h]
    \centering
    \includegraphics[scale=0.8]{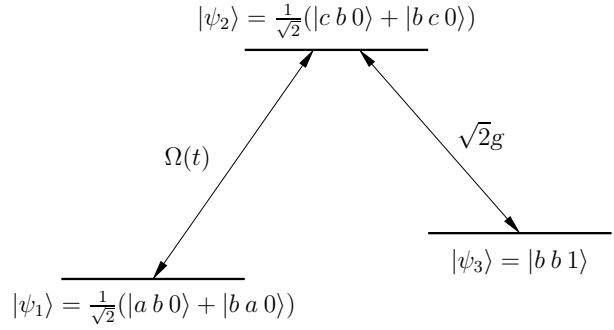}
    \caption{The associated three-level $\Lambda$ system driven by a classical
    field with Rabi frequency $\Omega(t)$ and interacting with a quantum field
    of strength $\sqrt{2}\textsl{g}$.}
    \label{Fig:EIT}
\end{figure}

By adding Eqs.~\eqref{Eq:Eqs_motion_ab} and~\eqref{Eq:Eqs_motion_ba}, we obtain
\begin{equation}
    \frac{d}{dt}\bigl[ C_{ab} + C_{ba} \big] =
        - \frac{\Omega(t)}{2} \bigl[\widetilde{C}_{cb}
        + \widetilde{C}_{bc}  \bigr].
    \label{Eq:Symmetric_Cab}
\end{equation}
Then, by adding Eqs.~\eqref{Eq:Eqs_motion_cb} and~\eqref{Eq:Eqs_motion_bc}, we
have
\begin{equation}
    \frac{d}{dt}\bigl[ \widetilde{C}_{cb} + \widetilde{C}_{bc} \big] =
        - \frac{\Gamma}{2} \bigl[\widetilde{C}_{cb} + \widetilde{C}_{bc} \bigr]
        + \frac{\Omega(t)}{2} \bigl[C_{ab} + C_{ba} \big] + \textsl{g}\,C_{bb}.
    \label{Eq:Symmetric_Ccb}
\end{equation}
And finally, from Eq.~\eqref{Eq:Eqs_motion_bb}, we get
\begin{equation}
    \frac{d}{dt}\,C_{bb} = - \frac{\textsl{g}}{2}
        \bigl[ \widetilde{C}_{cb} + \widetilde{C}_{bc}\bigr].
    \label{Eq:Symmetric_Cbb}
\end{equation}
If we assume that the evolution of the system takes place in the Hilbert space
spanned by the \textit{symmetric} eigenvectors
\begin{subequations}
\begin{align}
    \ket{\psi_1} & = \frac{1}{\sqrt{2}}
        \bigl(\ket{a\:b\:0} + \ket{b\:a\:0}\bigr),  \label{Eq:psi_1} \\
    \ket{\psi_2} & = \frac{1}{\sqrt{2}}
        \bigl(\ket{c\:b\:0} + \ket{b\:c\:0} \bigr), \label{Eq:psi_2} \\
    \ket{\psi_3} & = \ket{b\:b\:1},                 \label{Eq:psi_3}
\end{align}
\end{subequations}
then we define the state vector of the system as the linear superposition
\begin{equation}
    \ket{\Psi_S(t)} = B_1(t) \ket{\psi_1} + B_2(t) \ket{\psi_2}
        + B_3(t) \ket{\psi_3}.
    \label{Eq:symmetricVector}
\end{equation}
Here $B_1$, $B_2$, and $B_3$ represent the probability amplitudes of finding the
system in the symmetric states $\ket{\psi_1}$, $\ket{\psi_2}$, and
$\ket{\psi_3}$, respectively. These coefficients can be expressed in terms of
the individual probability amplitudes~\eqref{Eq:ProbabilityAmplitudes} as
follows:
\begin{subequations}
\label{Eq:Symmetric_C}
\begin{align}
    B_1(t) & = \braket{\psi_1}{\Psi_S(t)} \notag \\
           & = \frac{1}{\sqrt{2}} \bigl[ \braket{a\:b\:0}{\Psi_S(t)}
             + \braket{b\:a\:0}{\Psi_S(t)} \bigr] \notag \\
           & = \frac{1}{\sqrt{2}} \bigl[C_{ab} + C_{ba} \bigr].
\intertext{Similarly}
    B_2(t) & =\frac{1}{\sqrt{2}} \bigl[\widetilde{C}_{cb}
             + \widetilde{C}_{bc} \bigr], \\
    B_3(t) & = C_{bb}.
\end{align}
\end{subequations}
On inserting Eqs.~\eqref{Eq:Symmetric_C} into Eqs.~\eqref{Eq:Symmetric_Cab},
\eqref{Eq:Symmetric_Ccb}, and~\eqref{Eq:Symmetric_Cbb}, we obtain
\begin{subequations}
\label{Eq:EIT_system}
\begin{align}
    \dot{B}_1 & = - \frac{\Omega(t)}{2}\,B_2, \\
    \dot{B}_2 & = + \frac{\Omega(t)}{2}\,B_1 - \frac{\Gamma}{2}\,B_2
                  + \frac{\textsl{g}}{\sqrt{2}}\,B_3, \\
    \dot{B}_3 & = - \frac{\textsl{g}}{\sqrt{2}}\,B_2.
\end{align}
\end{subequations}
These equations represent the dynamics of a three-level $\Lambda$ system with
states $\ket{\psi_1}$ and $\ket{\psi_3}$ coupled to an excited state
$\ket{\psi_2}$ via, respectively, a classical laser field $\Omega(t)$ and a
cavity mode field $\sqrt{2}\textsl{g}$ (see Fig.~\ref{Fig:EIT}).

\begin{figure}[htbp]
\centering
\mbox{
\subfigure[$\Gamma=0.00$]{\includegraphics[scale=1.0]{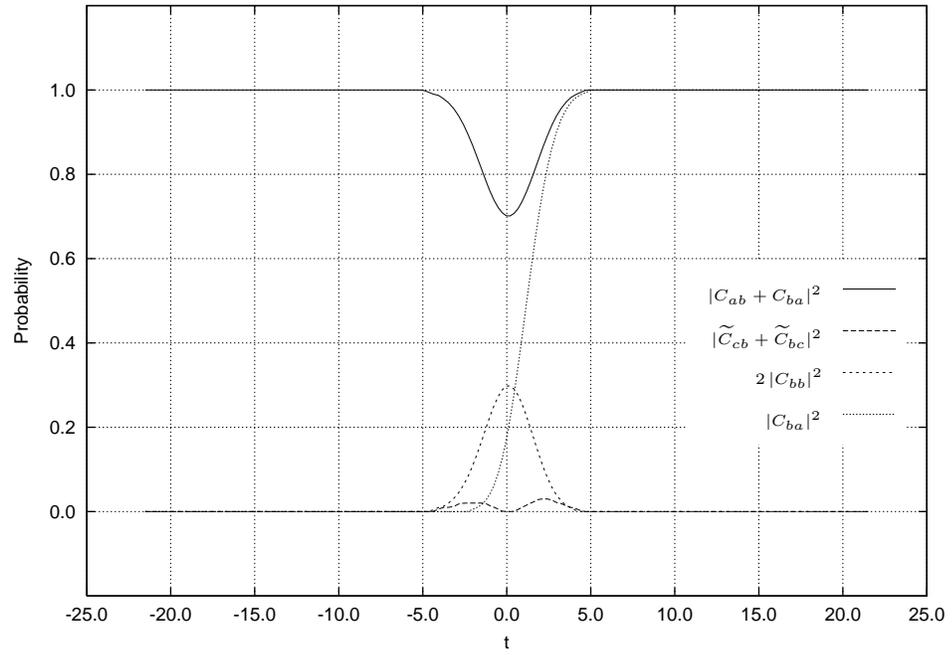}}
}
\mbox{
\subfigure[$\Gamma=0.10$]{\includegraphics[scale=1.0]{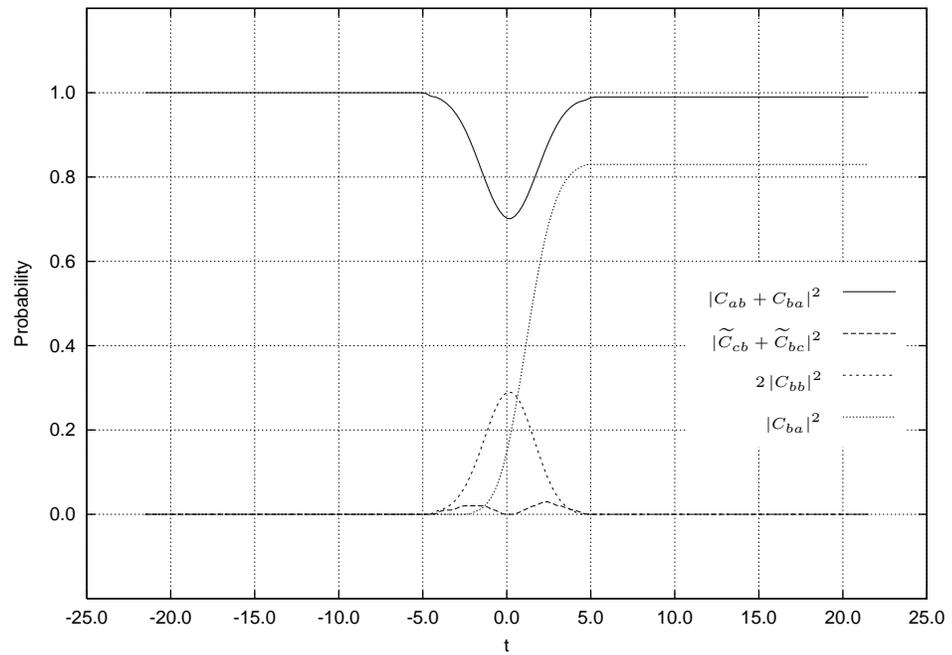}}
}
\caption{Transferring of coherence in the associated three-level system.}
\label{Fig:Associated_Three-level}
\end{figure}

The evolution of system~\eqref{Eq:EIT_system} is controlled by the non-Hermitian
Hamiltonian
\begin{align}
    H_S = \frac{\hbar}{2}
        \begin{bmatrix}
               0   &      -\Omega        &        0           \\
            \Omega &      -\Gamma        & \sqrt{2}\textsl{g} \\
                0  & -\sqrt{2}\textsl{g} &        0
        \end{bmatrix},
\end{align}
which has a \textit{dark} state given by
\begin{equation}
    \ket{D} = \frac{-\sqrt{2}\textsl{g}\,\ket{\psi_1} + \Omega\ket{\psi_3}}
                   {\sqrt{2\textsl{g}^2 + \Omega^2}}.
    \label{Eq:Dark4.2}
\end{equation}
By defining the mixing angle as
\begin{equation}
    \Phi(t) = \tan^{-1} \biggl[ \frac{\sqrt{2}\textsl{g}}{\Omega(t)} \biggr],
\end{equation}
we can write Eq.~\eqref{Eq:Dark4.2} in the form
\begin{equation}
    \ket{D} = - \sin\Phi(t) \ket{\psi_1} + \cos\Phi(t) \ket{\psi_3}. \\
\end{equation}
As before, we can transfer population between states $\ket{\psi_1}$ and
$\ket{\psi_3}$ without populating the leaking state $\ket{\psi_2}$ by employing
adiabatic passage via the dark state of the system. The process of coherence
transfer in this system is illustrated in Fig.~\ref{Fig:Associated_Three-level}.

We note that, if $\sqrt{2}\textsl{g} \gg \Omega$, then the state $\ket{D}$
corresponds almost identically to the state $\ket{\psi_1}$. This means that a
single photon excitation is shared among the atoms, favoring the transfer of
coherence. In addition, the effect of cavity decay is reduced, since the state
with a cavity photon is not populated. In this limit, a superposition given by
the dark state contains only a very small component of the single-photon state
$\ket{b\:b\:1}$. This increases the lifetime of the combined atom-cavity system
and is the essential feature of intracavity electromagnetically induced
transparency (EIT)~\cite{Fleischhauer:OC179}.

\subsection{The associated two-level system:
            $\boldsymbol{2}\boldsymbol{\pi}$-pulse coherence transfer}

\begin{figure}[h]
    \centering
    \includegraphics[scale=0.8]{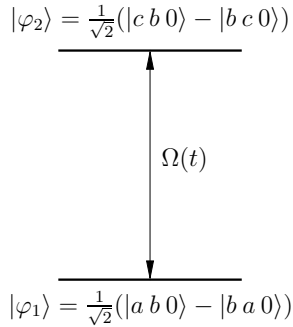}
    \caption{The associated two-level atom driven by a classical field with Rabi
    frequency $\Omega(t)$.}
    \label{Fig:2levelCoherence}
\end{figure}

On the other hand, by subtracting Eq.~\eqref{Eq:Eqs_motion_ba}
from~\eqref{Eq:Eqs_motion_ab}, we get
\begin{equation}
    \frac{d}{dt}\bigl[ C_{ab} - C_{ba} \big] =
        - \frac{\Omega}{2} \bigl[\widetilde{C}_{cb} - \widetilde{C}_{bc} \bigr];
    \label{Eq:Antisymmetric_Cab}
\end{equation}
and by subtracting Eq.~\eqref{Eq:Eqs_motion_bc} from~\eqref{Eq:Eqs_motion_cb},
we obtain
\begin{equation}
    \frac{d}{dt}\bigl[ \widetilde{C}_{cb} - \widetilde{C}_{bc} \big] =
        - \frac{\Gamma}{2} \bigl[\widetilde{C}_{cb} - \widetilde{C}_{bc} \bigr]
        - \frac{\Omega}{2} \bigl[C_{ab} - C_{ba} \big].
    \label{Eq:Antisymmetric_Ccb}
\end{equation}
Now, if we assume that the evolution of the system takes place in the Hilbert
space spanned by the \textit{antisymmetric} eigenvectors
\begin{subequations}
\begin{align}
    \ket{\varphi_1} & = \frac{1}{\sqrt{2}}
        \bigl[\ket{a\:b\:0} - \ket{b\:a\:0} \bigr], \label{Eq:phi_1} \\
    \ket{\varphi_2} & = \frac{1}{\sqrt{2}}
        \bigl[\ket{c\:b\:0} - \ket{b\:c\:0} \bigr], \label{Eq:phi_2}
\end{align}
\end{subequations}
then we define the state vector of the system as the linear superposition
\begin{equation}
    \ket{\Psi_A(t)} = A_1(t) \ket{\varphi_1} + A_2(t) \ket{\varphi_2}.
    \label{Eq:antisymmetricVector}
\end{equation}
Here $A_1$ and $A_2$ represent the probability amplitudes of finding the system
in the antisymmetric states $\ket{\varphi_1}$ and $\ket{\varphi_2}$,
respectively. As before, these coefficients can be expressed in terms of the
individual probability amplitudes~\eqref{Eq:ProbabilityAmplitudes} as follows:
\begin{subequations}
\label{Eq:Antisymmetric_A}
\begin{align}
    A_1(t) & = \braket{\varphi_1}{\Psi_A(t)} \notag \\
           & = \frac{1}{\sqrt{2}} \bigl[ \braket{a\:b\:0}{\Psi_A(t)}
             - \braket{b\:a\:0}{\Psi_A(t)} \bigr] \notag \\
           & = \frac{1}{\sqrt{2}} \bigl[C_{ab} - C_{ba} \bigr]. \\
\intertext{Similarly}
    A_2(t) & = \frac{1}{\sqrt{2}} \bigl[\widetilde{C}_{cb}
             - \widetilde{C}_{bc} \bigr].
\end{align}
\end{subequations}
On inserting Eqs.~\eqref{Eq:Antisymmetric_A} into
Eqs.~\eqref{Eq:Antisymmetric_Cab} and~\eqref{Eq:Antisymmetric_Ccb}, we obtain
\begin{subequations}
\label{Eq:2pi-inversion}
\begin{align}
    \dot{A}_1 & = - \frac{\Omega(t)}{2}\,A_2, \\
    \dot{A}_2 & = + \frac{\Omega(t)}{2}\,A_1 - \frac{\Gamma}{2}\,A_2.
\end{align}
\end{subequations}
These equations represent the evolution of a two-level system with ground state
$\ket{\varphi_1}$ and excited state $\ket{\varphi_2}$ coupled by a classical
laser field $\Omega(t)$, as illustrated in Fig.~\ref{Fig:2levelCoherence}.

\begin{figure}[htbp]
    \centering

\mbox{\subfigure[$\Gamma=0.00$]{\includegraphics[scale=1.0]{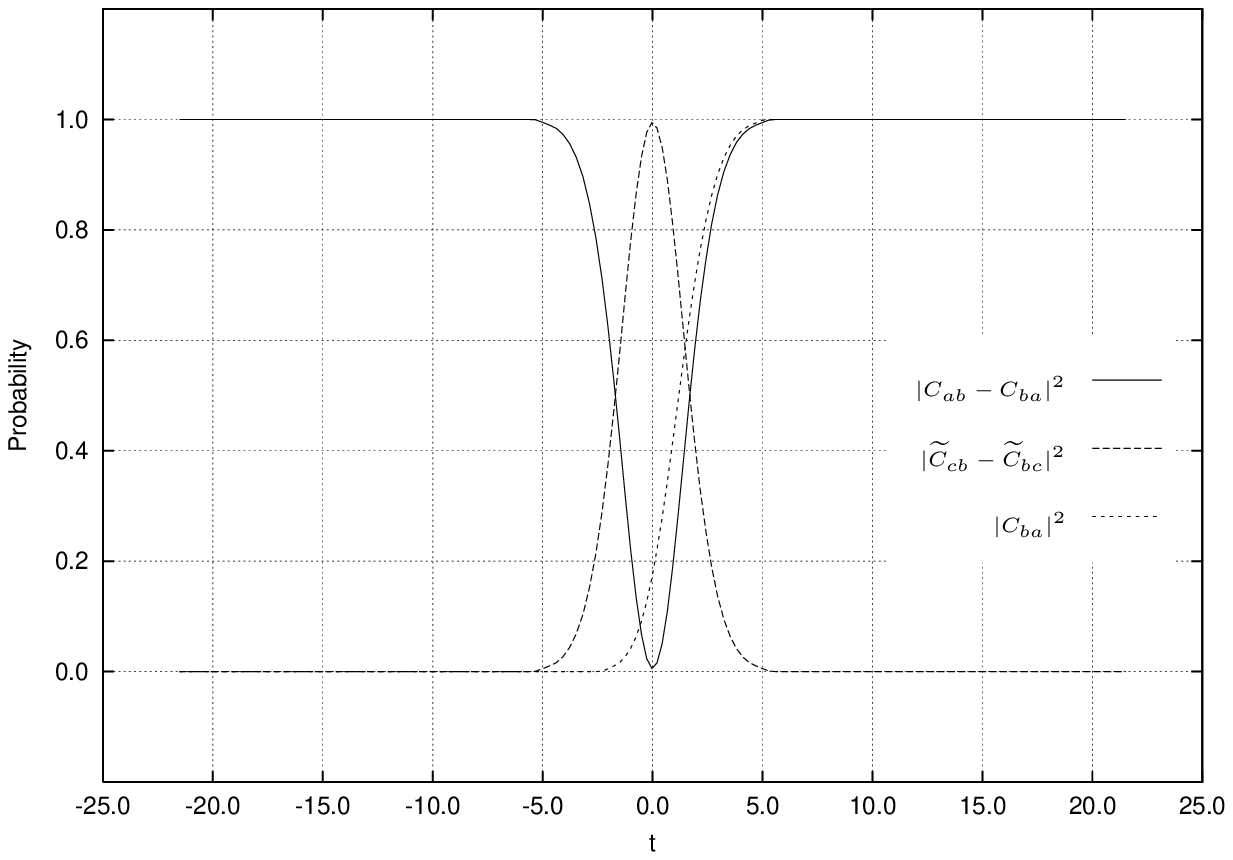}}}

\mbox{\subfigure[$\Gamma=0.10$]{\includegraphics[scale=1.0]{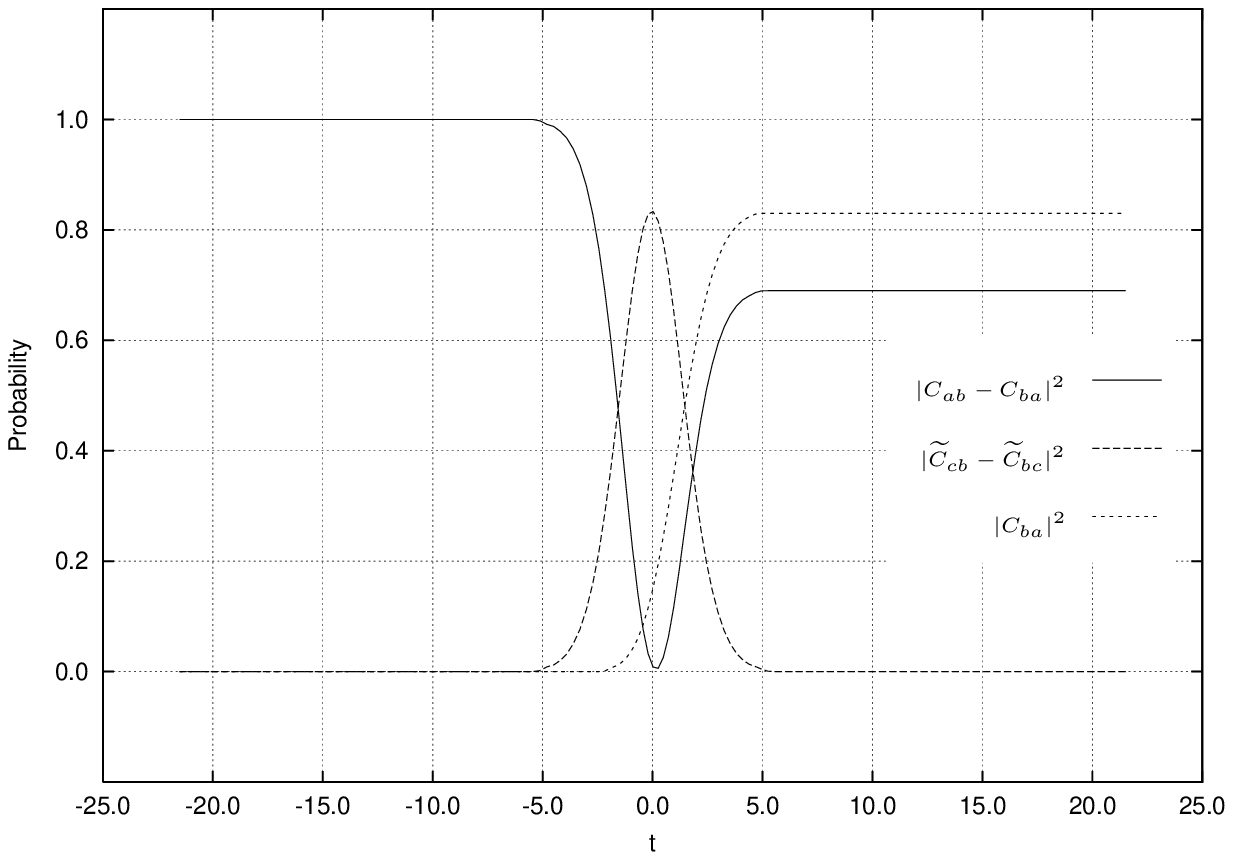}}}
    \caption{Transferring of coherence in the associated two-level system.}
    \label{Fig:Associated_Two-level}
\end{figure}

\noindent The evolution of system~\eqref{Eq:2pi-inversion} is governed by the
non-Hermitian Hamiltonian
\begin{align}
    H_A = \frac{\hbar}{2}
        \begin{bmatrix}
               0   &      -\Omega \\
            \Omega &      -\Gamma
        \end{bmatrix}.
\end{align}
Here the process of coherence transfer is achieved by a $2\pi$-pulse process, as
illustrated in Fig.~\ref{Fig:Associated_Two-level}.
%

\subsection{Transferring the coherence}

By using the symmetric and antisymmetric eigenvectors defined in
Eqs.~\eqref{Eq:psi_1} and~\eqref{Eq:phi_1}, the evolution of the system
$\ket{a\:b\:0} \to \ket{b\:a\:0}$ can be written in the form
\begin{equation}
    \frac{1}{\sqrt{2}}\bigl[\ket{\psi_1} + \ket{\varphi_1} \bigr]
    \to
    \frac{1}{\sqrt{2}}\bigl[\ket{\psi_1} - \ket{\varphi_1} \bigr].
\end{equation}
In this way, the transfer of atomic coherence can be viewed as a combination of
two different processes: an adiabatic passage acting on the symmetric eigenstate
$\ket{\psi_1}$, and a $2\pi$-pulse process acting on the antisymmetric
eigenstate $\ket{\varphi_1}$.

\chapter*{Conclusions}

In this work, we explored numerically two of the most useful methods for
transferring population: $\pi$-pulse and adiabatic passage schemes. In
particular, we applied these methods to transfer population in an atom + cavity
system, and to transfer the atomic coherence in a two-atom + cavity system. We
discussed some important characteristics, as well as advantages and
disadvantages of these two methods.

The numerical simulations revealed very interesting resonance-like features in
the adiabatic passage scheme. By choosing appropriately some parameters of the
system, like the ratio of the pulses and the pulse delay, we achieved very high
transfer efficiencies. To find out different ways of choosing the right values
for these parameters and reduce the failure probability was the main goal of
this dissertation.

It is well known that when a system undergoes \textit{no} spontaneous emission,
we can obtain transfer efficiencies of $100$\% by using $\pi$-pulses methods.
However, when the system undergoes spontaneous emission, the transfer efficiency
of the method decreases as the radiative decay rate increases. Our numerical
results showed that efficiencies near to $100$\% were possible to obtain by
using adiabatic passage methods for a particular set of parameters, with or
without radiative decay.

For the atom + cavity system and no spontaneous emission, we obtained
efficiencies of $99.9$\% transferring the population via adiabatic passage. The
Sech pulses showed to be more effective than the Gaussian pulses, since they
required less relative energy ($\Omega/\textsl{g}$) and less interaction time
($\Delta t/\sigma_{\textsl{g}}$). When we consider the effects of spontaneous
emission, the transfer efficiency of the $\pi$-pulses method decreased
accordingly to the increment in the rate of radiative decay. In particular, for
$\Gamma=0.10$, the Gaussian pulses had an efficiency of $93.5$\%, while the Sech
pulses presented a maximum efficiency of $92.6$\%. However, in the adiabatic
passage scheme was possible to find a set of parameters for which the transfer
efficiency achieved a $99.0$\%. This set was $\{\Omega=2.75, \:\sigma=3.09,
\:\textsl{g}=1.00, \:\sigma_{\textsl{g}}=2.48, \:\Delta t=5.29\}$. Gaussian
pulses proved to be more efficient this time than Sech pulses.

By comparing the ratio of the Gaussian pulses in
adiabatic passage, with the ratio of the Gaussian pulses in the $\pi$-pulses
method, and their respective interaction times, we observed that adiabatic
passage was more efficient transferring population than $\pi$-pulses. For Sech
pulses to achieve the same transfer efficiency, it is required to use more
energy and more time delay.

We also observed that, in adiabatic passage methods, the efficiency can be
optimized as much as we want by increasing the values of the Rabi frequencies,
or equivalently, the widths of the pulses. Thus, for bigger Rabi frequencies,
bigger efficiencies. Because this is something that has no experimental worth,
we examined here those values of the Rabi frequency and time delay that
can be reproduced in a laboratory, and for which very good transfer efficiency
values could be obtained. So we considered a \textit{reasonable} value of
$\log_{10}p=-2.00$ as a goal for our numerical simulations.

We obtained also interesting results transferring the coherence in the two-atom
+ cavity system. With no spontaneous emission, efficiencies of $99.9$\% were
achieved by using the adiabatic passage scheme with Gaussian and Sech pulses.
However this time, Gaussian pulses required less energy and less interaction
time than Sech pulses to obtain such transfer efficiency. When spontaneous
emission was included, Sech showed a better performance than Gaussian pulses
with a comparatively small ratio of the pulses. A transfer efficiency of
$99.1$\% was achieved for the set of parameters $\{\Omega_{10}=3.80,\:
\sigma_1=4.40,\: \Omega_{20}=1.00,\: \sigma_2=5.00,\: \textsl{g}=1.00,\:
\Delta t=11.48\}$.

A simple but very useful analytical model used to better understand the transfer
efficiencies in the adiabatic passage scheme was introduced in Ch.~\ref{Ch:03}.
When the widths of the pulses are unequal, we do not expect adiabatic passage to
work. However, we still found some cases for which a high transfer efficiency
were obtained. This model described ``qualitatively'' the dependence of the
failure probability on the product of parameters $\Omega\;T$. We confirmed that
for large values of $\Omega\;T$ the probability failure power-law decreased.

Finally, we examined the two-atom + cavity system by using an alternative and
simplified model based on the superposition of symmetric and antisymmetric
eigenstates. We found possible to qualitatively explain the transfer of the
Zeeman coherence between two atoms, by using an adiabatic passage method for the
symmetric state, and a $2\pi$-pulses process for the antisymmetric state.

\backmatter
\singlespacing

\begin{thebibliography}{OBRW96}

\bibitem[AE75]{Allen:Book}
Leslie Allen and J.~H. Eberly, \emph{Optical resonance and two-level atoms},
  Wiley, New York, 1975.

\bibitem[Blo46]{Bloch:PR70}
F.~Bloch, \emph{Nuclear induction}, Phys.\ Rev. \textbf{70} (1946), no.~7,
  460--474.

\bibitem[BS40]{Bloch:PR57}
F.~Bloch and A.~Siegert, \emph{Magnetic resonance for nonrotating fields},
  Phys.\ Rev. \textbf{57} (1940), 522--527.

\bibitem[DH98]{Drese:EPJD3}
K.~Drese and M.~Holthaus, \emph{Perturbative and nonperturbative processes in
  adiabatic population transfer}, Eur.\ Phys.\ J. D. \textbf{3} (1998), 73--86.

\bibitem[DP76]{Davis:JCP64}
Jon~P. Davis and Philip Pechukas, \emph{Nonadiabatic transitions induced by a
  time-dependent hamiltonian in the semiclassical/adiabatic limit: The
  two-state case}, J. Chem.\ Phys. \textbf{64} (1976), no.~8, 3129--3137.

\bibitem[Elg80]{Elgin:PL80A}
J.~N. Elgin, \emph{Semiclassical formalism for the treatment of three-level
  systems}, Phys.\ Lett. \textbf{80A} (1980), no.~2, 140--142.

\bibitem[EWG76]{Einwohner:PRA14}
T.~H. Einwohner, J.~Wong, and J.~C. Garrison, \emph{Analytical solutions for
  laser excitation of multilevel systems in the rotating-wave approximation},
  Phys.\ Rev.\ A \textbf{14} (1976), no.~4, 1452--1456.

\bibitem[FVH57]{Feynman:JAP57}
Richard~P. Feynman, Frank~L. Vernon, and Robert~W. Hellwarth, \emph{Geometrical
  representation of the schr\"odinger equation for solving maser problems}, J.\
  Appl.\ Phys. \textbf{28} (1957), no.~1, 49--52.

\bibitem[FYL00]{Fleischhauer:OC179}
M.~Fleischhauer, S.~F. Yelin, and M.~D. Lukin, \emph{How to trap photons?
  storing single-photon quantum states in collective atomic excitations}, Opt.\
  Commun. \textbf{179} (2000), 395--410.

\bibitem[HE81]{Hioe:PRL81}
F.~T. Hioe and J.~H. Eberly, \emph{\textit{N}-level coherence vector and higher
  conservation laws in quantum optics and quantum mechanics}, Phys.\ Rev.\
  Lett. \textbf{47} (1981), no.~12, 838--841.

\bibitem[HE82]{Hioe:PRA82}
\bysame, \emph{Nonlinear constants of motion for three-level quantum systems},
  Phys.\ Rev.\ A \textbf{25} (1982), no.~4, 2168--2171.

\bibitem[Hio83]{Hioe:PL83}
F.~T. Hioe, \emph{Theory of generalized adiabatic following in multilevel
  systems}, Phys.\ Lett. \textbf{99A} (1983), no.~4, 150--155.

\bibitem[KGHB89]{Kuklinski:PRA40}
J.~R. Kuklinski, U.~Gaubatz, F.~T. Hioe, and K.~Bergmann, \emph{Adiabatic
  population transfer in a three-level system driven by delayed laser pulses},
  Phys.\ Rev.\ A \textbf{40} (1989), no.~11, 6741--6744.

\bibitem[LS96]{Laine:PRA96}
Timo~A. Laine and Stig Stenholm, \emph{Adiabatic processes in three-level
  systems}, Phys.\ Rev.\ A \textbf{53} (1996), no.~4, 2501--2512.

\bibitem[Mes99]{Messiah}
Albert Messiah, \emph{Quantum mechanics}, Dover, Mineola, New York, 1999.

\bibitem[Mor64]{Morris:PR133}
R.~J. Morris, \emph{Theory of adiabatic rapid passage for three equally spaced
  levels}, Phys.\ Rev. \textbf{133} (1964), no.~3A, A740--A750.

\bibitem[MT95]{Marion}
Jerry~B. Marion and Stephen~T. Thornton, \emph{Classical dynamics of particles
  and systems}, fourth ed., Saunders College Pub., Fort Worth, 1995.

\bibitem[MYM94]{Morin:PRL73}
S.~E. Morin, C.~C. Yu, and T.~W. Mossberg, \emph{Strong atom-cavity coupling
  over large volumes and the observation of subnatural intracavity atomic
  linewidths}, Phys.\ Rev.\ Lett. \textbf{73} (1994), no.~11, 1489--1492.

\bibitem[OBRW96]{Oppo}
G-L Oppo, S.~M. Barnett, E.~Riis, and M.~Wilkinson (eds.), \emph{Quantum
  dynamics of simple systems: the forty-fourth scottish universities summer
  school in physics}, Scottish Universities Summer School in Physics,
  Edinburgh, 1996.

\bibitem[PGCZ95]{Pellizzari:PRL95}
T.~Pellizzari, S.~A. Gardiner, J.~I. Cirac, and P.~Zoller, \emph{Decoherence,
  continuous observation, and quantum computing: a cavity qed model}, Phys.\
  Rev.\ Lett. \textbf{75} (1995), no.~21, 3788--3791.

\bibitem[PMZ{\etalchar{+}}95]{Parkins:PRA51}
A.~S. Parkins, P.~Marte, P.~Zoller, O.~Carnal, and H.~J. Kimble,
  \emph{Quantum-state mapping between multilevel atoms and cavity light
  fields}, Phys.\ Rev.\ A \textbf{51} (1995), no.~2, 1578--1596.

\bibitem[PMZK93]{Parkins:PRL71}
A.~S. Parkins, P.~Marte, P.~Zoller, and H.~J. Kimble, \emph{Synthesis of
  arbitrary quantum states via adiabatic transfer of zeeman coherence}, Phys.\
  Rev.\ Lett. \textbf{71} (1993), no.~19, 3095--3098.

\bibitem[RRS54]{Rabi:RMP26}
I.~I. Rabi, N.~F. Ramsey, and J.~Schwinger, \emph{Use of rotating coordinates
  in magnetic resonance problems}, Rev.\ Mod.\ Phys. \textbf{26} (1954), no.~2,
  167--171.

\bibitem[SBK{\etalchar{+}}92]{Shore:PRA45}
Bruce~W. Shore, K.~Bergmann, A.~Kuhn, S.~Schiemann, J.~Oreg, and J.~H. Eberly,
  \emph{Laser-induced population transfer in multistate systems: A comparative
  study}, Phys.\ Rev.\ A \textbf{45} (1992), no.~7, 5297--5300.

\bibitem[Shi63]{Shirley:JAP34}
Jon~H. Shirley, \emph{Some causes of resonant frequency shifts in atomic beam
  machines. i. shifts due to other frequencies of excitation}, J.\ Appl.\ Phys.
  \textbf{34} (1963), no.~4, 783--788.

\bibitem[SZ97]{Scully:book}
Marlan~O. Scully and M.~Suhail Zubairy, \emph{Quantum optics}, University
  Press, Cambridge, 1997.

\bibitem[TRK92]{Thompson:PRL68}
R.~J. Thompson, G.~Rempe, and H.~J. Kimble, \emph{Observation of normal-mode
  splitting for an atom in an optical cavity}, Phys.\ Rev.\ Lett. \textbf{68}
  (1992), no.~8, 1132--1137.

\bibitem[VS96]{Vitanov:OC96}
N.~V. Vitanov and S.~Stenholm, \emph{Non-adiabatic effects in population
  transfer in three-level systems}, Opt.\ Commun. \textbf{127} (1996),
  215--222.

\end{thebibliography}
\newcommand{\etalchar}[1]{$^{#1}$}
\providecommand{\bysame}{\leavevmode\hbox to3em{\hrulefill}\thinspace}
\providecommand{\MR}{\relax\ifhmode\unskip\space\fi MR }
\providecommand{\MRhref}[2]{%
  \href{http://www.ams.org/mathscinet-getitem?mr=#1}{#2}
}
\providecommand{\href}[2]{#2}

\appendix
\doublespacing

\chapter{\label{App:changeVariables} Appendix A: Change of variables}

Let us consider the system of three coupled linear differential equations
\begin{subequations}
\label{Eq:C_dots2(A)}
\begin{align}
    \dot{C}_a & = \frac{i}{2} \,\Omega_1(t)
        e^{i\alpha}e^{i\Delta_1 t} \,C_c,               \label{Eq:Ca_dot2(A)} \\
    \dot{C}_b & = \frac{i}{2} \,\Omega_2(t)
        e^{i\beta} e^{i(\Delta_1 + \Delta_2)t} \,C_c,   \label{Eq:Cb_dot2(A)} \\
    \dot{C}_c & = \frac{i}{2} \,\Omega_1^*(t)
        e^{-i\alpha} e^{-i\Delta_1 t} \,C_a + \frac{i}{2} \,\Omega_2^*(t)
        e^{-i\beta} e^{-i(\Delta_1 + \Delta_2)t} \,C_b. \label{Eq:Cc_dot2(A)}
\end{align}
\end{subequations}
It is possible to eliminate the exponentials from Eqs.~\eqref{Eq:C_dots2(A)} by
making the following change of variables
\begin{equation}
    C_a = e^{i\phi_a t} \,\widetilde{C}_a, \quad
    C_b = e^{i\phi_b t} \,\widetilde{C}_b, \quad
    C_c = e^{i\phi_c t} \,\widetilde{C}_c.
    \label{Eq:newVar}
\end{equation}
By differentiating with respect to time, we obtain
\begin{subequations}
\label{Eq:newVardots}
\begin{align}
    \dot{C}_a & = e^{i\phi_a t}  \,\dot{\widetilde{C}}_a
        + i \phi_a e^{i\phi_a t} \,\widetilde{C}_a, \\
    \dot{C}_b & = e^{i\phi_b t}  \,\dot{\widetilde{C}}_b
        + i \phi_b e^{i\phi_b t} \,\widetilde{C}_b, \\
    \dot{C}_c & = e^{i\phi_c t}  \,\dot{\widetilde{C}}_c
        + i \phi_c e^{i\phi_c t} \,\widetilde{C}_c.
\end{align}
\end{subequations}
Now plugging Eqs.~\eqref{Eq:newVar} and~\eqref{Eq:newVardots} into
Eqs.~\eqref{Eq:C_dots2(A)} gives
\begin{subequations}
\label{Eq:C_tildes(A)}
\begin{align}
    \dot{\widetilde{C}}_a & = -i \phi_a \widetilde{C}_a
        + i \,\frac{\Omega_1(t)}{2} \,e^{i\alpha}
        e^{i(\Delta_1 + \phi_c - \phi_a) t} \,\widetilde{C}_c,
    \label{Eq:Ca_tilde_dot(A)} \\
    \dot{\widetilde{C}}_b & = -i \phi_b \widetilde{C}_b
        + i \,\frac{\Omega_2(t)}{2} \,e^{i\beta}
        e^{i(\Delta_1 + \Delta_2 + \phi_c - \phi_b) t} \,\widetilde{C}_c,
    \label{Eq:Eq:Cb_tilde_dot(A)} \\
    \dot{\widetilde{C}}_c & = -i \phi_c \widetilde{C}_c
        + i \,\frac{\Omega_1^*(t)}{2} \,e^{-i\alpha}
        e^{-i(\Delta_1 + \phi_c - \phi_a) t} \,\widetilde{C}_a \notag \\
        & \qquad + i \,\frac{\Omega_2^*(t)}{2} \,e^{-i\beta}
        e^{-i(\Delta_1 + \Delta_2 + \phi_c - \phi_b) t} \,\widetilde{C}_b.
    \label{Eq:Eq:Cc_tilde_dot(A)}
\end{align}
\end{subequations}
Because we want the phases to cancel out, they must satisfy the conditions
\begin{subequations}
\begin{align}
               \Delta_1 + \phi_c - \phi_a  = 0, \\
    \Delta_1 + \Delta_2 + \phi_c + \phi_b  = 0.
\end{align}
\end{subequations}
Solving this system of equations for $\phi_a$, $\phi_b$, and $\phi_c$ gives
\begin{equation}
    \phi_a = \Delta_1 + \phi_c,            \qquad
    \phi_b = \Delta_1 + \Delta_2 + \phi_c, \qquad
    \phi_c \text{ arbitrary}.
    \label{Eq:phases(A)}
\end{equation}
By choosing $\phi_c = 0$, and plugging these phases into
Eqs.~\eqref{Eq:C_tildes(A)}, we get
\begin{subequations}
\label{Eq:C_deltas(A)}
\begin{align}
    i\dot{C}_a(t) & = \Delta_1 C_a(t) -
                      \frac{\Omega_1(t)}{2} e^{i\alpha} \,C_c(t),   \\
    i\dot{C}_b(t) & = (\Delta_1 + \Delta_2) C_b(t) -
                      \frac{\Omega_2(t)}{2} e^{i\beta}\,C_c(t),   \\
    i\dot{C}_c(t) & = -\frac{\Omega_1^*(t)}{2} e^{-i\alpha} \,C_a(t)
                      -\frac{\Omega_2^*(t)}{2} e^{-i\beta} \,C_b(t).
\end{align}
\end{subequations}
Finally, the phases $\alpha$ and $\beta$ allow us to flip the signs of the
terms by choosing suitable values. If $\alpha=\beta=\pi$,
Eqs.~\eqref{Eq:C_deltas(A)} become
\begin{subequations}
\begin{align}
    i\dot{C}_a(t) & = \Delta_1 C_a(t) +
                      \frac{\Omega_1(t)}{2} \,C_c(t),   \\
    i\dot{C}_b(t) & = (\Delta_1 + \Delta_2) C_b(t) +
                      \frac{\Omega_2(t)}{2} \,C_c(t),   \\
    i\dot{C}_c(t) & = \frac{\Omega_1^*(t)}{2} \,C_a(t) +
                      \frac{\Omega_2^*(t)}{2} \,C_b(t).
\end{align}
\end{subequations}

\chapter{\label{App:eigensystem} Appendix B: The instantaneous Hamiltonian
eigenstates}

Let us find the eigenvalues and eigenvectors of the instantaneous RWA
Hamiltonian ($\hbar=1$)
\begin{equation}
    H(t) = \half\,
        \begin{bmatrix}
                 0      &      0      & \Omega_1(t) \\
                 0      &      0      & \Omega_2(t) \\
            \Omega_1(t) & \Omega_2(t) &       0
        \end{bmatrix}.
    \label{Eq:Hmatrix02(A)}
\end{equation}
Solving the characteristic equation
\begin{align}
    \det(H-\omega I) =
    \begin{vmatrix}
         -\omega      &       0       & \half\Omega_1 \\
             0        &   -\omega     & \half\Omega_2 \\
        \half\Omega_1 & \half\Omega_2 &   -\omega
    \end{vmatrix}
                     = 0,
    \label{Eq:detH(A)}
\end{align}
or
\begin{align}
    \omega^3 - \frac{1}{4}(\Omega_1^2 + \Omega_2^2) \omega = 0,
    \label{Eq:charPol(A)}
\end{align}
we find the eigenvalues
\begin{equation}
    \omega^+ = +\half \sqrt{\Omega_1^2 + \Omega_2^2}, \qquad
    \omega^0 =  0, \qquad
    \omega^- = -\half \sqrt{\Omega_1^2 + \Omega_2^2},
    \label{Eq:eigenvalues(A)}
\end{equation}
For $\omega^+$ the eigenvector can be determined from the system
\begin{align}
                   \Omega_1 \,z & = \Omega \,x, \notag \\
                   \Omega_2 \,z & = \Omega \,y, \\
    \Omega_1 \,x + \Omega_2 \,y & = \Omega \,z, \notag
\end{align}
where $\Omega = \sqrt{\Omega_1^2 + \Omega_2^2}$. Thus
\begin{equation}
    x = \frac{\Omega_1}{\Omega} z, \quad
    y = \frac{\Omega_2}{\Omega} z,   \quad
    z \text{ arbitrary}.
\end{equation}
Assuming the representation
\begin{equation}
    \ket{a} = \begin{bmatrix} 1 \\ 0 \\ 0 \end{bmatrix}, \quad
    \ket{b} = \begin{bmatrix} 0 \\ 1 \\ 0 \end{bmatrix}, \quad
    \ket{c} = \begin{bmatrix} 0 \\ 0 \\ 1 \end{bmatrix},
    \label{Eq:rep(A)}
\end{equation}
the eigenvector associated with $\omega^+$ can be written as
\begin{subequations}
\label{Eq:eigenvectors(A)}
\begin{align}
    \ket{W^+} & = \frac{1}{\sqrt{2}} \biggl[ \frac{\Omega_1}{\Omega} \ket{a}
        + \frac{\Omega_2}{\Omega} \ket{b} + \ket{c} \biggr].
    \label{Eq:W+(A)}
\intertext{Similarly, for $\omega^0$ and $\omega^-$,
        the associated eigenvectors are}
    \ket{W^0} & = \frac{\Omega_2}{\Omega} \ket{a}
        - \frac{\Omega_1}{\Omega} \ket{b},
    \label{Eq:W0(A)} \\
    \ket{W^-} & = \frac{1}{\sqrt{2}} \biggl[ \frac{\Omega_1}{\Omega} \ket{a}
        + \frac{\Omega_2}{\Omega} \ket{b} - \ket{c} \biggr],
    \label{Eq:W-(A)}
\end{align}
\end{subequations}
respectively. By using the trigonometric relations
\begin{equation}
    \sin\Phi(t) = \frac{\Omega_1(t)}{\Omega(t)}, \quad
    \cos\Phi(t) = \frac{\Omega_2(t)}{\Omega(t)}, \quad
    \tan\Phi(t) = \frac{\Omega_1(t)}{\Omega_2(t)},
\end{equation}
we can rewrite Eqs.~\eqref{Eq:eigenvectors(A)} as follows
\begin{subequations}
\begin{align}
    \ket{W^+} & = \frac{1}{\sqrt{2}} \bigl[ \sin\Phi(t) \ket{a}
        + \cos\Phi(t) \ket{b}  + \ket{c} \bigr],
    \label{Eq:W+trig(A)} \\
    \ket{W^0} & = \cos\Phi(t) \ket{a} - \sin\Phi(t) \ket{b},
    \label{Eq:W0trig(A)} \\
    \ket{W^-} & = \frac{1}{\sqrt{2}} \bigl[ \sin\Phi(t) \ket{a}
        + \cos\Phi(t) \ket{b}  - \ket{c} \bigr].
    \label{Eq:W-trig(A)}
\end{align}
\end{subequations}

\chapter{\label{App:FHO} Appendix C: The forced harmonic oscillator}

The equation of motion for a particle of mass $m$ moving under the combined
influence of a linear restoring force $-kx$, a resisting force $-b\dot{x}$, and
an external driving force $F(t)$ is given by
\begin{equation}
    m \ddot{x} + b \dot{x} + kx = F(t). \label{Eq:FHO(A)}
\end{equation}
The most general solution to this differential equation is composed of the
\textit{complementary} and \textit{particular} solutions (see~\cite{Marion}):
\begin{equation}
    x(t) = x_c(t) + x_p(t). \label{Eq:FHO_Solution(A)}
\end{equation}

\section*{Complementary solution}

The complementary solution has the general form
\begin{equation}
    x_c(t) = e^{-\gamma \,t} \biggl[ C_1 \,\exp\biggl( { i\sqrt{\omega_0^2
        - \gamma^2} \,t} \biggr) +   C_2 \,\exp\biggl( {-i\sqrt{\omega_0^2
        - \gamma^2} \,t} \biggr) \biggr],
    \label{Eq:cSolution1(A)}
\end{equation}
where
\begin{equation}
    \gamma   = \frac{b}{2m}, \qquad \text{and} \qquad
    \omega_0 = \sqrt{\frac{k}{m}}.
    \label{Eq:parameters1(A)}
\end{equation}
There are three general cases of interest:
\begin{alignat}{2}
    &\text{Underdamping:}    & \quad \gamma^2 &< \omega_0^2 \notag \\
    &\text{Critical damping:}& \quad \gamma^2 &= \omega_0^2 \notag \\
    &\text{Overdamping:}     & \quad \gamma^2 &> \omega_0^2 \notag
\end{alignat}
For the case of underdamped motion, the exponents in the brackets of
Eq.~\eqref{Eq:cSolution1(A)} are imaginary, and the solution can be written as
\begin{equation}
    x_c(t) = \bigl[A\,\sin\omega_1 t + B\,\cos\omega_1 t \bigr],
    \label{Eq:underD(A)}
\end{equation}
where
\begin{equation}
    \omega_1 = \sqrt{\omega_0^2 - \gamma^2}. \label{Eq:omega1}
\end{equation}
For the case of critical damping, the roots of the auxiliary equation ($r^2 +
2\gamma\,r + \omega_0^2 r = 0$) are equal, and the complementary function must
be written in the form
\begin{equation}
    x_c(t) = e^{-\gamma \,t} \bigl( A + B t). \label{Eq:criticalD(A)}
\end{equation}
Finally, for the case of overdamped motion, the exponents in the brackets of
Eq.~\eqref{Eq:cSolution1(A)} become real quantities:
\begin{equation}
    x_c(t) = e^{-\gamma \,t} \bigl[ C_1 e^{ \omega_1 t}
        + C_2 e^{-\omega_1 t} \bigr].
    \label{Eq:overD(A)}
\end{equation}

\section*{Particular solution}

Now we seek a particular solution to the inhomogeneous equation in the form
\begin{equation}
    x_p(t) = \int_{-\infty}^{\infty} F(t')G(t,t') \,dt',
    \label{Eq:pSolution1(A)}
\end{equation}
where $F(t')$ is the applied external force (inhomogeneity), and $G(t,t')$ is
the Green's function for Eq.~(\ref{Eq:FHO(A)}). We define
\begin{equation}
    G(t,t') = \begin{cases}
                  \frac{1}{m\omega_1} \,e^{-\gamma\,(t-t')}
                  \sin[\omega_1(t-t')] & \quad \text{if $t \geq t'$}, \\
                  0                    & \quad \text{if $t   <  t'$}.
              \end{cases}
    \label{Eq:GreenFunction}
\end{equation}
Then, the particular solution can be expressed as
\begin{equation}
    x_p(t) = \int_{-\infty}^t \frac{F(t')}{m\omega_1}\, e^{-\gamma\,(t-t')}
        \sin[\omega_1(t-t')] \,dt'.
    \label{Eq:FHO_pSolution2(A)}
\end{equation}

\section*{The simple analytical model}

For the simple analytical model studied in Sec.~\ref{Sec:populationTransfer}, we
obtained a second-order linear differential equation of the form
\begin{equation}
    \frac{d\,^2}{d\tau^2}\delta\theta[\tau] + \delta\theta[\tau]
        = -\frac{d}{d\tau}\Phi[\tau].
    \label{Eq:deltasFHO(A)}
\end{equation}
On comparing Eqs.~\eqref{Eq:FHO(A)} and~\eqref{Eq:deltasFHO(A)}, we have
\begin{equation}
    m = 1, \quad b = 0, \quad k = 1, \quad F(\tau) = -\frac{d}{d\tau}\Phi.
    \label{Eq:parameters2(A)}
\end{equation}
Upon substitution of these parameters into Eq.~\eqref{Eq:parameters1(A)} and
Eq.~\eqref{Eq:omega1}, we get
\begin{equation}
    \omega_0 = 1, \quad \gamma = 0, \quad \omega_1 = 1.
    \label{Eq:parameters3(A)}
\end{equation}
For this case we have $\omega_0^2 > \gamma^2$ (underdamping). Therefore the most
general solution to Eq.~\eqref{Eq:deltasFHO(A)} can be expressed as
\begin{equation}
    \delta\theta[\tau] = A \,\sin[\tau] + B \,\cos[\tau]
         - \int_{-\infty}^{\tau} \sin[\,\tau-\tau'\,]
         \frac{d\Phi(\tau')}{d\tau'} \,d\tau'.
    \label{Eq:deltasFHO_solution(A)}
\end{equation}

\end{document}